\documentclass[]{imag-ms-template}

\usepackage{graphicx, caption, subcaption}
\usepackage{multirow}

\newtheorem{proposition}{Proposition}

\graphicspath{{figures/}}

\title{ Detecting Mild Traumatic Brain Injury with MEG Scan Data: One-vs-K-Sample Tests} 

\author{Jian Zhang,$^{1\ast}$ Gary Green,$^{2}$ \\
{\small $^{1}$School of Mathematics, Statistics and Actuarial Science, University of Kent,}\\
{\small Canterbury, Kent CT2 7NF, UK}\\
{\small $^{2}$Department of Psychology, University of York, York YO10 5DD, UK}\\
{\small and Innovision IP Ltd, 50 Seymour Street, London W1H 7JG, UK}\\
{\small $^\ast$Correspondence:  j.zhang-79@kent.ac.uk}
}

\newcommand{\Bigskip}{\vspace{0.3 in}}
\newcommand{\bX}{\mbox{\bf X}}

\newcommand{\bY}{\mbox{\bf Y}}
\newcommand{\bR}{\mbox{\bf R}}
\newcommand{\bG}{\mbox{\bf G}}
\newcommand{\bB}{\mbox{\bf B}}
\newcommand{\bF}{\mbox{\bf F}}
\newcommand{\bH}{\mbox{\bf H}}
\newcommand{\bQ}{\mbox{\bf Q}}
\newcommand{\bx}{\mbox{\bf x}}
\newcommand{\by}{\mbox{\bf y}}

\newcommand{\be}{\mbox{\bf e}}

\newcommand{\argmax}{\mbox{argmax}}

\def\bvep{{\bf{\varepsilon}}}

\def\bY{{\boldsymbol{Y}}}

\newcommand{\R}{\mathbb{R}}

\begin{document} 

\maketitle


\begin{abstract}
 Magnetoencephalography (MEG) scanner has been shown to be more accurate than other medical devices in detecting mild traumatic brain injury (mTBI). However, MEG scan data in certain spectrum ranges can be skewed, multimodal and heterogeneous which can mislead the conventional case-control analysis that requires the data to be homogeneous and normally distributed within the control group. To meet this challenge, we propose a flexible one-vs-K-sample testing procedure for detecting brain injury for a single-case versus heterogeneous controls. The new procedure begins with source magnitude imaging using MEG scan data in frequency domain, followed by region-wise contrast tests for abnormality between the case and controls.  The critical values for these tests are automatically determined by cross-validation. We adjust the testing results for heterogeneity effects by similarity analysis.   An asymptotic theory is established for the proposed test statistic. By simulated and real data analyses in the context of neurotrauma, we show that the proposed test outperforms commonly used nonparametric methods in terms of overall accuracy and ability in accommodating data non-normality and subject-heterogeneity.  
\end{abstract}

\section{Introduction}
  Around 8 to 12 percent of the global population have been estimated to live with traumatic brain injury (TBI) (Frost et al., 2013; James et al., 2019). Although mild TBIs, which include concussions, account for $70–90\%$ of TBI cases, there is no generally accepted standard for diagnosing one. The early identification of mTBI and accurate assessment of recovery after a treatment is vital to ensuring the best treatment and rehabilitation outcomes.  The state of the art in finding a neural signature of mTBI and classifying patterns of neural damages that determine behavioural recovery from early post-injury to sub-acute outcome, is at an early stage of investigation. As clinical assessment tools, such as the Glasgow Coma Scale, which scores a person’s verbal and motor responses, as well as eye opening, are subjective, clinicians often turn to imaging techniques.  The scanners currently used to diagnose these injuries, structural magnetic resonance imaging (sMRI) and computerised tomography (CT), have a less than 10\% detection accuracy and  are not sensitive enough to identify the microscopic damage that is characteristic of mTBIs. 
In contrast, MEG scanner can detect subtle pathology that often goes undetected in individuals with mTBI when using sMRI and CT (Huang et al., 2014).  Huang et al. (2020) and Allen et al. (2021) provided a recent review of use of MEG to assess a brain injury.
 However, MEG scan data in certain spectrum ranges can be skewed, multimodal and heterogeneous which raises a critical issue to the conventional case-control analysis that requires the data to be homogeneous and normally distributed.
In this paper we address the above issue through MEG-based one-vs-K-sample (OK) hypothesis tests for a single case compared to a group of healthy controls. 

 MEG is a non-invasive functional brain-mapping device that detects magnetic fields induced by neuronal electrical activity, with millisecond time scale resolution (Schwartz et al., 2010).
Statistical modelling of MEG scan data can be found, for example, in Zhang et al. (2014).
The MEG scans for brain injury are obtained when a subject is in a resting state with eyes open and eyes closed, and ideally repeated twice with sample rate 1 kHz for $8$ minutes. Prior to acquisition, empty room data are acquired so that correct noise removal procedures can be exploited. The voxel-wise MEG source magnitude images were obtained using a high-resolution inverse imaging method called Fast-VESTAL (Huang et al., 2012).
Brain activity is often described in terms of the amount of oscillatory activity in different
frequency bands, for example, the delta band describes slow waves with frequencies between 1 and 4 Hz while the gamma band is for the spectrum ranging from $30$ to $80$ Hz. The bands with other ranges include theta (5-7 Hz), alpha (8-12 Hz) and beta (15-29 Hz).
Huang et al. (2012) measured functional changes using MEG in both civilian and military personnel with mTBI showing an increase in delta power after head injury at both the group and individual levels. 
These changes in low frequency, considered pathologic in otherwise healthy adults, have been associated with brain lesions, Parkinson's disease, hypoxia, schizophrenia and states associated with abnormal or damaged brain tissue, in addition to mTBI (Knyazev, 2012). These facts suggest MEG measured delta wave power is a compelling diagnostic and prognostic tool for concussion in human brain (Davenport et al., 2022). Although non-invasive detection of gamma-band activity is challenging since coherently active source areas are small at such frequencies, Huang et al. (2020) revealed abnormal resting-state gamma activity in mTBI by using MEG scans.

 In the context of multiple-sample studies, statistical
 tests such as two-sample t-test, Crawford-Garthwaite p-value test, two-sample or K-sample Anderson-Darling (AD) test and Disco analysis can be employed to infer the presence of cognitive impairments in a patient (Crawford and Garthwaite, 2007; Scholz and Stephens, 1987; Rizzo and Gabor, 2010; Huang et al., 2016). The Crawford-Garthwaite test is a Bayesian t-test for a mean shift in a case compared to controls under a normality assumption. The two-sample and K-sample AD tests involve determining whether multiple samples are each drawn from the same distribution. Disco analysis extended the classical multivariate analysis of variance (MANOVA) with multiple samples. Like many non-parametric tests although the AD and Disco have not made strong distributional assumptions, tests based on specific distributional assumptions are generally believed to be more powerful than non-parametric techniques if the distributional assumptions can be validated. Researchers have witnessed a lot of development in finite mixture and non-parametric modelling in the context of one-sample tests, for example, the EM tests of Chen and Li (2009) and Chen et al. (2012), goodness of fit tests of Wichitchan et al. (2019), among others. However, not all these tests are applicable directly to the problem of OK testing in the context of a single case against multiple controls. 

\begin{center}
\begin{figure}[htp]
\centering
\includegraphics[height=1.in,width=0.5\textwidth]{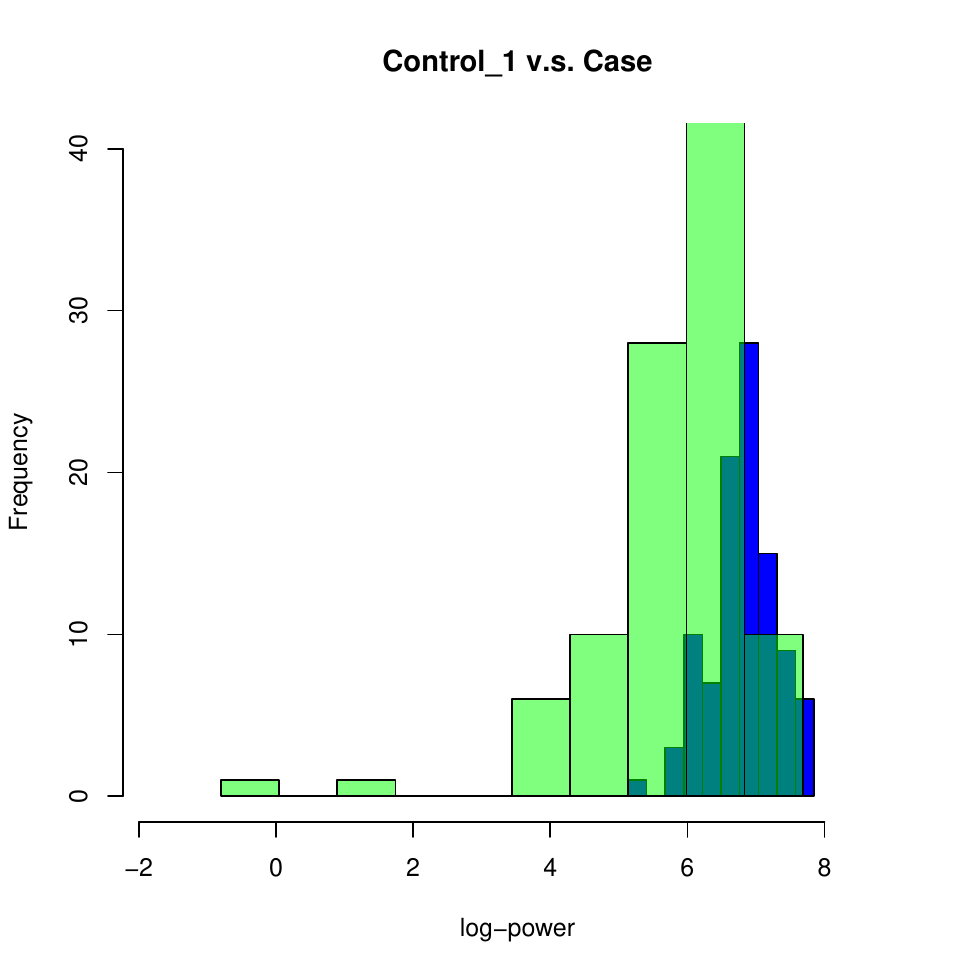}\hfill
\includegraphics[height=1.in,width=0.5\textwidth]{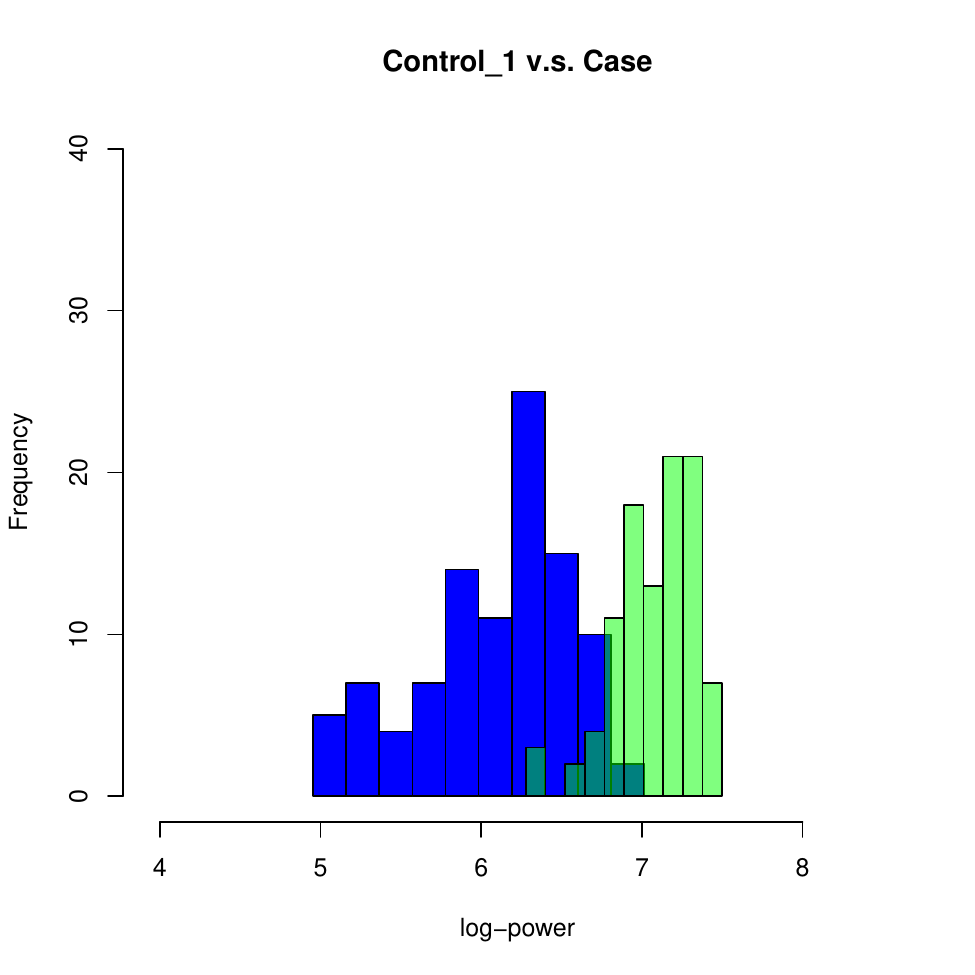}\\
\includegraphics[height=1.in,width=0.5\textwidth]{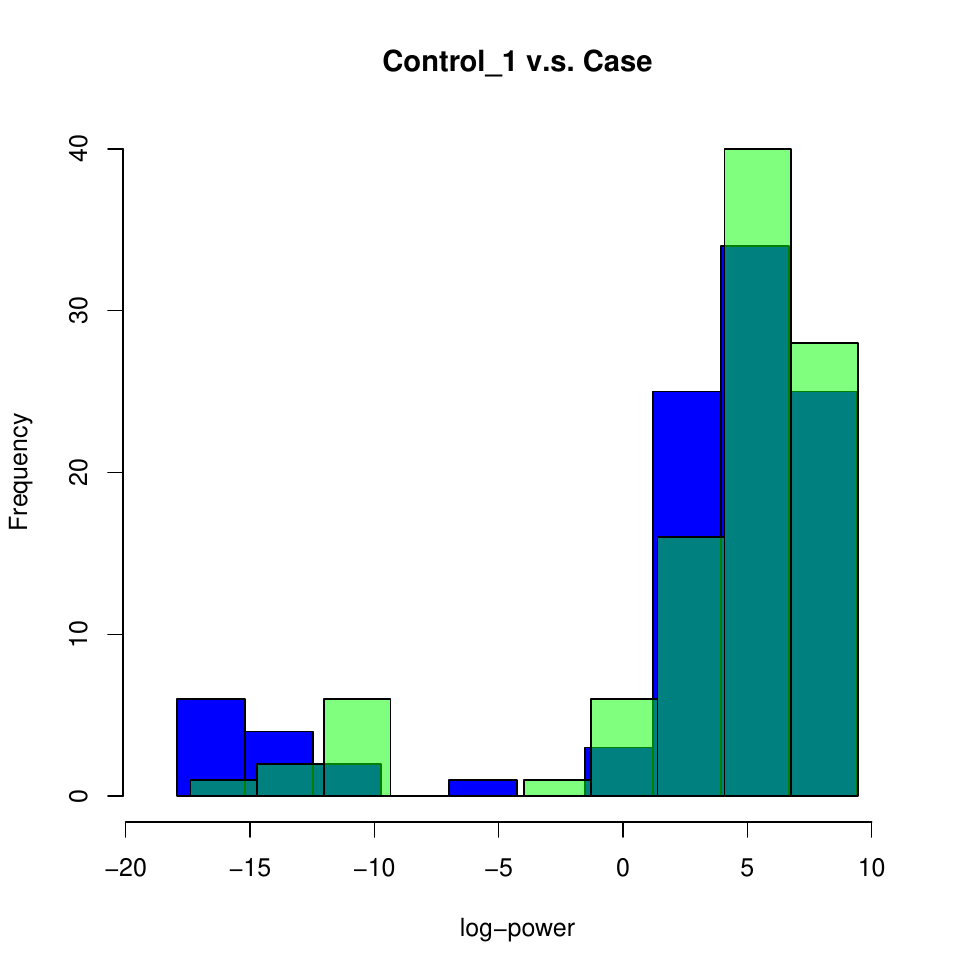}\hfill
\includegraphics[height=1.in,width=0.5\textwidth]{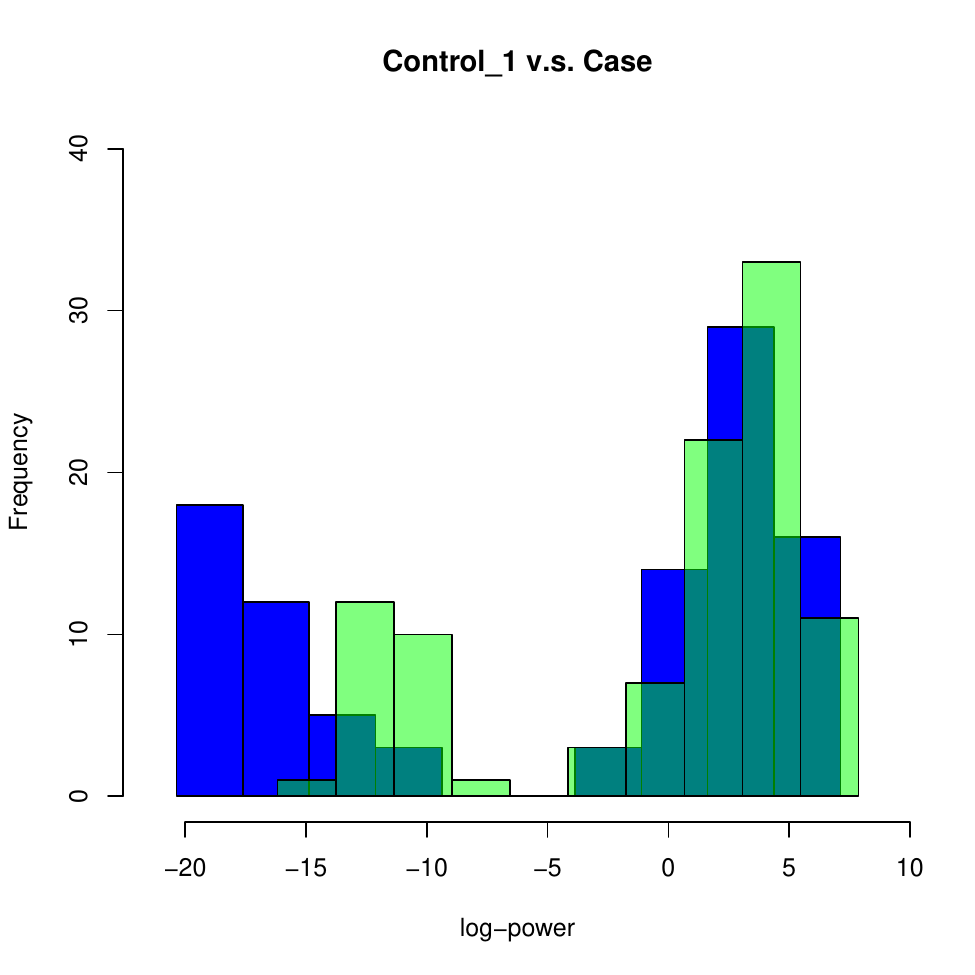}\\
\includegraphics[height=1.in,width=0.5\textwidth]{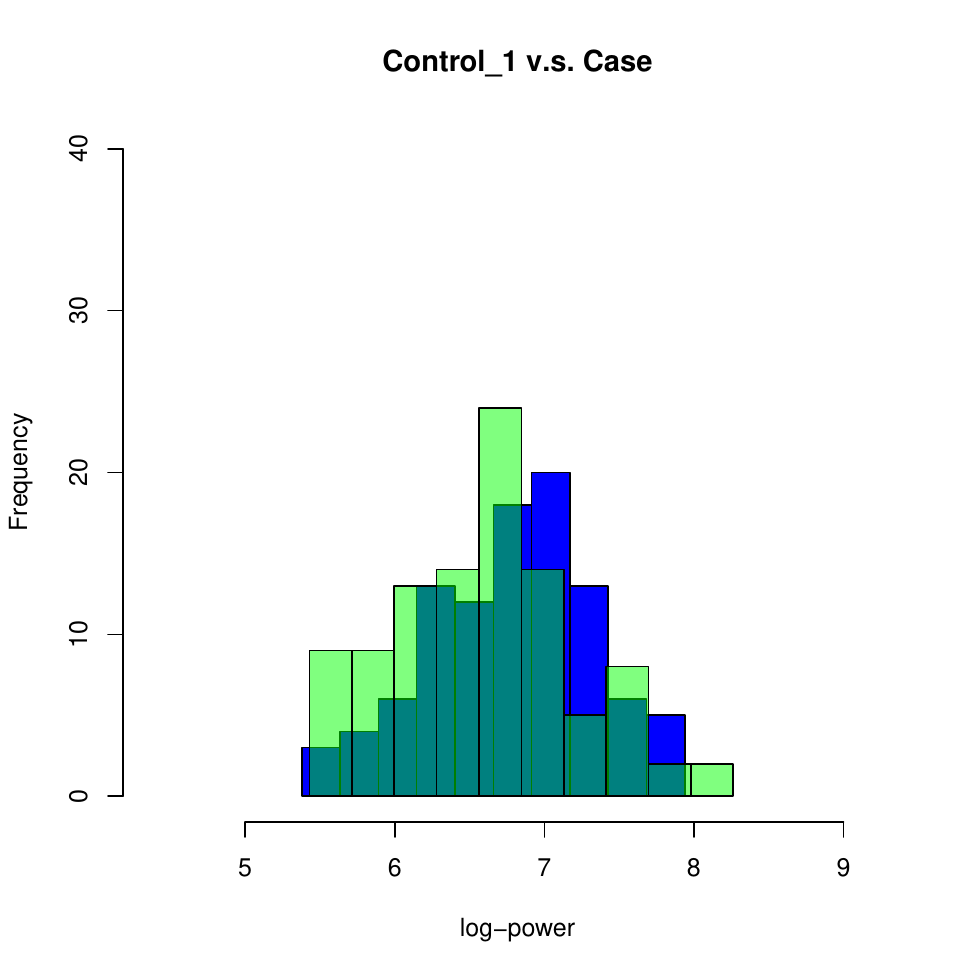}\hfill
\includegraphics[height=1.in,width=0.5\textwidth]{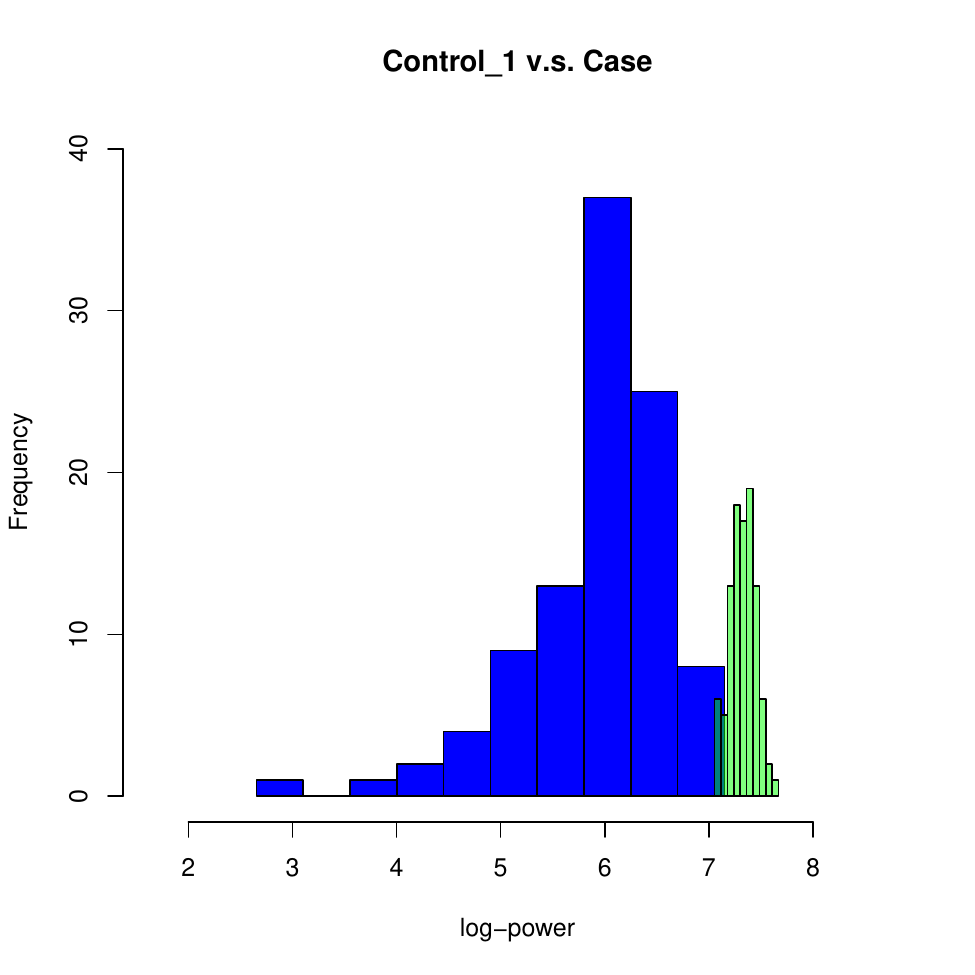}\\
\includegraphics[height=1.in,width=0.5\textwidth]{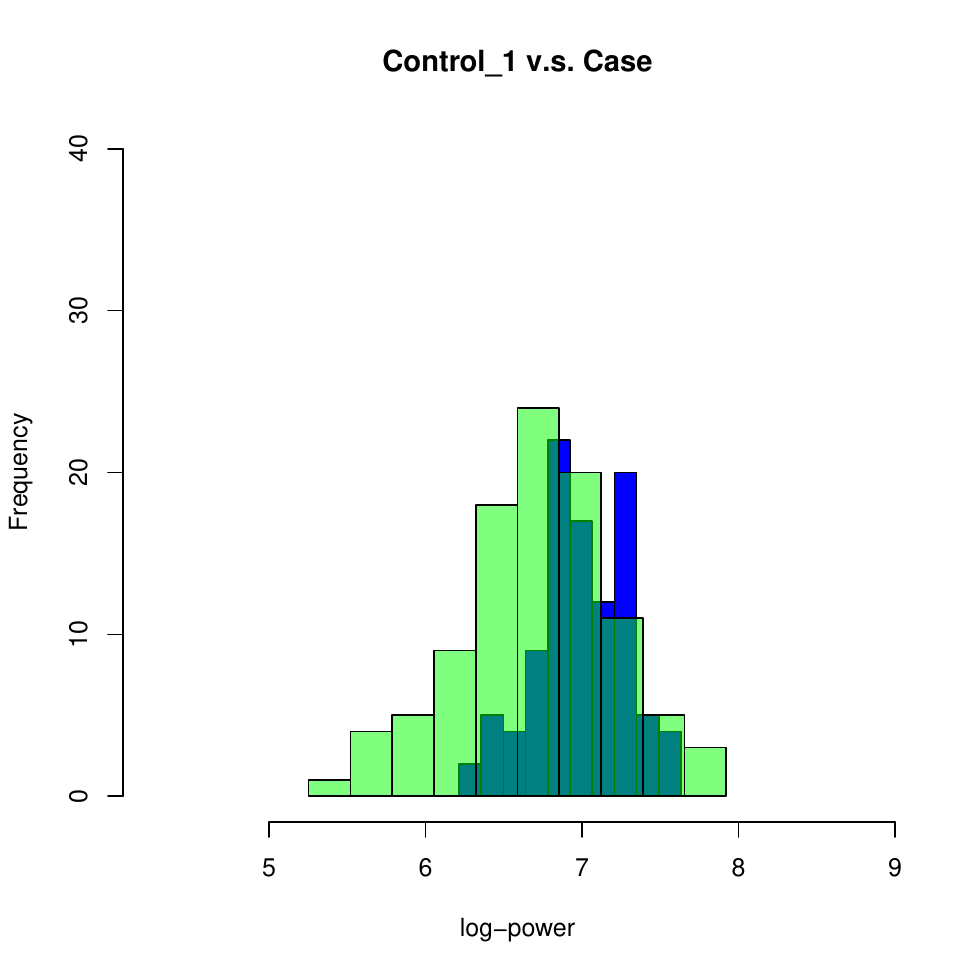}\hfill
\includegraphics[height=1.in,width=0.5\textwidth]{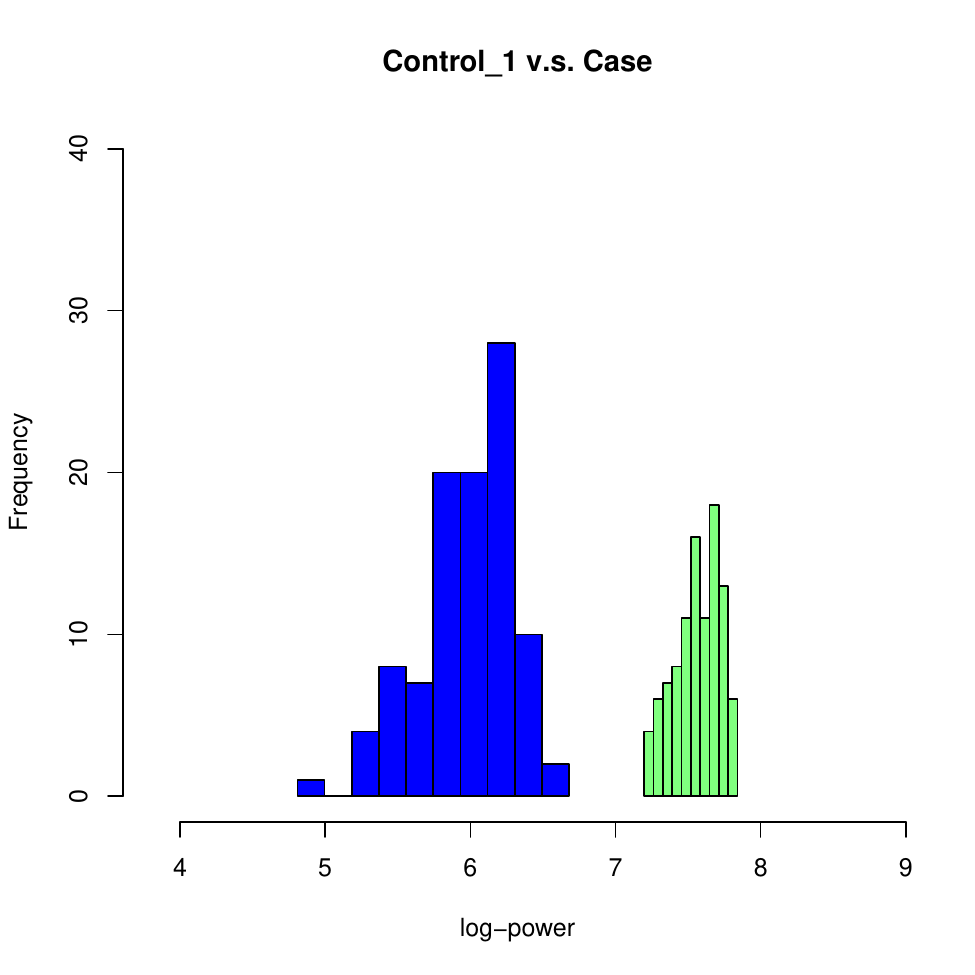}\\
\caption{\small{ Histogram plots for a case and one of controls in the delta and gamma bands in areas 3 (ctx-lh-caudalmiddlefrontal), 6 (ctx-lh-frontalpole), 9 (ctx-lh-inferiortemporal) and 12 (ctx-lh-lateraloccipital). Row 1: delta and gamma bands in area 3. Row 2: delta and gamma bands 
in area 6. Row 3: delta and gamma bands in area 9. Row 4: delta and gamma bands in area 12.
 The blue and green coloured histograms are for the control subject 1 and the case respectively.}}
\label{Histogram}
\end{figure}
\end{center}

\begin{center}
\begin{figure}[htp]
\includegraphics[height=1.3in,width=0.5\textwidth]{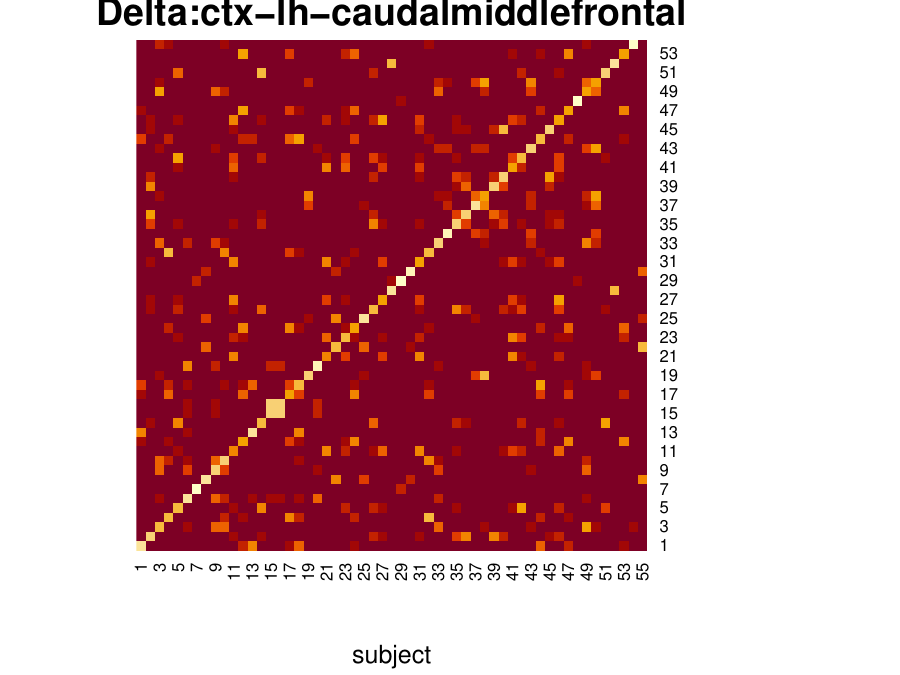}
\hfill
\includegraphics[height=1.3in,width=0.5\textwidth]{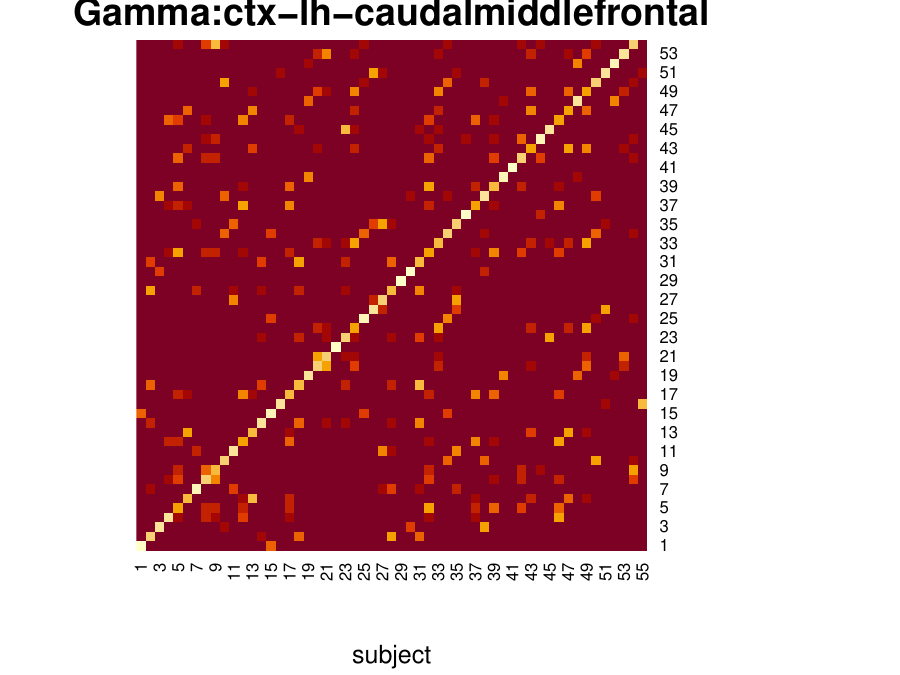}\\
\includegraphics[height=1.3in,width=0.5\textwidth]{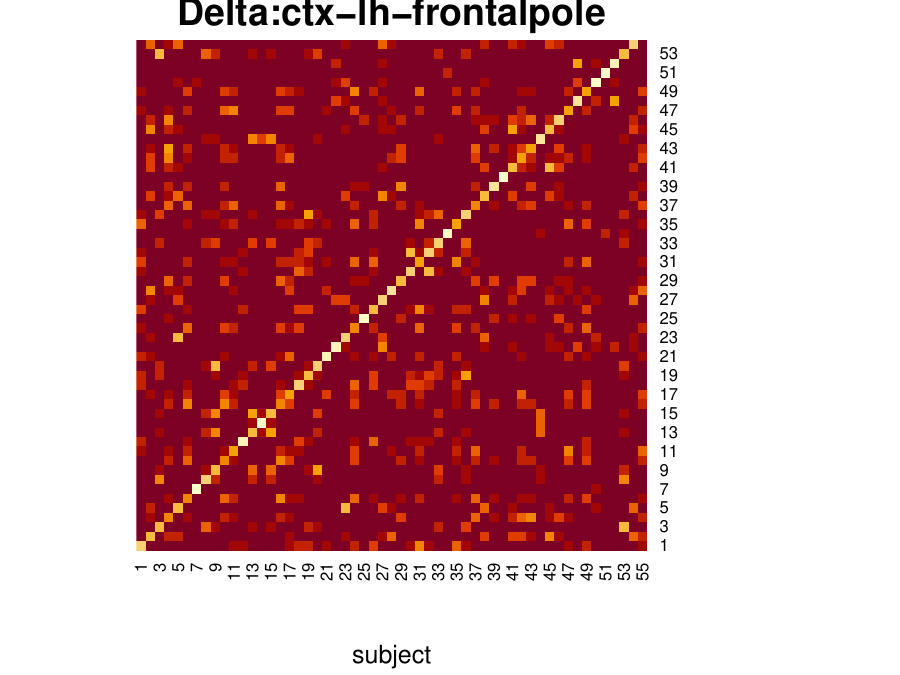}\hfill
\includegraphics[height=1.3in,width=0.5\textwidth]{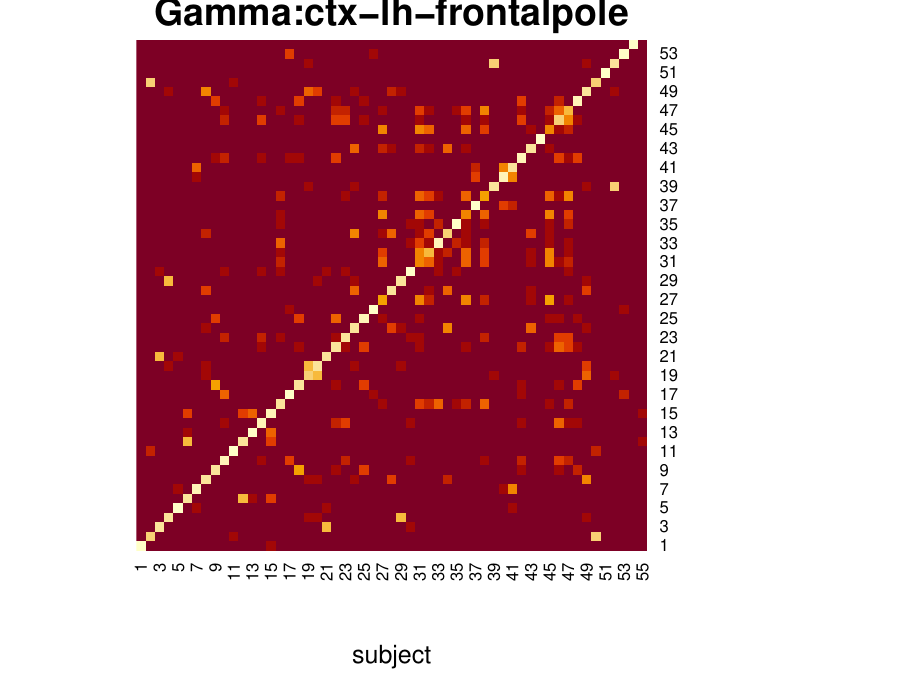}\\
\includegraphics[height=1.3in,width=0.5\textwidth]{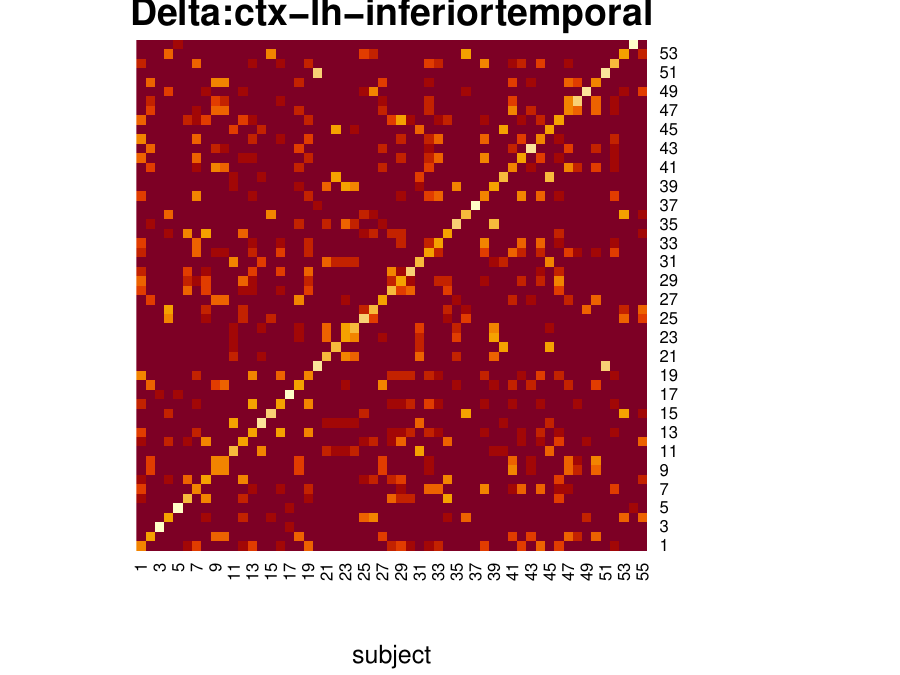}\hfill
\includegraphics[height=1.3in,width=0.5\textwidth]{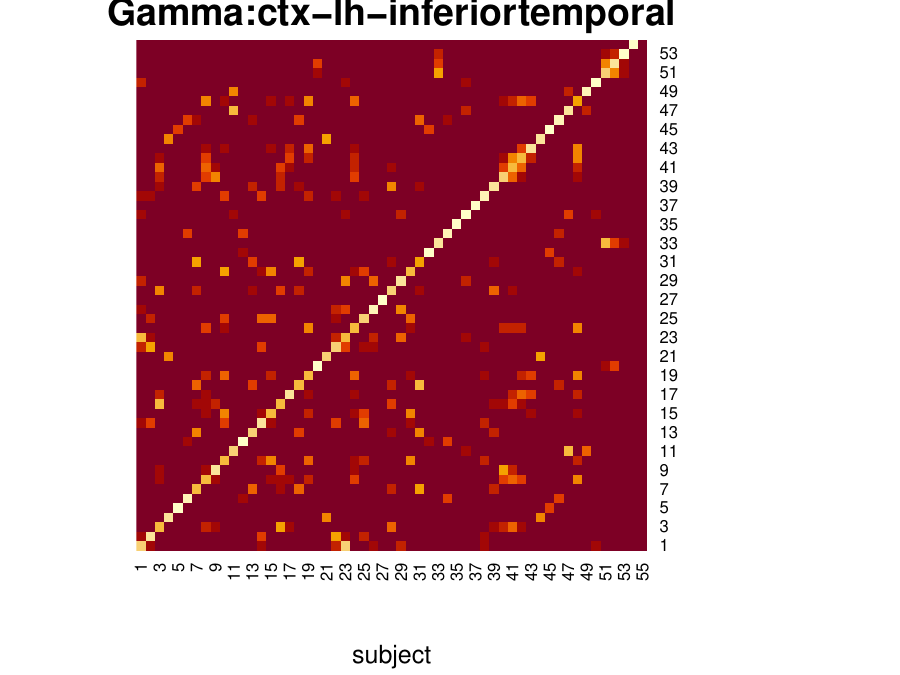}\\
\includegraphics[height=1.3in,width=0.5\textwidth]{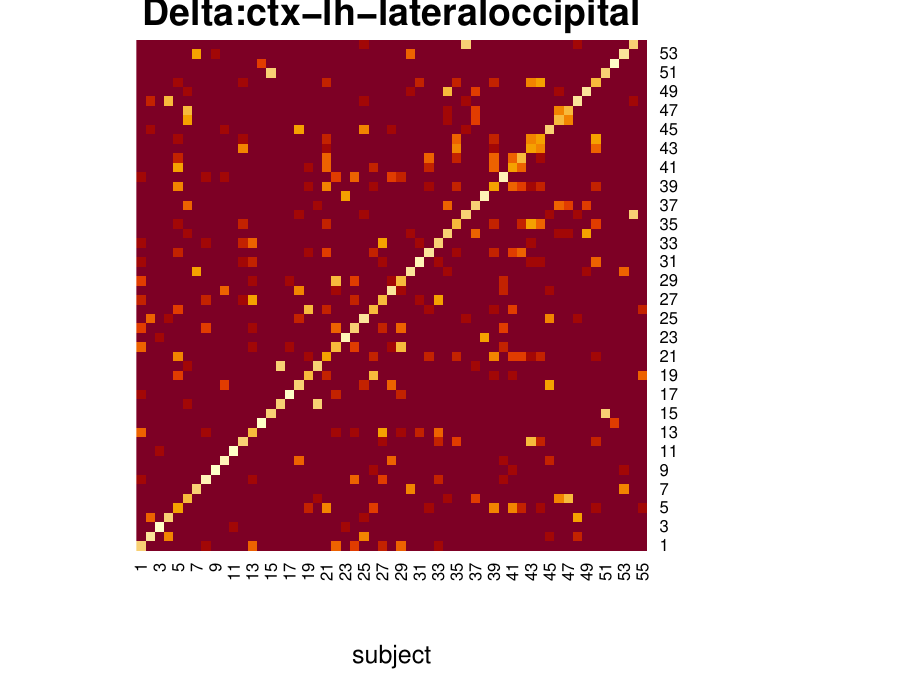}\hfill
\includegraphics[height=1.3in,width=0.5\textwidth]{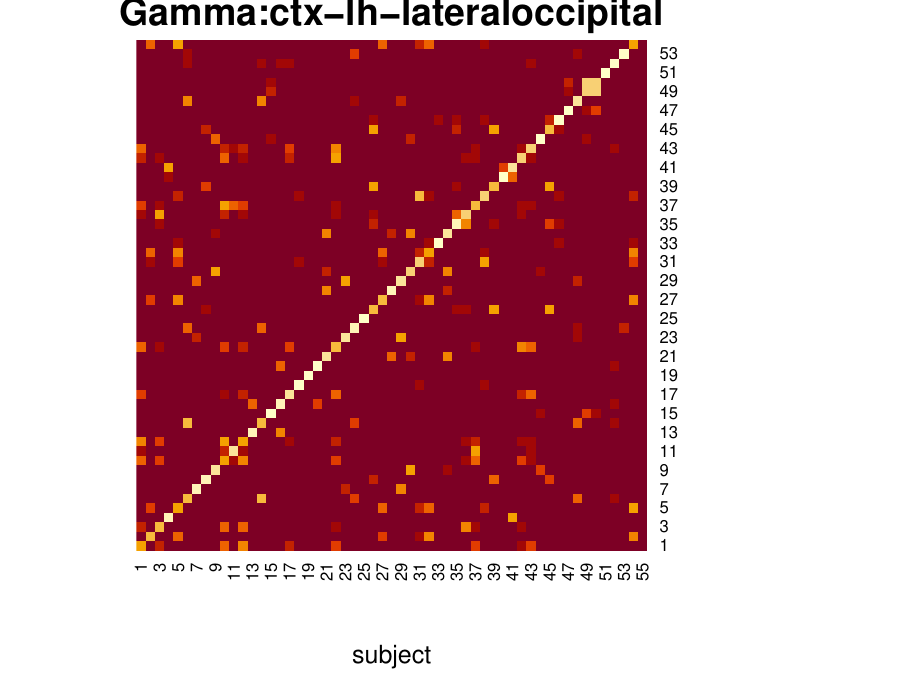}\\
\caption{\small{\noindent Heatmaps for (1-p)-values of pairwise Anderson-Darling tests between the controls and between the controls and the case. The controls are indexed by $1$ to $54$ while the case is indexed by $55$. Rows 1 to 4 are respectively for areas 3, 6, 9 and 12 as described in Figure \ref{Histogram}.  Each row displays the heatmaps of the (1-p)-values of pairwise Anderson-Darling tests for the delta and gamma bands from the left to the right. The color changes from dark red to light white as (1-p)-value decreases from $1$ to $0.$
Dark red color indicates most significant p-values while the light colors stand for less significant p-values. }}
\label{AndersonDarling}
\end{figure}
\end{center}

In a step toward understanding how delta- or gamma-band neural responses to brain injury, the present study concerns distribution changes of this adaptation in the context of source magnitudes/band powers with MEG scan data recently acquired by the Innovision IP Ltd. The data consist of MEG scans for a single testing subject and $54$ age-matched control subjects. In the data, according to the Desikan-Killiany Atlas, the brain was divided into $A=68$ functional regions of interest, indexed by 1, 2,\ldots,34 for the areas in the left hemisphere and by 35,\ldots,68 for the mirror areas in the right hemisphere (Desikan et al., 2006). Voxel-wise, MEG source magnitude/power data over grid points in different spectrum bands were calculated with these scans using the Fast-VESTAL. The average delta and gamma band powers were then calculated in each area and in each epoch for individual subjects, generating multiple $A\times N$ data matrices. 
Our exploratory data analysis raises the following questions for a further statistical analysis, necessitating the development of flexible and adaptable methodologies for the OK testing.
 First, the band power distributions are skewed and multimodal as shown by histogram plots in Figure \ref{Histogram}.
This raises the concern of robustness of the Crawford-Garthwaite test when the underlying distribution family is deviated from normals. Secondly, unlike the traditional case-control studies that requires the assumption of within-group homogeneity, in the current study, we compare a testing subject to a heterogeneous control group. The heterogeneity in the controls is manifested by pairwise AD tests for distributional shift displayed in Figure \ref{AndersonDarling}, where the deep red colour highlights these p-values close to the low limit $0$ whereas the white colour marks these p-values close to the up limit $1$. The p-values increases from $0$ to $1$ in the brightness of colour.
The first $54$ columns in the map demonstrate p-values derived from pairwise AD tests for each control against the remaining controls respectively, while the last column shows p-values for the case against the controls. It can be seen from
 Figure \ref{AndersonDarling} that for most of the pairs of subjects, the p-values derived from AD tests are close to $0$ as the corresponding squares in Figure \ref{AndersonDarling} have colors close to the red rather than yellow or white. 
In addition, for squares near diagonal line, they become white coloured, which mean subjects have p-values of $1$ when compared to themselves.
These facts imply that some control subjects behave very similar to the case when they are tested against the remaining controls. For example, in Figure 2, the Online Supplementary Materials, taking these pairwise p-values as similarity scores, we perform average linkage hierarchical clustering on $55$ subjects. The case will be expected to be the last subject merged into the dendrogram if the target brain area does differ the case and controls.  In the cortical area $9$ and in the gamma band, the case subject is the last to be merged in the dendrogram and well separated from most of the controls. However, the case behaves similar to the controls $35$, $27$ and $54$ in the above area. This finding is not by coincidence as similar phenomena are revealed in other cortical areas. The details are omitted. Furthermore, as anticipated, Figure $2$, the Online Supplementary Materials indicates grouping structures in the controls in the areas $9$ and $12$. 
 Such a heterogeneity occurs in other brain conditions as the assumption of within-group homogeneity is reflected neither in clinical populations nor in the heterogeneous pathological nature of neurodegenerative diseases (Verdi et al., 2021). 
Thirdly, permutation tests are increasingly being used as a reliable method for inference in neuroimaging (Winkler et al., 2016). For example, combining permutation techniques with AD test of a single case against multiple controls, we first pool the individual control samples into a single sample under the homogeneity assumption and then draw multiple random subsets of the same size as the case sample,  against which the AD test is conducted for the case sample, obtaining multiple p-values. The testing result is claimed significant if the average of these p-values is. Unfortunately,
the heterogeneity makes this permutation-based null model biased and causes the AD test over-sensitive to individual differences in the controls.
This demonstrates that
the failure to incorporate heterogeneity in inference may have a negative effect on the accuracy of diagnosis of brain conditions. In particular,
for these group average-based studies (e.g. average case versus average control), effects of heterogeneity have not been taken into account. 
For the current data, averaging non-diagonal entries for each column in the above heatmap shows that the average (($1$-p)-value)/discrepancy between the case and the controls is larger than the within-group discrepancy of the controls. This suggests a possibility of adjusting the p-values in a single-case study to improve the accuracy of diagnosis.
Fourthly, the concept of p-value has been widely used to measure the degree of discrepancy between the data and the null model. 
The significance of a traditional p-value is determined on a uniform scale as the p-value has a uniform null-distribution. 
However, this uniformity deteriorates when the controls are heterogeneous. To remedy this difficulty, we need to develop a robust testing procedure that can automatically adjust the critical value when the controls are heterogeneous. For this purpose, we impose some penalty on non-uniformity of p-values when determining the critical value for the above diagnostic test.
  Finally, it is notoriously difficult to determine the null distribution of a mixture likelihood ratio test statistic as the classical Wilks' asymptotic theory may not hold for mixture likelihood ratios. To overcome the difficulty, Dacunha-Castella and Gassiat (1999) developed a local conic parametrisation approach for deriving the asymptotic distribution of a likelihood ratio test statistic. However, finding an explicit asymptotic distribution for a finite normal mixture model of unknown number of components is still an open problem. 

Here, we develop a robust and flexible modelling strategy in which the case density is charaterised by a finite normal mixture and the control density by a double mixture, which contains two layers; the first layer is for modeling subject-heterogneity while the second layer is nested in the first layer for modeling heterogeneity within subjects. These models, allowing for different model dimensions in the case and controls, can explore the skewness and multi-modality in the data. Based on these models, we construct a novel likelihood ratio test for difference between the individual case and its age-matched control group. The critical value is automatically determined by imposing a bootstrap cross-validated penalty on p-value. We develop an asymptotic theory to support the proposed testing procedure.  We apply the proposed procedure to the MEG scan data, reporting a list of brain damage areas for the patient and improving our understanding of the neuronal mechanism underpinning brain injury. To evaluate the performance of the new procedures, we compare the proposed likelihood ratio procedure to the average pairwise AD test, the pairwise permutated AD test and the AD mean test by simulations. Overall, the proposed method shows its advantage over these AD methods, improving the existing methods by reducing not only false positive rate but also false negative rate even when the underlying distributions substantially deviate from normal mixtures. 

\centerline { [Put Figures \ref{Histogram} and \ref{AndersonDarling}  here.]}

The remaining paper is organised as follows.  The details of the proposed methodology are provided in Section 2. The applications of the proposed methods to the brain injury dataset and synthetic data are presented in Section 3.  The asymptotic theory is developed in Section 4. Discussion and conclusions are made in Section 5.
\newpage

\newpage

\section {Methodology}
An OK test can be implemented in two steps: We begin with a source magnitude imaging in frequency domain followed by performing case-control contrast tests. The testing results are adjusted for effects of heterogeneity by similarity analysis based on hierarchical clustering.

\subsection{MEG source magnitude imaging}
Let $N$ be the total number of epochs and $s$ the total number of sensors considered in the study. For epoch $n, 1\le n\le N$, consider $J$ time points. Let $B_{nij}$ denote the measurement of sensor $i$ at the $j$th time point,
and $\bB_{nj}=(B_{n1j},...,B_{nsj})^T$ be all measurements at the $j$th time point in epoch $n$,  $1\le j\le J$. Let $\bQ_{nj}$ $=(Q_{n1j},...,Q_{npj})^T $  be the magnitude vector of the candidate sources at grids $\{r_1,...,r_p\}$ in the brain and $\{Q_{nkj}: 1\le j\le J\}$ the source magnitude time-course at location $r_k$ and epoch $n$.
Following Zhang and Su (2015), assume that the true sources are approximately located on the grids when they are sufficiently dense (i.e., $p$ is sufficiently large).
  Let $\bG=(\bG_1,...,\bG_p)$ denote the $s\times p$ gain matrix derived from unit inputs. Sarvas (1987) showed that the contribution of an individual source to $\bB_j$ can be numerically calculated by the use of an Maxwell's equation-based forward model and that the contributions of multiple sources can be summed up linearly. 
 Accordingly, we have the source model
$
\bB_{nj}=\bG\bQ_{nj}+\bvep_{nj},\quad 1\le j\le J,
$                                         
where $1\le p< \infty$, $\bvep_{nj}$ is the background noise vector of the $s$ sensors at time $j$. As pointed out before, brain activity is evidenced by the amount of oscillatory activity in different frequency bands. Therefore, it is necessary to transform source signals into frequency bands (Huang et al., 2012). For this purpose, we perform discrete Fourier transformation on both sides of the above equation in frequency band $m$, obtaining
\begin{eqnarray}\label{meg1}
\bF_{nm}=\bG\bH_{nm}+\be_{nm}
\end{eqnarray}
with $p$-vector
\begin{eqnarray*}
\bF_{nm}&=&\sum_{j=1}^J\bB_{nj}\exp(-i2\pi mj/J),\quad \bH_{nm}=\sum_{j=1}^J\bQ_{nj}\exp(-i2\pi mj/J),\\
\be_{nm}&=&\sum_{j=1}^J\bvep_{nj}\exp(-i2\pi mj/J),
\end{eqnarray*}
where  $i=\sqrt{-1} $ is a unit complex number.
When $p$ is much larger than the number of sensors, the model estimation becomes challenging as there are a diverging number of candidate models which can fit to the data. To circumvent the problem, Huang et al. (2012) developed the Fast-VESTAL MEG source imaging procedure by imposing $L_1$ restraints on the magnitude vector in (\ref{meg1}).
For epoch $n$, each area and each band, calculate the average magnitude over the grids in the region and over the spectra in the band. 

Let $\bY=(y_{ij})=(\by_1,...,\by_{A})^T\in {\mathbb R}^{A\times N},$
$\bX_{k}=(\bx_{ak})_{1\le a\le A}\in {\mathbb R}^{A\times J}, k=1,...,K$ are log-transformed band power data for a single case and $K$ controls respectively.
Let $\bX_a=(\bx_{ak})_{1\le k\le K}.$ Suppose that for region $a$, $(y_{aj})_{1\le j\le N}$ is a sample drawn from the case density $f(y|\psi_{a0})$ and $(x_{ajk})_{1\le j\le J}$ a sample drawn from the control density $f_a(x|\psi_{ak})$, $k=1,...,K,$ where $\psi_0$ and
$\psi_{ak}$ are unknown parameters. Then, for region $a$,  our research problem can be formulated as testing the hypotheses
\begin{eqnarray}\label{H0}
H_{a0}: f(\cdot|\psi_{a0})\in\{f(\cdot|\psi_{ak}), 1\le k\le K\} \mbox { v.s. } H_{a1}:  f(\cdot|\psi_{a0})\not\in\{f(\cdot|\psi_{ak}), 1\le k\le K\}.
\end{eqnarray}
We consider the following four OK contrast tests.

\subsection{Likelihood ratio test in frequency-domain}

As the likelihood ratio test is the most powerful test of a simple null hypothesis against a simple alternative hypothesis, the first OK test we proposed is the likelihood ratio test. In our problem setting, both the null density $f(\cdot|\psi_{a0})$ and the alternative $f(\cdot|\psi_{ak})$ are unknown. We are unable to use the likelihood ratio test directly. Note that histograms in Figure \ref{Histogram} have already demonstrated that $f_a(y|\psi_{a0})$ and $f_a(x|\psi_{ak})$, $k=1,...,K,$  can be well approximated by finite mixtures of normals. So, we can use the data to estimate these unknown likelihoods. Here, for each subject, using the R-package Mclust (Scrucca at al., 2023), we fit a finite mixture of normals to the data with order being estimated via Bayesian Informatic Criterion (BIC), estimating the maximum log-likelihood under the null and alternative hypotheses respectively. Let $l_a(\hat{\psi}_{a0}|\by_a)$ and $l_a(\hat{\psi}_{ak}|\bx_{ak})$, $k=1,...,K$ be the estimated maximum log-likelihoods corresponding to the case and controls respectively. 
To incorporate the null hypothesis in the test statistic, we consider the following frequency-band log-likelihood ratio test statistic
\begin{eqnarray*}
l_{a0k}=\max_{\psi_{a0}=\psi_{ak}}l(\psi_{a0},\psi_{ak}|\by_a,\bx_{ak})-\max_{\psi_{a0}}l(\psi_{a0}|\by_a)-\max_{\psi_{ak}}l(\psi_{ak}|\bx_{ak}),
\end{eqnarray*}
where $l(\psi_{a0},\psi_{ak}|\by_a,\bx_{ak})=l(\psi_{a0}|\by_a)+l(\psi_{ak}|\bx_{ak})$, $l(\psi_{a0}|\by_a)$ and 
$l(\psi_{ak}|\bx_{ak})$ are respectively the joint and the individual log-likelihood functions based on the samples $\by_a$ and $\bx_{ak}$. The larger the $\mbox{l}_{a0k}$, the higher the chance that $\psi_{a0}$ is equal to $\psi_{ak} $. The p-value can then be estimated by
\begin{eqnarray*}
\mbox{p}_{l}(\by_a,\bx_a)=\frac 1K\sum_{k=1}^KI(l_{a0k}\ge \log(1 -c_0)),
\end{eqnarray*}
where the critical value $c_0$ is determined either by an approximate null distribution of $\mbox{l}_{a0k}$  if the asymptotic null distribution is available.  Note that the asymptotic null distribution of the above normal-mixture based test is unknown and may depend on the underlying null models. This makes it hard to set the critical value $c_0$. To tackle the issue, we cross-validate the above average p-values by use of the following one-out-of-K scheme. For each $1\le k\le K$, we perform the above likelihood ratio test on $\bx_{ak}$ against the remaining samples, obtaining p-values
$\mbox{p}_{flr}(\bx_{ak},\bx_{am}, 1\le m\not= k \le K).$ 
The cross-validated pairwise likelihood ratio p-value ($\mbox{cp}_{flr}(\by_a,\bx_a)$) is then calculated through counting the proportion of $\mbox{p}_{flr}(\bx_{ak},\bx_{am}, 1\le m\not= k \le K)$ being larger than or equal to $\mbox{p}_{flr}(\by_a,\bx_a).$ We choose $c_0$ by minimising $\mbox{p}_{l}(\by_a,\bx_a)+\mbox{cp}_{flr}(\by_a,\bx_a)$ with respect to $c_0$.
However, it is hard to develop an asymptotic theory as  given $(\by_a,\bx_a),$ $\mbox{p}_{flr}(\bx_{ak},\bx_{am}, 1\le m\not= k \le K)$ are not conditionally independent. To fix this, we use the bootstrap resampling to cross-validate $\mbox{p}_{l}(\by_a,\bx_a)$, obtaining the cross-validated p-value $\mbox{cp}_{l}(\by_a,\bx_a).$ The details are as follows.

For each area $a$ and each control subject $0\le k\le K$, we generate a bootstrap sample $\bx^{(kb)}_a$ from estimated density $f_a(\cdot|\hat{\theta}_k)$. We calculating the p-value $\mbox{p}_l(\bx^{(kb)}_a,\bx_a)$ by performing the above likelihood ratio test on the bootstrap sample $\bx^{(kb)}_a $ against the controls. This provides a bootstrap estimate of the background scale for the observed p-value. We count the proportion of the cross-validated  p-values which are at least significant as the observed p-value of the case $\mbox{p}_l(\by_a,\bx_a)$, leading to the following cross-validated p-value
\begin{eqnarray*}
\mbox{cp}_l(\by_a,\bx_a)=\frac 1K\sum_{k=1}^KI\left(\mbox{p}_l(\bx^{(kb)}_a,\bx_a)\le \mbox{p}_l(\by_a,\bx_a) \right).
\end{eqnarray*}
We choose the critical value $c_0$ by minimising $\mbox{p}_{l}(\by_a,\bx_a)+\mbox{cp}_{l}(\by_a,\bx_a)$ with respect to $c_{min}\le c_0\le c_{max}.$ Here, we pre-choose $c_{min}$ and $c_{max}$ so that the size of the test at a pre-specified level. We apply the Benjamini-Hochberg procedure to control false discovery rate for multiple testing.

\subsection{Modified Anderson-Darling tests in frequency-domain}
We are testing multiple hypotheses in (\ref{H0}). Unlike before, we will not pre-specify the distrubtions $f_{ak}, 0\le k\le K, 1\le a\le A.$ We consider the following nonparametric tests.

{\it Pairwise AD (PAD) test for a distributional shift.} For region $a$, $1\le a\le A$, based on the R-package ``two-samples" (Dowd, 2023) , we perform the AD
two-sample test of the case versus each control, obtaining $K$ p-values. Denote the average of these p-values as $\mbox{p}_{adp}(\by_a,\bx_{ak}, 1\le k \le K).$ We reject the null hypothesis $H_{a0}$ if the resulting p-value is less than or equal to a pre-specified level. We cross-validate the above average p-values by use of the following one-out-of-K scheme. For each $1\le k\le K$, we perform the Anderson-Darling test on $\bx_{ak}$ against the remaining samples, obtaining p-values
$\mbox{p}_{pad}(\bx_{ak},\bx_{am}, 1\le m\not= k \le K).$ 
The cross-validated pairwise Anderson-Darling p-value $\mbox{cp}_{pad}$ is then calculated through counting the proportion of $\mbox{p}_{pad}(\bx_{ak},\bx_{am}, 1\le m\not= k \le K)$ being larger than or equal to
$\mbox{p}_{pad}(\by_a,\bx_a).$ 

{\it AD permutation (PMAD) test for a distributional shift under the assumption of population homogeneity.} 
For region $a$, $1\le a\le A$, we randomly draw $N$ subsets, each of size $J$, from the pooled control samples $\{x_{akj}:
 1\le j\le J, 1\le k\le K\}.$ 
We perform the two-sample AD test for each subset versus the case sample, obtaining $N$ p-values. Denote the average of these p-values by 
 $\mbox{p}_{ad}(\by_a,\bx_{ak}, 1\le k \le K).$ As usual, we apply the Benjamini-Hochberg procedure to control false discovery rate for multiple testing.
In this test, we implicitly use the homogeneity assumption for the control sample. We reject the null hypothesis $H_{a0}$ if the resulting p-value is  less than or equal to a pre-specified level. Similarly, we can calibrate these p-values by using the one-out-of-K scheme.

{\it AD test for a  mean shift.} The AD, applied to sample means, can be viewed as a non-parametric t-test. For region $a$, $1\le a\le A$, we first calculate the averages $\bar{y}_a$ and $\bar{x}_{ak},$ $1\le k\le K$ of the case and control samples over epochs. We perform the two-sample AD test for $\bar{y}_a$ versus $\bar{x}_{ak},$ $1\le k\le K.$ We reject the null hypothesis $H_{a0}$ if the resulting p-value  is less than or equal to a pre-specified level.

For all the above tests, As usual, we apply the Benjamini-Hochberg procedure to control false discovery rate for multiple testing.

\subsection{Correction and visualisation for heterogeneity effects}
In the proposed FLR and PAD procedures, the p-value of an OK test, defined by averaging the p-values of the corresponding pairwise tests of the case against individual controls, can be affected by heterogeneity of controls. Failing to adjust for such effects may lead to biased diagnosis and wrong conclusions.
 To remove these effects, we group controls for each area, i.e., clustering $K$ $n$-dimensional vectors, which is ill-posed if $K<n.$ The conventional methods such as $k$-means clustering and model-based clustering (Scrucca et al., 2023)) may miss double mixture structures in the data. Here, we consider a hierarchical clustering (HC) strategy below: We first, for each area in the brain, calculate p-values for both case-control pairs and control-control pairs. This results in a p-value based $(K+1)\times (K+1)$ similarity matrix for $K+1$ subjects. We use a bottom-up approach to create an upside-down clustering tree called dendrogram: At the bottom, each subject starts in its own cluster. We repeat the following two-steps until reaching the top hierarchy: (i) Calculate average similarity score for each pair of clusters. (ii) Find a pair of clusters with the maximum average similarity score, merge them as one and moves up the hierarchy. The case is expected to have the highest hierarchy in the dendrogram when the case is significantly different from the control group in terms of p-values. We claim the areas that the case has the highest hierarchy as HC-approved areas. Combining HC with FLR and PAD respectively, we have heterogeneity-adjusted tests, FLR-HC and PAD-HC. We can visualise multiple testing with the dendrograms.

\section {Numerical results}

In this section, using simulations and real data analysis, we evaluate the performance of the proposed likelihood ratio procedure FLR and compare it to the non-parametric competitors PAD, PMAD and ADM in testing multiple hypotheses: For region $a$, $1\le a\le A$,  we test $H_{a0}: f_{a0}\in \{f_{ak}, 1\le k\le K\}$ v.s. the alternative  $H_{a1}: f_{a0}\not\in \{f_{ak}, 1\le k\le K\}$.

\subsection {Synthetic data}
 Let the control group have the size $K=54$ and the epoch/sample size $N=100$ and $150$. For each case-control setting, we generated $50$ independent datasets. Each dataset contains a case sample of size $N$ drawn from  $f(x|\psi_{0})$, and $K$ control samples of size $N$ drawn from $f(x|\psi_{k}), 1\le k\le K$ respectively. 
Denote by $\phi(x|\mu,\sigma^2)$ the normal density with mean $\mu$ and variance $\sigma^2.$ Taking into account the patterns of skew, two-modes and heterogeneity in real MEG scan data displayed in Figure \ref{Histogram}, we consider three settings. In Setting 1, we assess the proposed procedure in a favourable situation, where the underlying distributions belong to a normal-mixture distribution family. In Settings 2 and 3, we evaluate the performance of proposed procedures when the underlying distributions are miss-specified, that is, they are outside the family of normal-mixture distributions. We employ performance metrics, precision, recall and $F$ scores, to compare the FLR with the other tests.
Precision is the fraction of true $H_1$ instances among the claimed instances (i.e., among instances of p-value less than $0.05$), where $1$- precision is equal to false discovery rate. Recall (also known as sensitivity) is the fraction of claimed $H_1$ instances among all $H_1$ instances. $F$ scores are measures that combines precision and recall. For example, the traditional $F_1$ is the harmonic mean of precision and recall. In general, we define $F_{w}=(1+w^2)( w^2/\mbox{recall}+1/\mbox{precision})^{-1}$, where the weight $0\le w\le 1$ is chosen such that recall is considered $w$ times as important as precision. Two commonly used values for 
$w$ are 2, which weighs recall higher than precision, and $0.5$, which weighs recall lower than precision. See Saito and Rehmsmeier (2015). 

{\it Setting 1 (Heterogeneous normal-mixtures)}: {\it Controls}: $f(x|\psi_{k})=0.2\phi(x|0,1)+0.8\phi(x|1,1), 1\le k\le 10, $ and
$f(x|\psi_{k})=0.4\phi(x|0,1)+0.6\phi(x|1,1), 11\le k\le K=54. $ In this setting, there are around $18\%$ controls drawn from a two-component normal mixture and $82\%$ controls drawn from a slightly different two-component normal mixture. The scenario imitates an empirical fact observed in Figures \ref{Histogram} that there may be two skew sub-populations in the controls, one with a relatively smaller size. Consider the following three case settings respectively. 
 {\it Case 1.1}: $f(x|\psi_{0})=0.2\phi(x|0,1)+0.8\phi(x|2,1).$
{\it Case 1.2}: $f(x|\psi_{0})=0.4\phi(x|0,1)+0.6\phi(x|1,1).$
 {\it Case 1.3}: $f(x|\psi_{0})=0.1\phi(x|0,1.5)+0.9\phi(x|1,1).$
{\it Case 1.4}: $f(x|\psi_{0})=0.4\phi(x|0,2)+0.6\phi(x|1,2).$
{\it Case 1.5}: $f(x|\psi_{0})=0.2\phi(x|-1,1)+0.8\phi(x|3,1).$

{\it Cases 1.1 and 1.2} are used to calculate the type I error rate of the test, where the null hypothesis $H_0$ is true, whereas {\it Cases 1.3, 1.4 and 1.5} are used to show the power of the test for a range of shifts. We consider the shifts in one of component variances and one of mixture weights in {\it Case 1.3}, in both component-variances in {\it Case 1.4}, and in both component means in {\it Case 1.5}.

\begin{center}
\begin{figure}[htp]
\includegraphics[height=1.3in,width=0.4\textwidth]{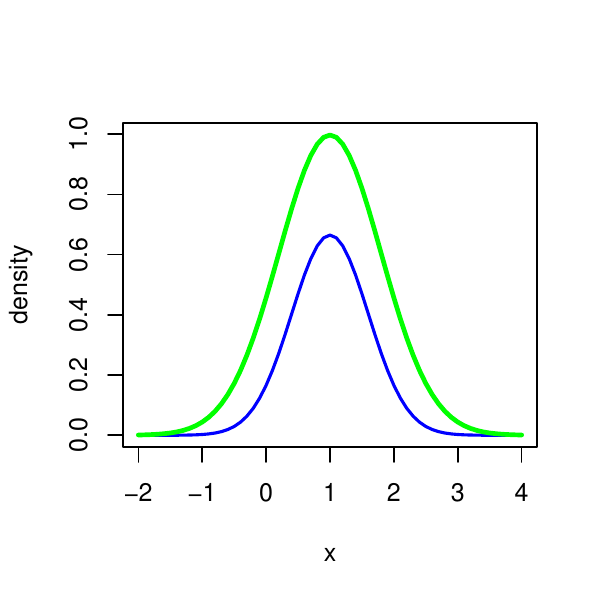}\hfill
\includegraphics[height=1.3in,width=0.6\textwidth]{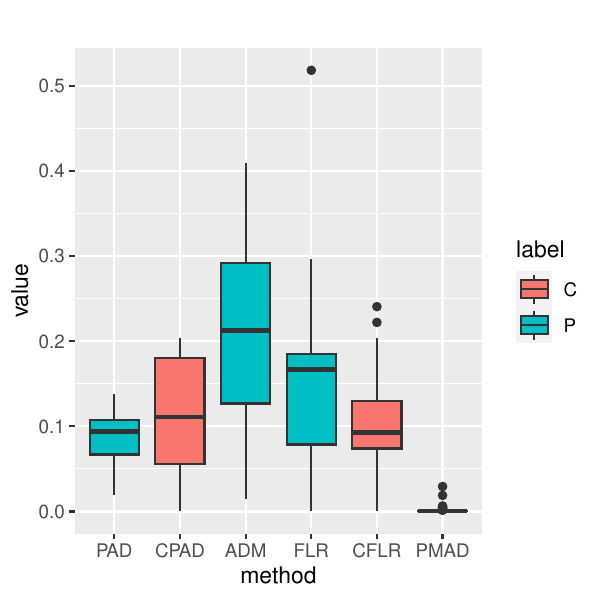}\\
\includegraphics[height=1.3in,width=0.4\textwidth]{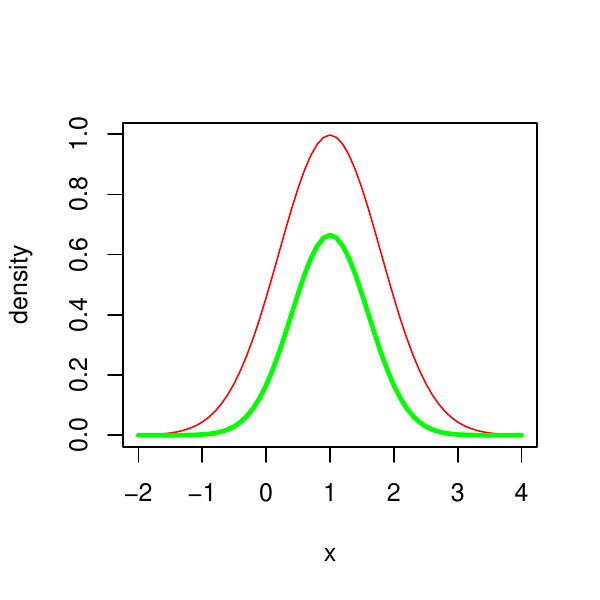}\hfill
\includegraphics[height=1.3in,width=0.6\textwidth]{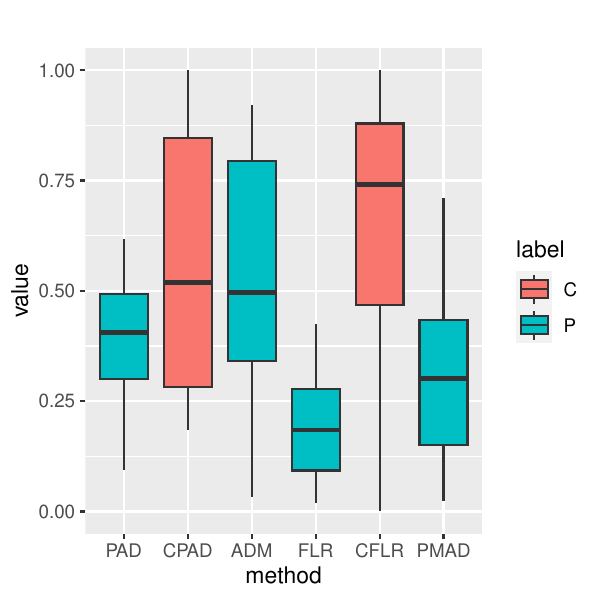}\\
\includegraphics[height=1.3in,width=0.4\textwidth]{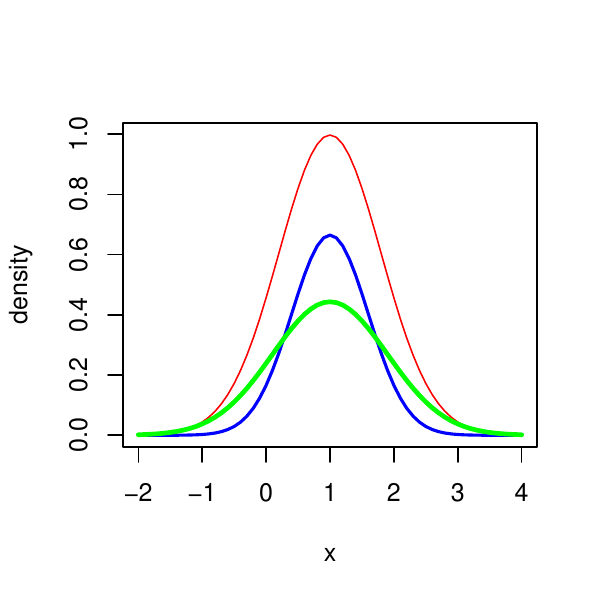}
\hfill
\includegraphics[height=1.3in,width=0.6\textwidth]{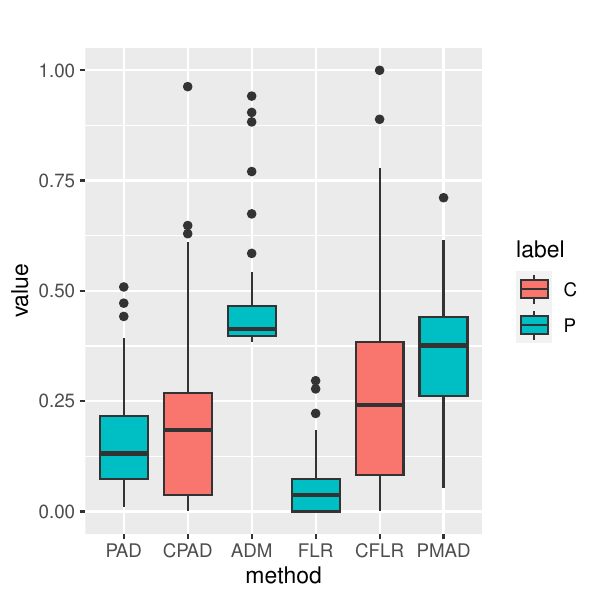}\\
\includegraphics[height=1.3in,width=0.4\textwidth]{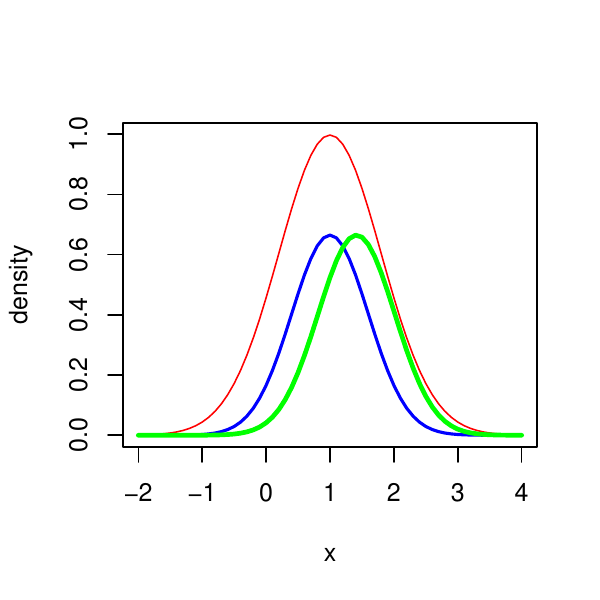}
\hfill
\includegraphics[height=1.3in,width=0.6\textwidth]{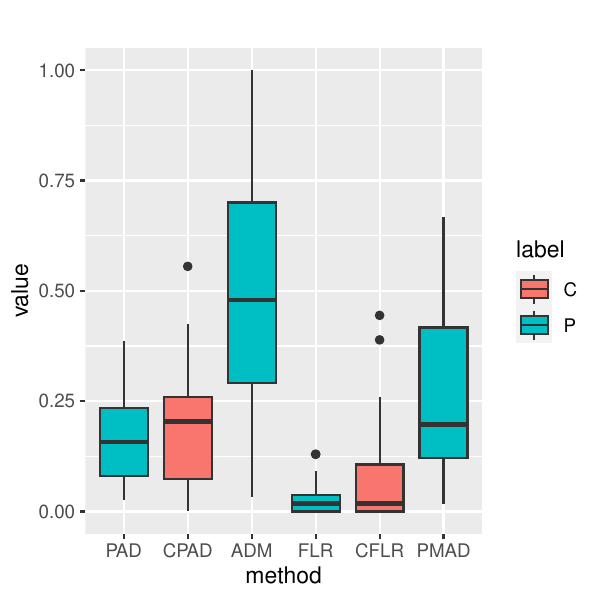}\\
\includegraphics[height=1.3in,width=0.4\textwidth]{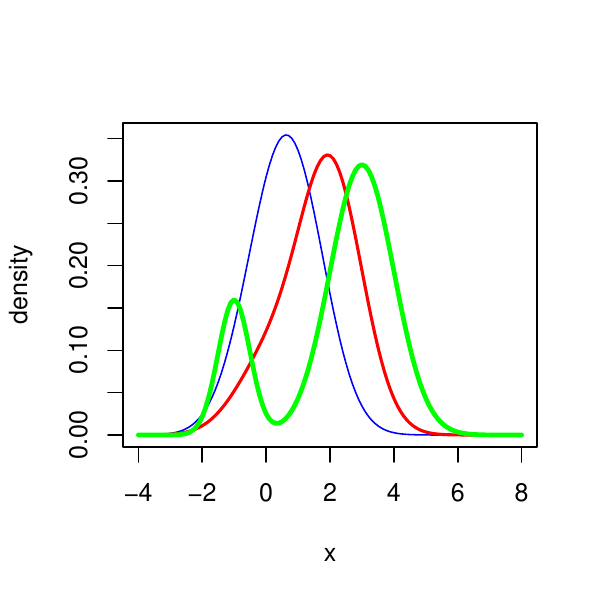}
\hfill
\includegraphics[height=1.3in,width=0.6\textwidth]{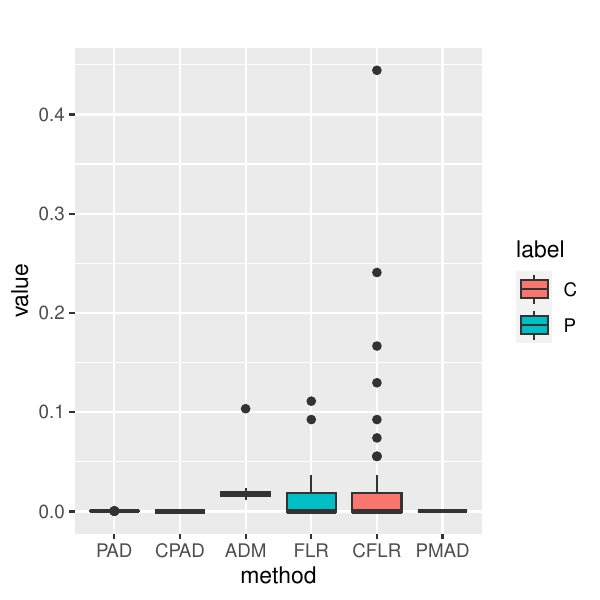}\\
\caption{\small{\noindent Heterogeneous normal mixtures (Setting 1). Left and right columns contain component density plots 
and p-value plots respectively. Rows 1 to 5 are corresponding to testing each of 5 cases against the Controls respectively. Estimated sizes of test: 0.16 for the PAD,  0.06 for the ADM, 0.12 for the FLR, and 1 for the PMAD. 
}}
\label{Set1}
\end{figure}
\end{center} 

{\it Setting 2 (Homogenous lognormal)}: {\it Controls}: $f(x|\psi_{k})=\phi(x|0,1), 1\le k\le K. $  We consider three scenarios respectively. 
 {\it Case 2.1}: $f(x|\psi_{0})=\phi(\ln(x)|0,1)/x, x>0.$
 {\it Case 2.2}: $f(x|\psi_{0})=\phi(\ln(x)|0.5,1)/x,$ $ x>0.$
{\it Case 2.3}: $f(x|\psi_{0})=\phi(\ln(x)|1,1)/x, x>0.$
Similar to Setting 1, {\it Case 2.1} is used to calculate the size of the test, where the null hypothesis $H_0$ is true, whereas {\it Cases 2.2 and 2.3} are used to show the power of testing for the location shifts from $0$ to $0.5$ and from $0$ to $1$ respectively.

\begin{center}
\begin{figure}[htp]
\includegraphics[height=1.2in,width=0.4\textwidth]{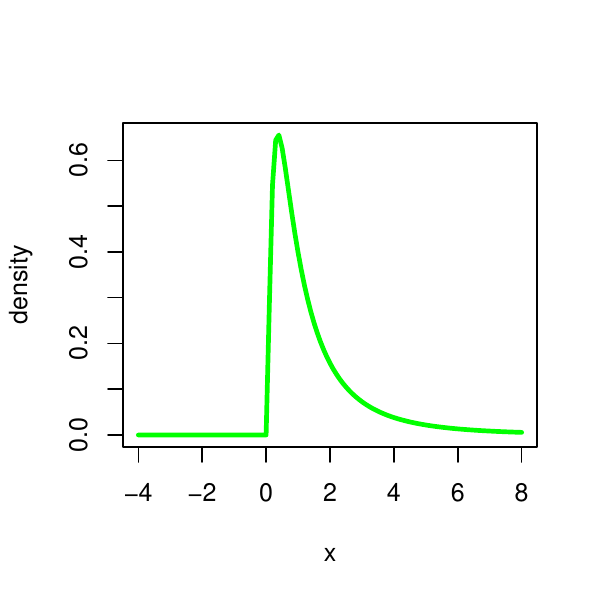}\hfill
\includegraphics[height=1.2in,width=0.6\textwidth]{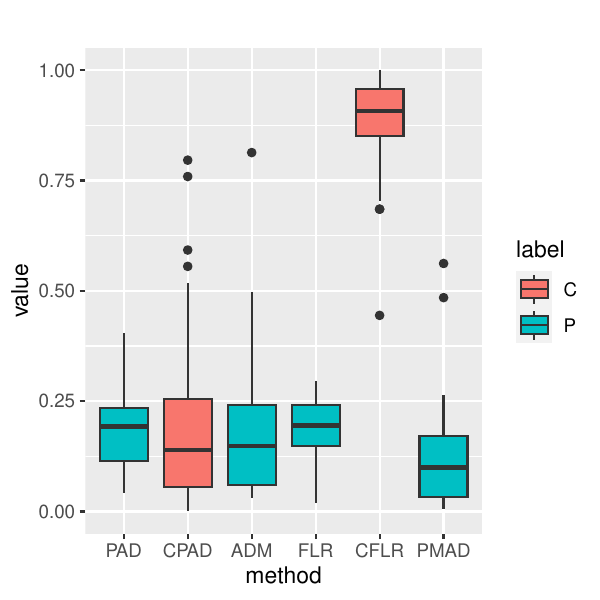}\\
\includegraphics[height=1.2in,width=0.4\textwidth]{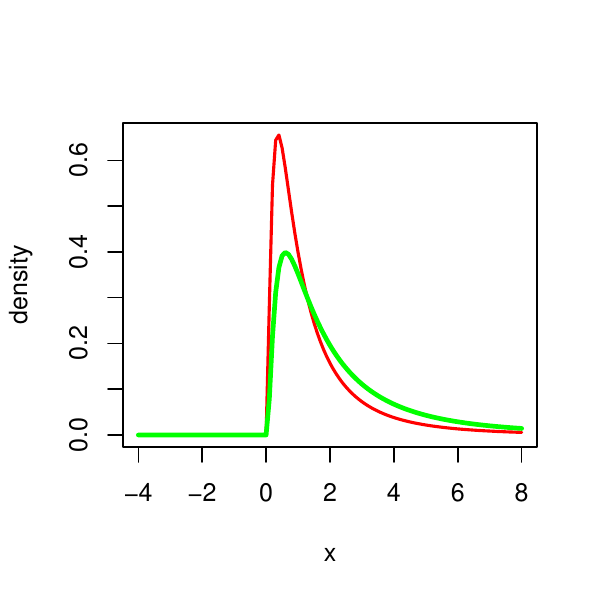}\hfill
\includegraphics[height=1.2in,width=0.6\textwidth]{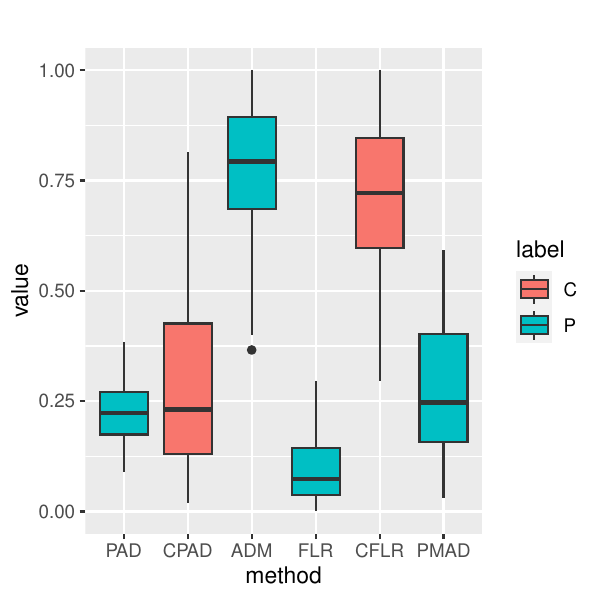}\\
\includegraphics[height=1.2in,width=0.4\textwidth]{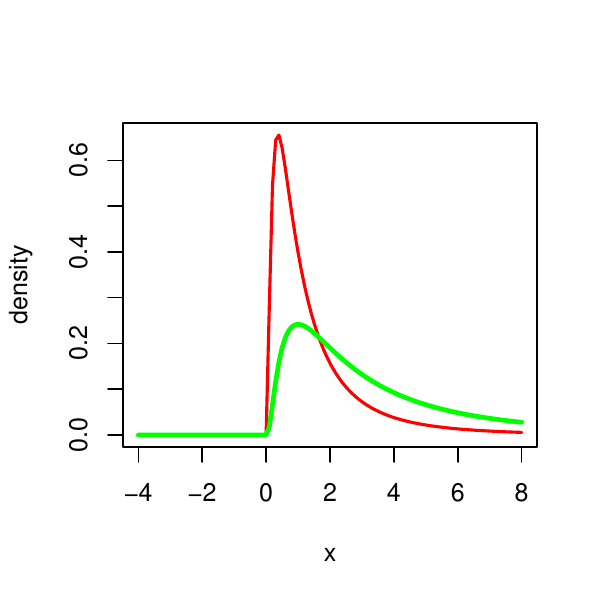}
\hfill
\includegraphics[height=1.2in,width=0.6\textwidth]{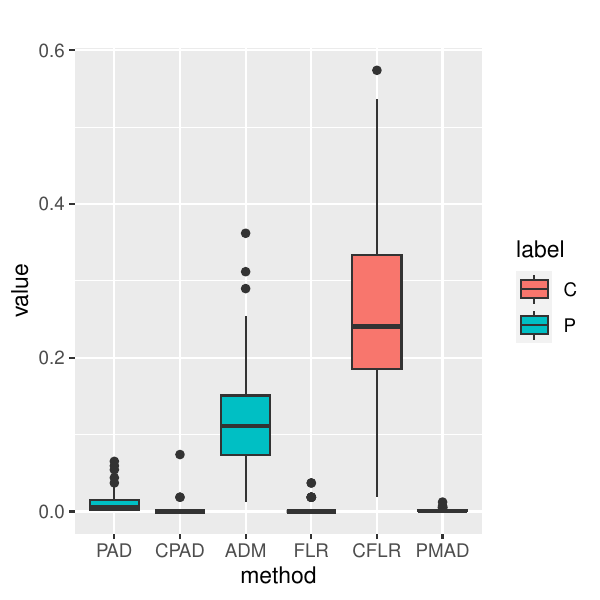}\\
\includegraphics[height=1.2in,width=0.4\textwidth]{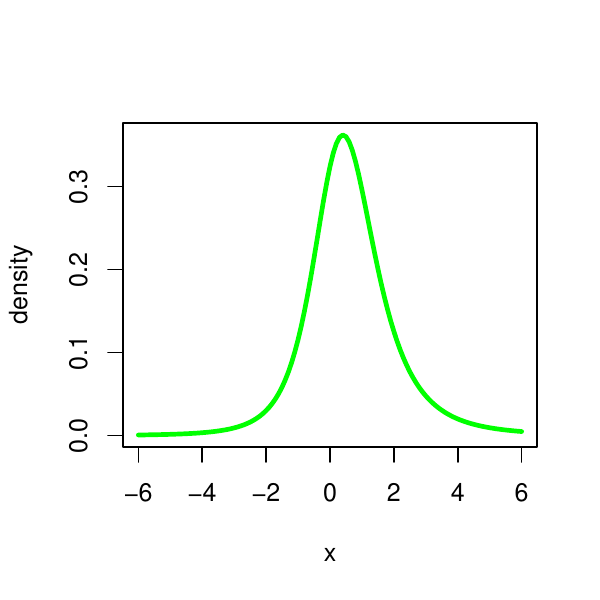}\hfill
\includegraphics[height=1.2in,width=0.6\textwidth]{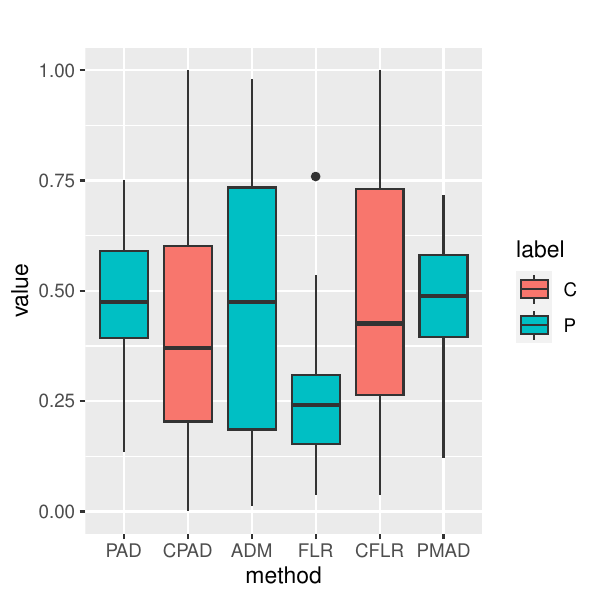}\\
\includegraphics[height=1.2in,width=0.4\textwidth]{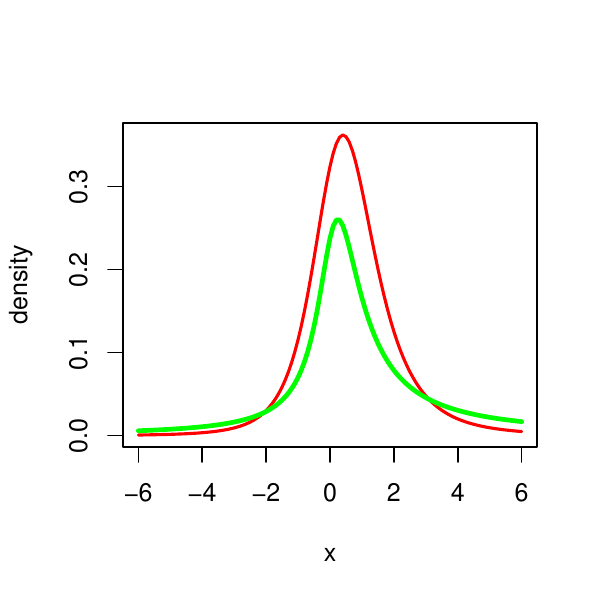}\hfill
\includegraphics[height=1.2in,width=0.6\textwidth]{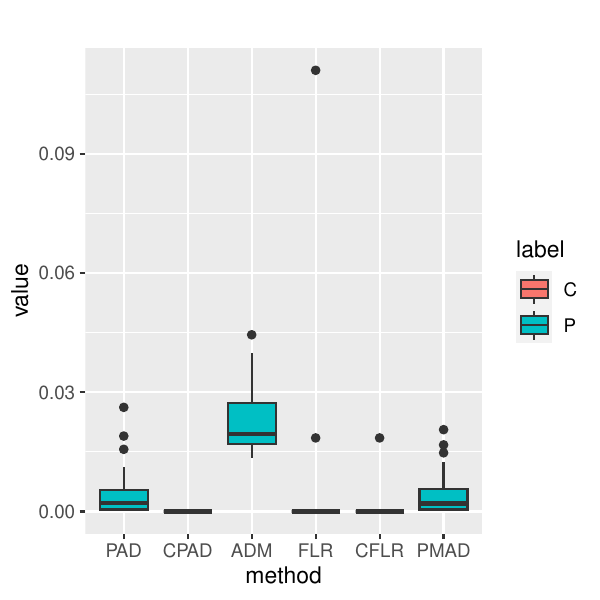}\\
\includegraphics[height=1.2in,width=0.4\textwidth]{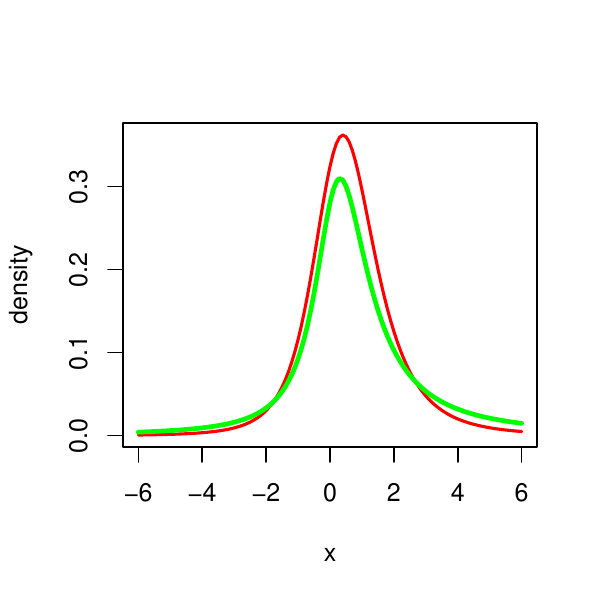}
\hfill
\includegraphics[height=1.2in,width=0.6\textwidth]{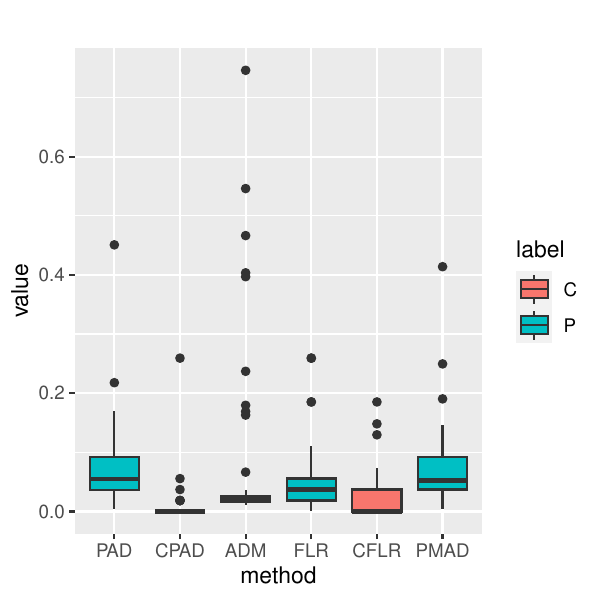}\\
\caption{\small{\noindent Homogenous log-normals (Setting 2) and homogenous noncentral t-distributions (Setting 3) . Left and right columns contain component density plots 
and p-value plots respectively. Rows 1 to 3 for Setting 2 while Rows 4 to 6 for Setting 3 are corresponding to testing each of 3 cases against the Controls respectively.
}}
\label{Set23}
\end{figure}
\end{center}

\begin{table}[htbp]
    \centering
    \caption{Percentages of instances with p-value less than $0.05$ and metrics: $N=100$.}
    \begin{tabular}{cccccccc}
        \toprule
   Setting/Metrics    & $ $ &  &  & Method &  &   &\\
      \cline{1-8}
        &  & FLR & CFLR  & PAD & CPAD & PMAD  & ADM\\
      \cline{3-8}
        \multirow{1}{*}{1.1} 
        &  & 0.12 &  0.2 &  0.16 &  0.22 &  1&0.06\\
        \cmidrule{3-8}
        \multirow{1}{*}{1.2}
        &  & 0.12 &  0.02 &  0 &  0  &  0.06& 0.08\\
        \cmidrule{3-8}
        \multirow{1}{*}{1.3}
        &  & 0.6 &  0.2&  0.2 &  0.28 & 0& 0\\
        \cmidrule{3-8}
        \multirow{1}{*}{1.4}
        &  & 0.78 &  0.58 &  0.12 &  0.2 &  0.08&0.08\\
       \cmidrule{3-8}
        \multirow{1}{*}{1.5}
        &  & 0.96    &  0.82    &  1         &  1         &  1        & 0.98\\
 \cmidrule{3-8}
Precision &  & {\bf 0.91} & 0.88  & 0.89 & 0.87 & 0.50  & 0.88\\
Recall &  &{\bf 0.78} & 0.53  & 0.44 & 0.49 & 0.36  & 0.35\\
$F_1$ &  & {\bf 0.84} & 0.66  & 0.63 & 0.63 & 0.42  & 0.50\\
$F_{0.5}$ &  & {\bf 0.88} & 0.78  & 0.86 & 0.75 & 0.46  & 0.80\\
$F_{2}$ &  & {\bf 0.80} & 0.58  & 0.50 & 0.54 & 0.38  & 0.40\\
   \cline{1-8}
        \multirow{1}{*}{2.1}
        &  & 0.04 &  0.2&  0.02 &  0.2 & 0.32& 0.2\\
        \cmidrule{3-8}
        \multirow{1}{*}{2.2}
        &  & 0.32&  0 &  0 &  0.08 &  0.2.3\\
       \cmidrule{3-8}
        \multirow{1}{*}{2.3}
        &  & 1        &  0.02    &  0.94     &  0.98     &  1        & 0.18\\
 \cmidrule{3-8}
Precision &  &{\bf  0.97} & 0.09  & 0.82 & 0.84 & 0.79  & 0.47\\
Recall &  & {\bf 0.66} & 0.01  & 0.47 & 0.53 & 0.60  & 0.09\\
$F_1$ &  & {\bf 0.79} & 0.02  & 0.60 & 0.65 & 0.68 & 0.15\\
$F_{0.5}$ &  & {\bf 0.89} & 0.03  & 0.71 & 0.75 & 0.74  & 0.25\\
$F_{2}$ &  & {\bf 0.71} & 0.01  & 0.51 & 0.57 & 0.63  & 0.11\\
 \cmidrule{1-8}
        \multirow{1}{*}{3.1}
        &  & 0.04 &  0.04&  0 &  0.06 & 0& 0.06\\
        \cmidrule{3-8}
        \multirow{1}{*}{3.2}
        &  & 0.98&  1&  1 &  1 &  1&1\\
       \cmidrule{3-8}
        \multirow{1}{*}{3.3}
        &  & 0.66        &  0.88    &  0.4     &  0.96     &  0.44        & 0.8\\
 \cmidrule{3-8}
Precision &  &{\bf 0.98} & 0.98  & 1 & 0.97 & 1  & 0.97\\
Recall &  & 0.82 & 0.94  & 0.70 &{\bf 0.98} & 0.72  & 0.90\\
$F_1$ &  & 0.89 & 0.96  & 0.82 & {\bf 0.97} & 0.84 & 0.93\\
$F_{0.5}$ &  & 0.84 & {\bf 0.97}  & 0.92 &{\bf 0.97} & 0.93  & 0.96\\
$F_{2}$ &  & 0.85 & 0.95  & 0.74 & {\bf 0.98} & 0.76 & 0.91\\
        \bottomrule
    \end{tabular}
\label{tab:simulation}
\end{table}

 The resultant $50$ p-values are plotted in Figures \ref{Set1} and  \ref{Set23}. The estimated percentages of p-values being less than or equal to $0.05$ are calculated in Table \ref{tab:simulation} and Table 1 in the Online Supplementary Materials, where outliers have been screened out by the boxplots. These estimated precisions and recalls are quite robust. These numerical results show that the FLR achieves the best overall performance ($F_1$ score) among six tests and the PAD ranks the second place. Compared to the PAD, the FLR performs much better in terms of recall but slightly worse in terms of precision.
In Setting 1, Figure \ref{Set1} shows that the FLR performed substantially better than the PAD, PMAD and ADM: 
In terms of $F_1$ score, the FLR improved the PAD by $33\%$ for $N=100$ and $14\%$ for $N=150$, the PMAD by $100\%$ for $N=100$ and $78\%$ for $N=150$, and the ADM by $68\%$ for $N=100$ and $64\%$ for $N=150$. In terms of $F_{0.5}$ score, the FLR improved the PAD by $2.3\%$ for $N=100$ and $7\%$ for $N=150$, the PMAD by $91\%$ for $N=100$ and $75\%$ for $N=150$, and the ADM by $10\%$ for $N=100$ $31\%$ for $N=150$.  In terms of $F_{2}$ score, the FLR improved the PAD by $60\%$ for $N=100$ and $19\%$ for $N=150$, the PMAD by $110\%$ for $N=100$ and $80\%$ for $N=150$, and the ADM by $100\%$ for $N=100$ and $119\% $ for $N=150$. 
In Setting 2, the FLR also outperformed the PAD, PMAD and ADM: In terms of $F_1$ score, the FLR improved the PAD by 32$\%$ , the PMAD by $16\%$ and the ADM by $427\%$ .
 In terms of $F_{0.5}$ score, the FLR improved the PAD by $25\%$, the PMAD by $20\%$ and the ADM by $256\%$.  In terms of $F_{2}$ score, the FLR improved the PAD by $39\%$, the PMAD by $13\%$ and the ADM by $545\%$. 
  In Setting 3, the PAD performed similar to the PAD and PMAD and better than the ADM, while the CPAD attains the best F-scores. However, the FLR performs better than the PAD, PMAD and ADM in terms of $F_1$ scores, while the FLR performs slightly worse than the PAD, PMAD and ADM in terms of $F_{0.5}$ scores.
Similar results hold for $N=150.$

To demonstrate the superior performance of the FLR-HC over the PAD-HC, we generate a dataset for each of Settings $1.1$, $1.2$ and $1.3$ . We apply the FLR-based HC,  FLR-HC and the PAD-based HC, PAD-HC to these datasets, drawing the corresponding dendrograms. The case is expected to be among the controls in Settings $1.1$ and $1.2$, whereas the case is expected to be outside the control group in Setting $1.3$. In all these settings, the FLR has clearly identified three subgroups in the controls if we horizontally cut the dendrograms at the hight of $0.99$, whereas the PAD can produce too many subgroups. Therefore, the results displayed Figures 2 and 3  in Section 3, the Online Supplementary Materials indicate that the FLR-HC is more powerful than the PAD-HC in correctly capturing hidden heterogeneity and in predicting the group identity of a case.

\subsection {Real MEG scan data}
As suggested by the simulation studies, the FLR performs better, in terms of $F$ scores, than both the PAD, PMAD and ADM in testing a single case versus multiple controls. In this subsection,  we applied the FLR, PAD, ADM and PMAD to the data discussed in the Introduction. The data consist of a single case against  $K=54$ gender and age matched controls.  We performed these tests for the $68$ areas (indexed by $1$ to $34$ in the left hemisphere and by $35$ to $68$ in the right hemisphere) simultaneously, obtaining a p-value for each area. For the FLR, we also resampled a bootstrap sample of size $N=100$ from the estimated mixture distribution of each control and used them to estimate the cross-validated p-values. To control the false discovery rate of multiple testing, we adjusted these $p$-values by using the Benjamini-Hochberg procedure (Benjamini and Hochberg, 1995). The significance level of these adjusted p-values is chosen at the level of $0.01$.
 

\centerline { [Put Tables \ref{tbl:delta} and \ref{tbl:gamma} here.]}

 {\it Delta band data analysis.} As pointed out before, we first apply the FLR procedure to the delta band data, followed by the Benjamini-Hochberg adjustment.  Thresholding these p-values by $0.01$ gives a list of significantly abnormal areas. We then correct this list for heterogeneity effects by the FLR-HC. See, for example, Figures $5\sim 8$,  the Online Supplementary Materials for details. The more details are omitted. According to the FLR-HC,  in areas 7, 12, 17, 19, 20, 23, 44, 46, 50, 54, 58, 62, the similar scores between the case and some controls are bigger than the average similarity score within the control group due to subject-heterogeneity. 
Filtering out these areas, the FLR-HC, in Table \ref{tbl:delta}, declares $13$  abnormal areas: Areas $3, 49$ and $63$ in the frontal lobe, $43$ and $67$ in the temporal lobe, $8, 32$ and $42$ in the parietal lobe,  $4$ and $60$ in the cuneus, $59$ in the paracentral area, $27$ in the cingulate, and $14$ in the lingual area. In some of these areas, their dendrograms do show subgrouping of the controls. For example, in the delta band and area 59, there are two subgroups in the controls if we cut the dendrogram at the height of $0.99$. 

 After the Benjamini-Hochberg procedure based adjustment, the CFLR gives abnormal areas $27$ and $61$ in the cingulate, $32$ in the parietal lobe, $60$ in the cuneus, and $63$ the frontal lobe. See Table \ref{tbl:delta}. 

 The PAD is applied to the delta band data, followed by the Benjamini-Hochberg adjustment. Thresholding these adjusted p-values at the level $0.01$ gives a list of significant abnormal areas. We then adjust this list for heterogeneity effects by PAD-HC. See Figures~5$\sim$8 in the Supplementary Materials. The results, summarised in Table \ref{tbl:delta}, show $5$ abnormal areas: Areas $17$ and $59$ in the paracentral areas, $27$ in the cingulate area, and $28$  and $63$ in the frontal lobe, where $27$, $59$ and $63$ are also identified by the FLR-HC. The CPAD, after the Benjamini-Hochberg procedure based correction, claims the following abnormal areas: the paracentral areas $17$, the frontal lobe area $28$, the cingulate areas $27$ and $61$. In the delta band and areas $17$ and $27$, there are at least three potential subgroups in the controls. 
 The results show that PAD-HC and CPAD identified a less number of abnormal areas than the FLR-HC and the CFLR.

In the delta band, the ADM has not found any abnormal areas at the level $0.01$ after the Benjamini-Hochberg adjustment for multiple testing.
On other hand, after the Benjamini-Hochberg adjustment,  the PMAD finds $53$ abnormal areas: $1-12, 14-34,$ $36-44, 46, 47, 49, 50, 52, 53,$ and $55-68,$ many more than found by the FLR-HC and the PAD-HC. This implies that the PMAD is too sensitive to subject-hetrogeneity than the FLR-HC and the PAD-HC. 

{\it Gamma band data analysis.}  The FLR is applied to the gamma band data, followed by the Benjamini-Hochberg adjustment.  Thresholding these p-values by $0.01$, the FLR gives a list of
significantly abnormal areas. We then correct this list for heterogeneity effects by the FLR-HC as before. See, for example, Figures $5\sim 8$, the Online Supplementary Materials for details. After the FLR-HC filtering,  in Table \ref{tbl:gamma},  $24$  areas are left: 
Areas $6$, $28$ and $40$ in the {\it frontal lobe}, $9, 43, 31$ and $34$ in the {\it temporal lobe}, $51, 57$ and $59$ the {\it central} areas,  $12$ in the {\it occipital}, $14$ and $48$ in the {\it lingual},  $7$ in the {\it fusiform},  $2, 11, 24, 27, 36, 45$ and $61$ in the {\it cingulate} areas, 18 in the parahippocampal, $5$ in the entorhinal, and $66$ in the supramarginal area. 
The CFLR claims abnormalities in $15$ areas: $6$ and $28$ in the frontal lobe areas, $2, 11, 24, 27, 36$ and $61$ in the cingulate, $9$ and $43$ in the temporal lobe, $12$ in the occipital, $14$ and $48$ in the lingual, $59$ in the central area, and $66$ in the supramarginal area. Again, in some of these areas, their dendrograms do show potential subgrouping of the controls. For example, in the delta band and areas $14$ and $27$, there are at least two potential subgroups in the controls. 

The PAD gives $33$ areas which are significant at the level of $0.01$ after  the Benjamini-Hochberg adjustment for multiple testing. Among them areas $1,$ $4$, $16$, $19$, $23$, $31$, $32$, $37$, $49$, $53$, $57$, $62$, $64$, $66$ and $67$ have been filtered out by the PAD-HC due to subject-heterogneity. Taking area $49$ as an example, PAD-HC dendrogram Figure 7, the Online Supplementary Materials demonstrates that compared to control-subjects $38$, $20,$ $17$ and $42$, subject $55$ is closer to the remaining $50$ controls although as a case it significantly differs from the controls overall. 
After the PAD-HC filtering, $18$ areas are left: Areas $28$, $30$ and $40$ in the {\it frontal lobe}, $9$ in the {\it temporal lobe}, $59$ in the {\it central} area, $12$ in the {\it occipital}, $14$ and $48$ in the {\it lingual}, $2, 24, 27, 36$ and $61$ in the {\it cingulate}, $38$ and $60$ in the precuneus and cuneus, $17$ and $25$ in the central areas, and $56$ in the pericalcarine. 
The details are omitted. 
The CPAD claims $8$ abnormal areas: $9, 17, 18, 24, 27, 31, 48$ and $56.$ The PMAD claims $63$ abnormal areas, many of which may be false positive due to being too sensitive to subject-heterogeneity in the controls. Similar to the FLR, in the gamma band and areas $14$ and $27$, the PAD can also show some potential subgroups in the controls.

\begin{center}
\begin{figure}[htp]
\includegraphics[height=1.2in,width=0.45\textwidth]{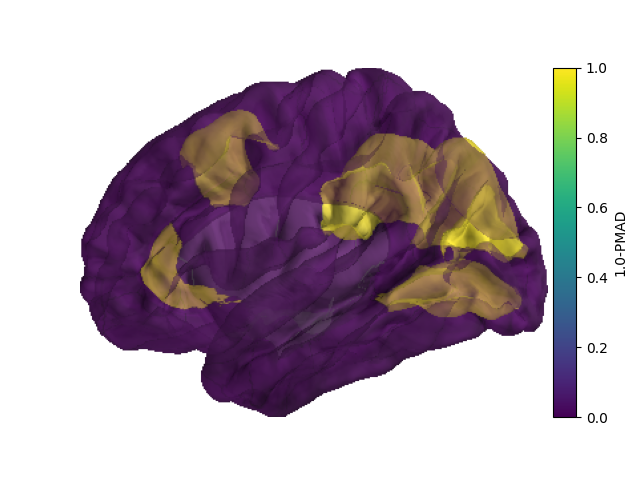}\hfill
\includegraphics[height=1.2in,width=0.45\textwidth]{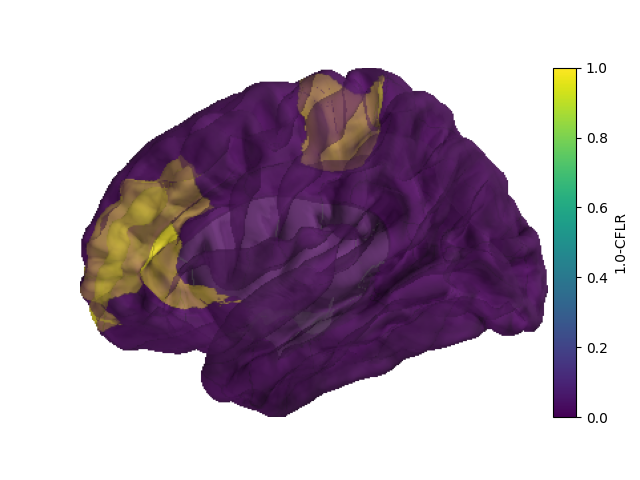}\\
\includegraphics[height=1.2in,width=0.45\textwidth]{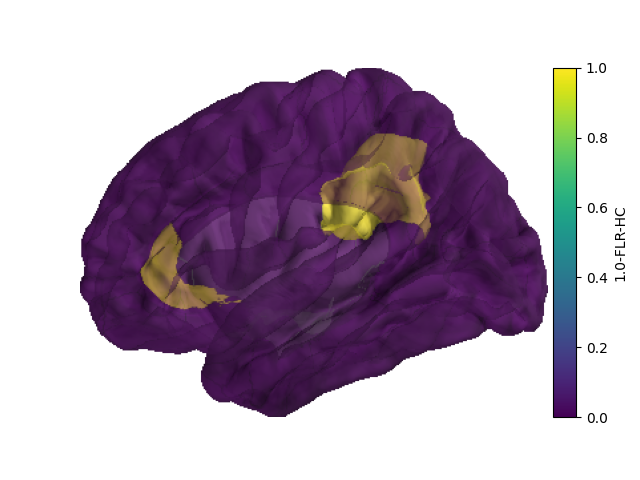}\hfill
\includegraphics[height=1.2in,width=0.45\textwidth]{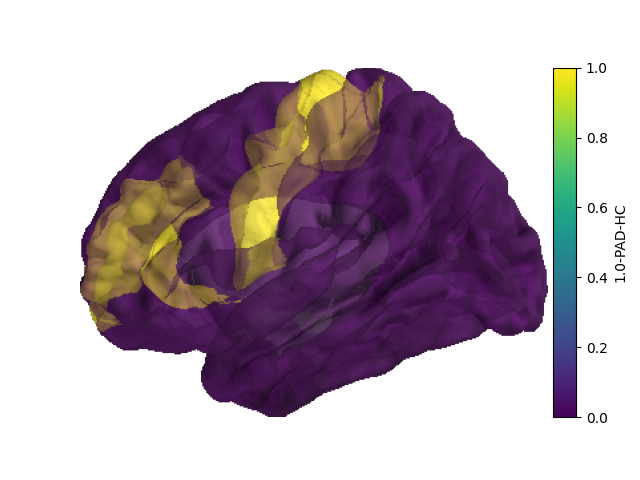}\\
\includegraphics[height=1.2in,width=0.45\textwidth]{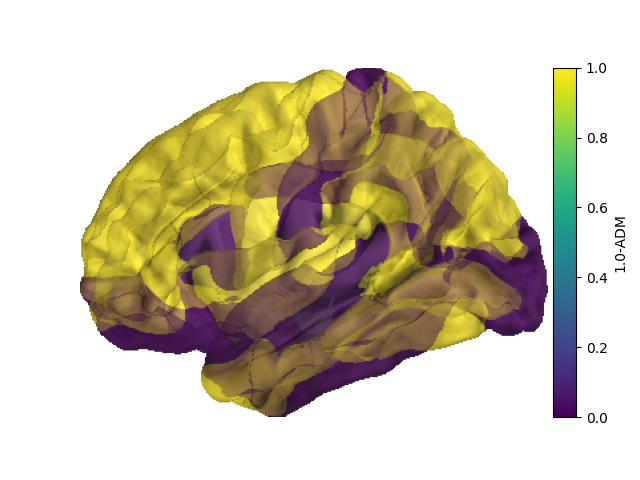}
\hfill
\includegraphics[height=1.2in,width=0.45\textwidth]{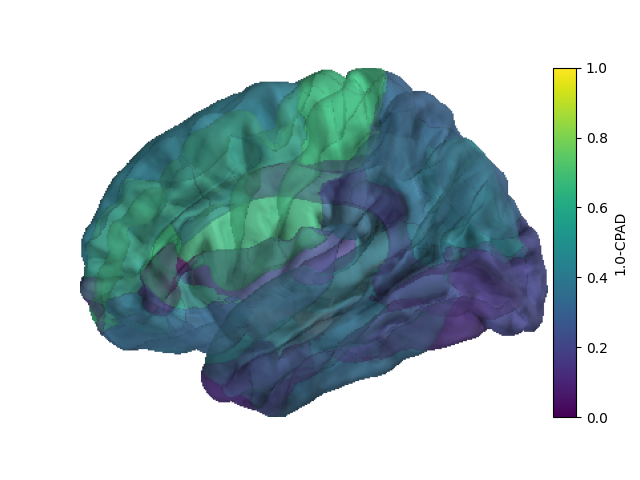}\\
\includegraphics[height=1.2in,width=0.45\textwidth]{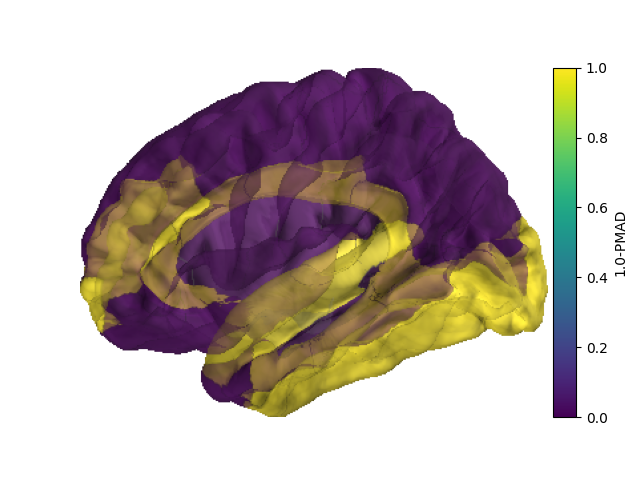}\hfill
\includegraphics[height=1.2in,width=0.45\textwidth]{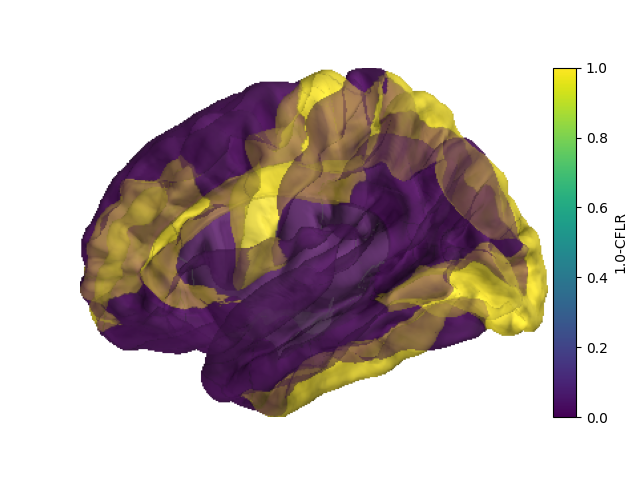}\\
\includegraphics[height=1.2in,width=0.45\textwidth]{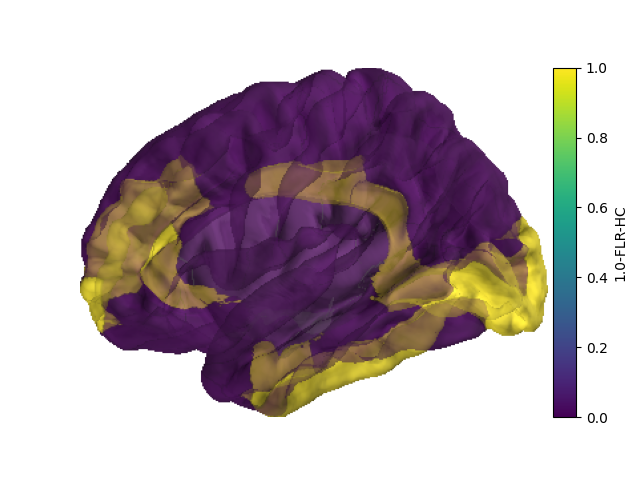}\hfill
\includegraphics[height=1.2in,width=0.45\textwidth]{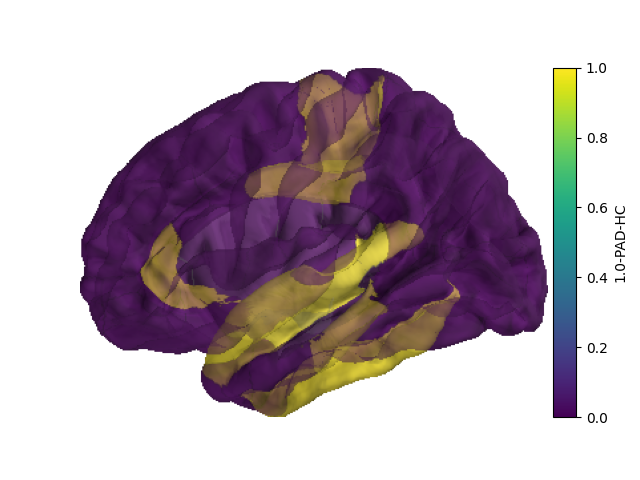}\\
\includegraphics[height=1.2in,width=0.45\textwidth]{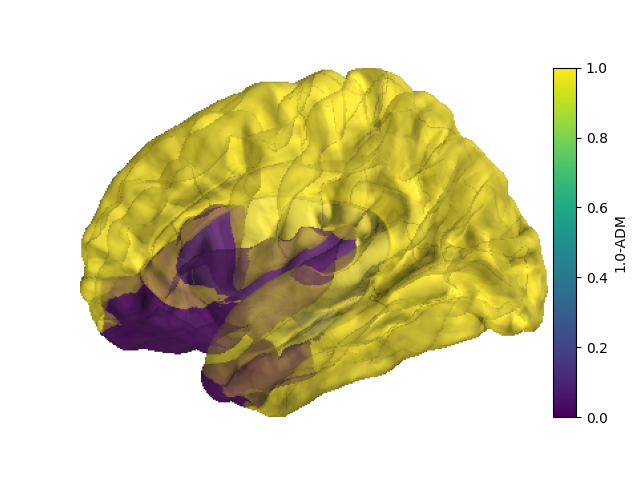}
\hfill
\includegraphics[height=1.2in,width=0.4\textwidth]{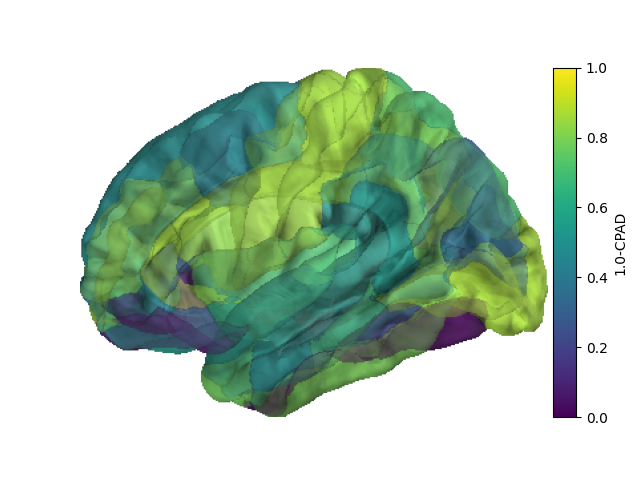}\\

\caption{\small{\noindent Plots of the adjusted (1-p)-values on left vertical areas for FLR-HC, PAD-HC, CFLR, CPAD, PMAD and ADM respectively. They can be divided into two blocks. Block 1 (rows 1 to 3) for the delta band while block 2 (rows 4 to 6) for for the gamma band. In each block, from the left to the right and the top to the bottom, the plots are made in the order FLR-HC, PAD-HC, CFLR, CPAD, PMAD and ADM.
}}
\label{Injurelh}
\end{figure}
\end{center}

\begin{center}
\begin{figure}[htp]
\includegraphics[height=1.2in,width=0.45\textwidth]{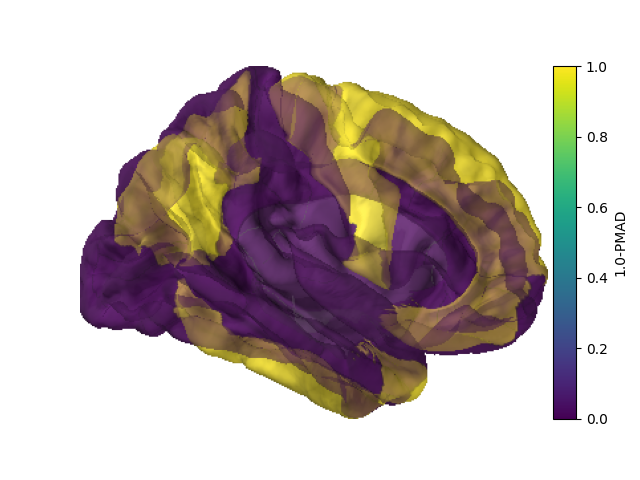}\hfill
\includegraphics[height=1.2in,width=0.45\textwidth]{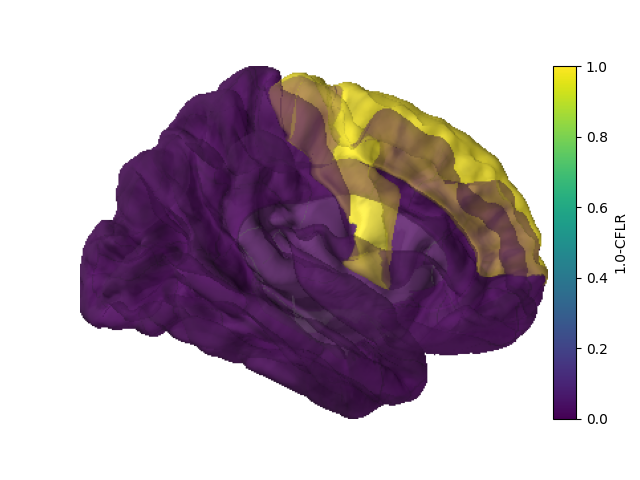}\\
\includegraphics[height=1.2in,width=0.45\textwidth]{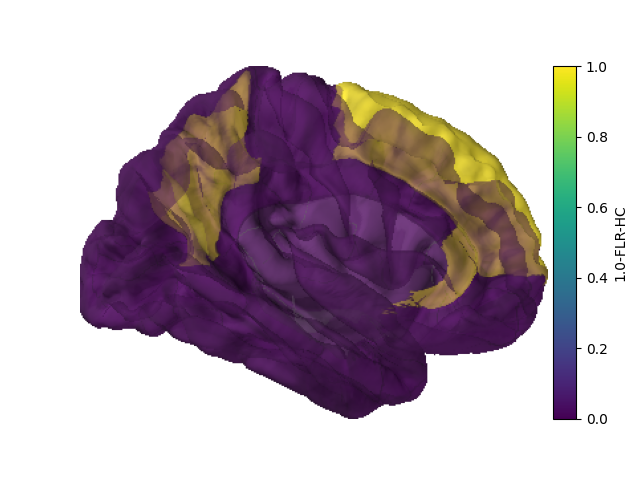}\hfill
\includegraphics[height=1.2in,width=0.45\textwidth]{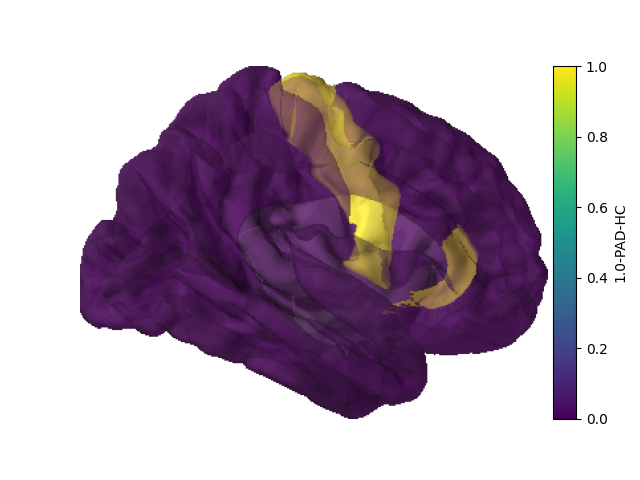}\\
\includegraphics[height=1.2in,width=0.45\textwidth]{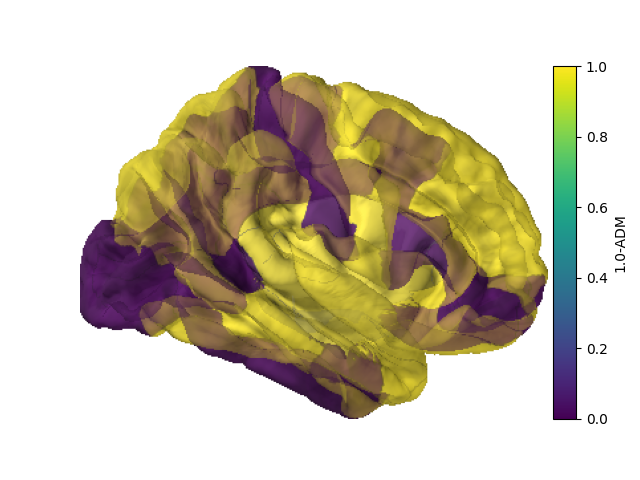}
\hfill
\includegraphics[height=1.2in,width=0.45\textwidth]{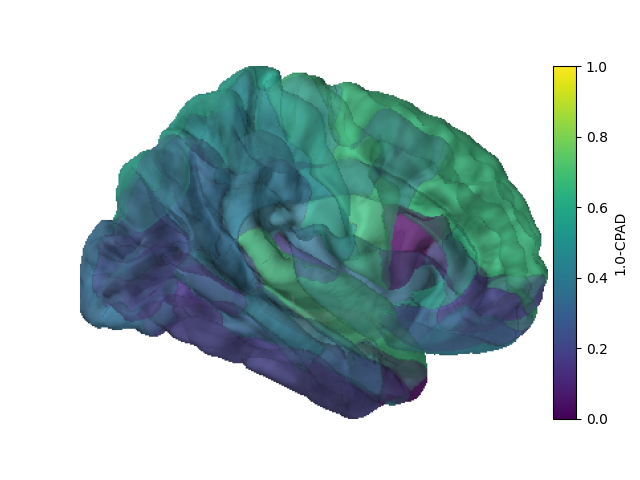}\\
\includegraphics[height=1.2in,width=0.45\textwidth]{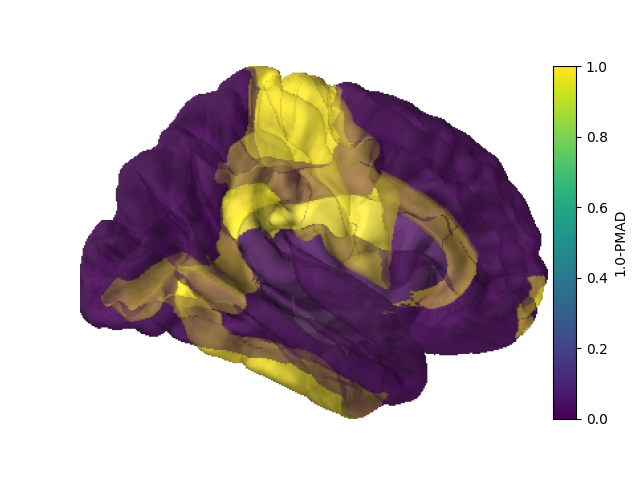}\hfill
\includegraphics[height=1.2in,width=0.45\textwidth]{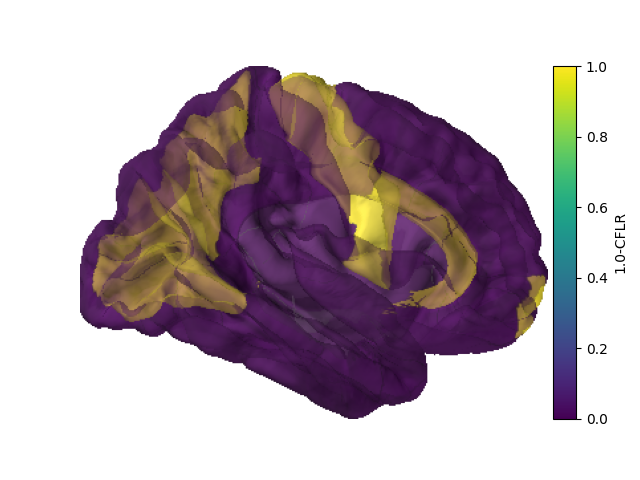}\\
\includegraphics[height=1.2in,width=0.45\textwidth]{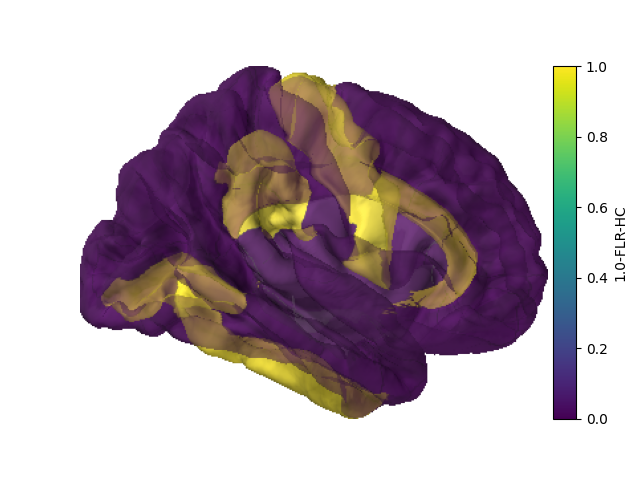}\hfill
\includegraphics[height=1.2in,width=0.45\textwidth]{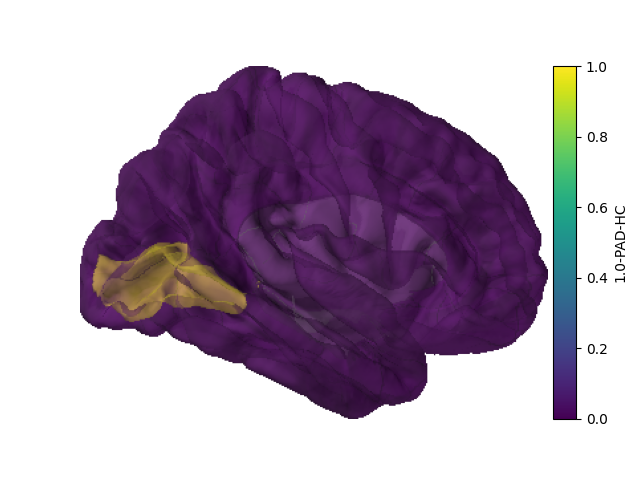}\\
\includegraphics[height=1.2in,width=0.45\textwidth]{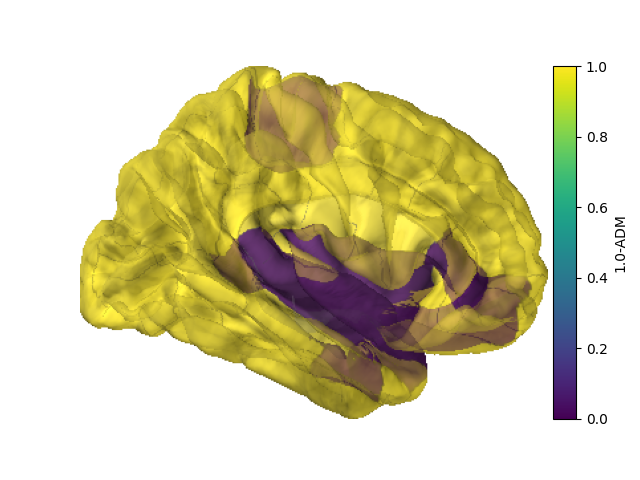}
\hfill
\includegraphics[height=1.2in,width=0.4\textwidth]{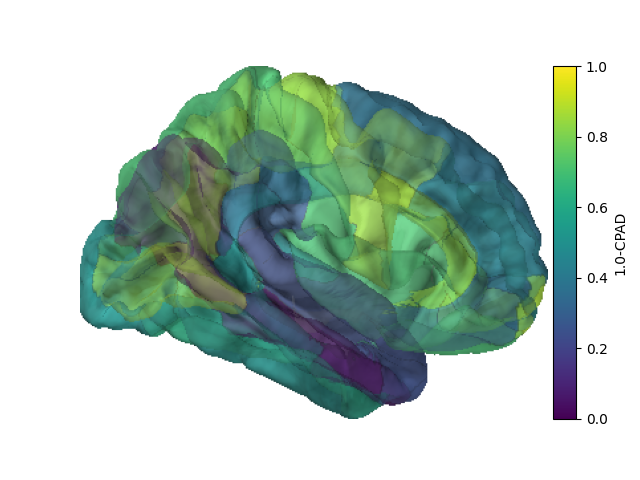}\\

\caption{\small{\noindent Plots of the adjusted (1-p)-values on right vertical areas for FLR-HC, PAD-HC, CFLR, CPAD, PMAD and ADM respectively. They can be divided into two blocks. Block 1 (rows 1 to 3) for the delta band while block 2 (rows 4 to 6) for for the gamma band. In each block, from the left to the right and the top to the bottom, the plots are made in the order FLR-HC, PAD-HC, CFLR, CPAD, PMAD and ADM.
}}
\label{Injurerh}
\end{figure}
\end{center}

 The above differences among the FLR, PAD, PMAD and ADM are clearly shown in Figures \ref{Injurelh} and \ref{Injurerh}, the $(1-p)$-value plots on the brain vertex.
Based on the delta and/or gamma band data, the above FLR-HC and PAD-HC analysis implies that when mTBI incurred, brain damages may be found in the frontal, occipital,  parietal and temporal lobes, and in cingulate gyrus, paracentral, precuneus, cuneus, lingual, fusiform, parahippocampal gyrus and entorhinal cortex. In particular, there are more damaged areas claimed in the gamma band than in the delta band. Many abnormal activities are detected in precuneus and cuneus in the delta band than in the gamma band. 
The frontal lobe, sitting at the front and top of the brain, is responsible for the highest levels of thinking and behaviour, such as planning, judgement, decision-making, impulse control, and attention. The frontal lobe contains the pars opercularis while paracentral is parts of both the frontal and parietal lobes.
The parietal lobe lying behind the frontal lobe takes in sensory information and helps an individual understand their position in their environment.
The temporal lobe in the lower front of the brain has strong links with visual memory, language, and emotion. The temporal lobe contains temporal hole, superior, middle and inferior temporal gyrus, parahippocampal/entorhinal gyri and fusiform gyrus. 
The occipital lobe at the back of the brain processes visual input from the eyes, which includes precuneus, cuneus, lingual gyrus and inferior occipital gyrus. The paracentral lobule has motor and sensory functions related to the lower limb. These facts suggest some expected changes in patient's behaviour when there were damages in these lobes.

The above findings have partially re-discovered what were found in
Huang et al. (2014) and Huang et al.(2023). Huang et al. (2014) showed that prefrontal, posterior parietal, inferior temporal, hippocampus, and cerebella areas were particularly vulnerable to brain trauma, and that MEG slow-wave generation in prefrontal areas positively correlated with personality change, trouble concentrating, affective lability, and depression symptoms. Huang et al. (2023) found that in both delta and gamma bands, the spatial differences in MEG activity in frontal and temporal lobes between a paediatric mTBI group and an orthopaedic injury control group were detected.
 
\begin{table}[htbp]
  \centering
\caption{The delta band data analysis.
}
  \begin{tabular}{ccccccc}
    \hline
    \multicolumn{1}{c}{Methods}  &\multicolumn{1}{c}{Hemisphere} &\multicolumn{4}{c}{Areas} &\multicolumn{1}{c}{Adj.p-values}\\
     \cline{3-7}
  \multirow{4}{*}{\rotatebox{90}{FLR}}\\
   \multicolumn{1}{c}{}& \multicolumn{1}{c}{lh} &\multicolumn{4}{c}{3, 4, 7, 8, 12, 14, 17, 19, 20, 23, 27, 32} &\multicolumn{1}{c}{$<0.01$}\\
  \multicolumn{1}{c}{}& \multicolumn{1}{c}{rh} &\multicolumn{4}{c}{ 42, 43, 44, 46, 49, 50, 54, 58, 59, 60, 62, 63, 67}&\multicolumn{1}{c}{$<0.01$} \\
  \multirow{4}{*}{\rotatebox{90}{FLR-HC}}\\
   \multicolumn{1}{c}{}& \multicolumn{1}{c}{lh} &\multicolumn{4}{c}{3, 4, 8, 14, 27, 32}&\multicolumn{1}{c}{$<0.01$} \\
  \multicolumn{1}{c}{}& \multicolumn{1}{c}{rh} &\multicolumn{4}{c}{42, 43, 49, 59, 60, 63, 67}&\multicolumn{1}{c}{$<0.01$} \\
  \multirow{4}{*}{\rotatebox{90}{CFLR}}\\
   \multicolumn{1}{c}{}& \multicolumn{1}{c}{lh} &\multicolumn{4}{c}{27, 32}&\multicolumn{1}{c}{$<0.01$} \\
  \multicolumn{1}{c}{}& \multicolumn{1}{c}{rh} &\multicolumn{4}{c}{60, 61, 63}&\multicolumn{1}{c}{$<0.01$} \\
 \multirow{4}{*}{\rotatebox{90}{PAD}}   \\
  \multicolumn{1}{c}{}& \multicolumn{1}{c}{lh} &\multicolumn{4}{c}{17, 25, 27, 28}&\multicolumn{1}{c}{$<0.01$} \\
 \multicolumn{1}{c}{}& \multicolumn{1}{c}{rh} &\multicolumn{4}{c}{59, 63}&\multicolumn{1}{c}{$<0.01$} \\
 \multirow{4}{*}{\rotatebox{90}{PAD-HC}}   \\
  \multicolumn{1}{c}{}& \multicolumn{1}{c}{lh} &\multicolumn{4}{c}{17,  27, 28}&\multicolumn{1}{c}{$<0.01$} \\
 \multicolumn{1}{c}{}& \multicolumn{1}{c}{rh} &\multicolumn{4}{c}{59, 63}&\multicolumn{1}{c}{$<0.01$} \\
 \multirow{4}{*}{\rotatebox{90}{CPAD}}   \\
  \multicolumn{1}{c}{}& \multicolumn{1}{c}{lh} &\multicolumn{4}{c}{17, 25, 27, 28}&\multicolumn{1}{c}{$<0.01$} \\
 \multicolumn{1}{c}{}& \multicolumn{1}{c}{rh} &\multicolumn{4}{c}{59, 61}&\multicolumn{1}{c}{$<0.01$} \\
 \multirow{4}{*}{\rotatebox{90}{PMAD}}   \\
  \multicolumn{1}{c}{}& \multicolumn{1}{c}{lh} &\multicolumn{4}{c}{1-12, 14-34}&\multicolumn{1}{c}{$<0.01$} \\
  \multicolumn{1}{c}{}& \multicolumn{1}{c}{rh} &\multicolumn{4}{c}{36-44, 46, 47, 49, 50, 52, 53, 55-68} &\multicolumn{1}{c}{$<0.01$}\\
 \multirow{4}{*}{\rotatebox{90}{ADM}}   \\
 \multicolumn{1}{c}{}& \multicolumn{1}{c}{lh} &\multicolumn{4}{c}{None}&\multicolumn{1}{c}{$<0.01$} \\
 \multicolumn{1}{c}{}& \multicolumn{1}{c}{rh} &\multicolumn{4}{c}{None}&\multicolumn{1}{c}{$<0.01$} \\
   \hline  
  \end{tabular}
  \label{tbl:delta}
\end{table}

\begin{table}[htbp]
  \centering
\caption{The gamma band data analysis.
}
  \begin{tabular}{ccccccc}
    \hline
    \multicolumn{1}{c}{Methods}  &\multicolumn{1}{c}{Hemisphere} &\multicolumn{4}{c}{Areas} &\multicolumn{1}{c}{Adj.p-values}\\
     \cline{3-7}
  \multirow{4}{*}{\rotatebox{90}{FLR}}\\
   \multicolumn{1}{c}{}& \multicolumn{1}{c}{lh} &\multicolumn{4}{c}{1, 2, 3, 5, 6, 7, 9, 11, 12, 14, 15, 18,19, 24, 27, 28, 29, 30, 31, 33, 34} &\multicolumn{1}{c}{$<0.01$}\\
  \multicolumn{1}{c}{}& \multicolumn{1}{c}{rh} &\multicolumn{4}{c}{35-37, 39, 40, 43, 45, 48-51, 57, 59, 61, 63, 65, 66}&\multicolumn{1}{c}{$<0.01$} \\
  \multirow{4}{*}{\rotatebox{90}{FLR-HC}}\\
   \multicolumn{1}{c}{}& \multicolumn{1}{c}{lh} &\multicolumn{4}{c}{2, 5, 6, 7, 9, 11, 12, 14, 18, 24, 27, 28,  31, 34}&\multicolumn{1}{c}{$<0.01$} \\
  \multicolumn{1}{c}{}& \multicolumn{1}{c}{rh} &\multicolumn{4}{c}{36, 40, 43, 45, 48, 51, 57, 59, 61, 66}&\multicolumn{1}{c}{$<0.01$} \\
  \multirow{4}{*}{\rotatebox{90}{CFLR}}\\
   \multicolumn{1}{c}{}& \multicolumn{1}{c}{lh} &\multicolumn{4}{c}{5, 6, 9, 11, 12, 14, 24, 27, 28}&\multicolumn{1}{c}{$<0.01$} \\
  \multicolumn{1}{c}{}& \multicolumn{1}{c}{rh} &\multicolumn{4}{c}{36, 43, 48, 59, 61, 66}&\multicolumn{1}{c}{$<0.01$} \\
\multirow{4}{*}{\rotatebox{90}{PAD}}   \\
  \multicolumn{1}{c}{}& \multicolumn{1}{c}{lh} &\multicolumn{4}{c}{1, 2, 4, 9, 12, 14, 16,17, 19, 23, 24, 25, 27, 28, 30-32}&\multicolumn{1}{c}{$<0.01$} \\
 \multicolumn{1}{c}{}& \multicolumn{1}{c}{rh} &\multicolumn{4}{c}{36-38, 40, 48, 49, 53, 56, 57, 59-62, 64, 66, 67}&\multicolumn{1}{c}{$<0.01$} \\
\multirow{4}{*}{\rotatebox{90}{PAD-HC}}   \\
  \multicolumn{1}{c}{}& \multicolumn{1}{c}{lh} &\multicolumn{4}{c}{2, 9, 12, 14, 17, 24, 25, 27,28, 30}&\multicolumn{1}{c}{$<0.01$} \\
 \multicolumn{1}{c}{}& \multicolumn{1}{c}{rh} &\multicolumn{4}{c}{36, 38, 40, 48, 56, 59, 60, 61}&\multicolumn{1}{c}{$<0.01$} \\
 \multirow{4}{*}{\rotatebox{90}{CPAD}}   \\
  \multicolumn{1}{c}{}& \multicolumn{1}{c}{lh} &\multicolumn{4}{c}{9, 17, 18, 24, 27, 31}&\multicolumn{1}{c}{$<0.01$} \\
 \multicolumn{1}{c}{}& \multicolumn{1}{c}{rh} &\multicolumn{4}{c}{48, 56}&\multicolumn{1}{c}{$<0.01$} \\
 \multirow{4}{*}{\rotatebox{90}{PMAD}}   \\
  \multicolumn{1}{c}{}& \multicolumn{1}{c}{lh} &\multicolumn{4}{c}{1-9, 11, 12,14-19, 21-33}&\multicolumn{1}{c}{$<0.01$} \\
 \multicolumn{1}{c}{}& \multicolumn{1}{c}{rh} &\multicolumn{4}{c}{36-68} &\multicolumn{1}{c}{$<0.01$}\\
 \multirow{4}{*}{\rotatebox{90}{ADM}}   \\
 \multicolumn{1}{c}{}& \multicolumn{1}{c}{lh} &\multicolumn{4}{c}{None}&\multicolumn{1}{c}{$<0.01$} \\
 \multicolumn{1}{c}{}& \multicolumn{1}{c}{rh} &\multicolumn{4}{c}{None}&\multicolumn{1}{c}{$<0.01$} \\
   \hline  
  \end{tabular}
  \label{tbl:gamma}
\end{table}


\section {Theory}
As the null distribution of the FLR test is difficult to calculate, the nominal significance level cannot be achieved precisely.
In this section, coupled with simulation studies, we carry out a theoretical study on its asymptotic null distribution. Under certain regularity conditions, we show that for each nominal significance level $\alpha$, a critical value $\log(1-c_0)$ can be identified to achieve the level asymptotically.

\subsection{Regular distribution families}

Log-normal distribution family and non-central t-distribution family are regular in the sense that the Fisher information matrix is positive-definite when the shape parameter is away from zero. 
Skew distribution families such as skew $t$-distribution family are also regular. See Azzalini and Capitanio (2014). Log-normal, non-central t-distribution and skew distribution families only allow for skewness but  not for multi-modality in the data. Nevertheless, we can state an asymptotic null-distribution for the likelihood ratio tests when both the case and controls are drawn from regular -distribution families as follows.

\subsection{Two-sample test}
Consider the following hypothesis
\begin{eqnarray}\label{H0a}
H_0: f(\cdot|\psi_{0})=f(\cdot|\psi_{1}) \mbox { v.s. } H_1:  f(\cdot|\psi_{0})
\not=f(\cdot|\psi_{1}),
\end{eqnarray}
where $ f(\cdot|\psi_{k})$ depends on d-dimensional parameter $\psi_{k},$ $ k= 0, 1.$
Suppose that we have an i.i.d. sample $\bX_k$ drawn from  $ f(\cdot|\psi_{k})$. Let 
$l_{x_{ki}}(\psi_k)=\log(f(x_{ki}|\psi_k))$ be the likelihood function. Suppose that the likelihood function has a second order derivative which is continuous and that the Fisher information matrix is strictly positive definite. Define the maximum likelihood ratio test statistic
\[
W_n(\bX_0,\bX_1)=\max_{\psi_0=\psi_1}\sum_{k=0}^1\sum_i l_{x_{ki}}(\psi_k)
-\max_{\psi_0}\sum_i l_{x_{0i}}(\psi_0)-\max_{\psi_1}\sum_i l_{x_{1i}}(\psi_1).
\]
 Then, the following proposition follows from a similar arguments used to prove Wilks' theorem
\begin{proposition}
Under the above regularity conditions, $-2W_n(\bX_0,\bX_1)$ converges to $ \chi^2_{d}$
in distribution as the sample size tends to infinite, where $\chi^2_{d} $ is a chi-squared distribution with $d$-degrees of freedom. 
\end{proposition}

\subsection{OK test}
The above result can be extended to the one-vs-K-sample likelihood ratio test under some regularity conditions.
For this purpose, consider the following hypothesis
\begin{eqnarray}\label{H0a}
H_0: f(\cdot|\psi_{0})\in \{f(\cdot|\psi_{k}):1\le k\le K\} \mbox { v.s. } H_1:  f(\cdot|\psi_{0})
\not\in\{f(\cdot|\psi_{k}): 1\le k\le K\},
\end{eqnarray}
where $ f(\cdot|\psi_{k})$ depends on d-dimensional parameter $\psi_{k},$ $  0\le k\le K$, and $\psi_{k}$, $1\le k\le K$ is an i.i.d. sample drawn from a hyperparameter distribution.
Suppose that we have an i.i.d. sample $\bX_k$ drawn from  $ f(\cdot|\psi_{k})$, $0\le k\le K.$ Then, the following proposition follows from a similar argument to prove Wilk's theorem.
The proposition provides a way to determine an asymptotic critical value at the level $\alpha.$

\begin{proposition}
Assume that the above regularity conditions holds. For a nominal level $0<\alpha<1,$ under $H_0$, for a large $K$, we have
\begin{eqnarray*}
W_n(\bX_0,\bX_k)&\to& \chi^2_{dk}, 1\le k\le K, \mbox{ in distribution}.\\
\hat{p}(c_0)&=&\sum_{k=1}^KI(W_n(\bX_0,\bX_k)\ge \log(1-c_0))/K\to \sum_{k=1}^KI(-0.5\chi^2_{dk}\ge \log(1-c_0))/K\\
&\approx& P(-0.5\chi^2_d\ge\log(1-c_0))=1-\alpha.
\end{eqnarray*}
\end{proposition}

\subsection{Normal-mixture distribution family}
Unlike skew distribution families, normal mixtures can allow for both multi-modality and skewness in the data.  However, Wilks' theorem no longer holds for normal mixtures, as they can be irregular in the sense that the Fisher information matrix is degenerate. In the following, using the technique of Dacunha-Castelle and Gassiat (1999), we develop the asymptotic null distribution for the proposed FLR test statistic and an empirical way to determine the critical value in the proposed test. We find that the asymptotic null distributions depend on the underlying order of mixture models. To ease the presentation, we begin with single-sample tests as follows.

\subsubsection{Single-sample test}
Let $\mathcal{G}_p$ denote the set of $p$-mixtures of normals, $g(x|\psi) $  in the form
$
g(x|\psi)=\sum_{i=1}^p\pi_i\phi(x|\eta_i), \quad 1\le p\le p_{max},
$
where parameters $\theta_i=(\mu_i,\sigma^2_i)\in \R\times \R^+, 1\le i\le p$, $\sum_{i=1}^p\pi_i=1$ and $\psi=(\pi_1,\ldots,\pi_{p},\eta_1,\ldots,\eta_p)$ is the vector of all the parameters in the mixture. Let $g_0(x)=g(x|\psi^0)
=\sum_{i=1}^p\pi^0_i\phi(x|\eta^0_i)$ denote the underlying mixture. Let $E_0$ denotes the expectation operator under density $g_0$.  Assume that
$$ \mbox{(C0): } g(x|\psi^0)\mbox{  is identifiable up to a permutation of components.}
$$
Suppose that we have a sample $\bX$ of size $n$, drawn from the unknown density $g(x|\psi).$ We want to test $H_0: g(x|\psi) =g_0(x)$ v.s. $H_1:g(x|\psi)\not =g_0(x).$
To tackle the model identifiability issue in $\mathcal{G}$, following Dacunha-Castelle and Gassiat (1999), we locally reparamametrise $g(x|\psi)$ around $\psi_0$ by a perturbation of $g_0$ in the form
\begin{eqnarray*}
g(x|\theta,\beta)=\sum_{i=1}^{p-p_0}\frac{\lambda_i\theta}{n(\beta)}\phi(x|\eta_i)
                        +\sum_{l=1}^{p_0}\left(\pi^0_i+\frac{\rho_l\theta}{n(\beta)}\right)\phi\left(x|\eta^0_l+\frac{\theta\delta_l}{n(\beta)}\right),
\end{eqnarray*}
where $\theta\in [0,\theta_g]\subset \bR^+$ is an identifiable parameter, while
 $\beta=(\lambda_1,\ldots,\lambda_{p-p_0},\delta_1,\ldots,\delta_{p_0},\rho_1,\ldots,\rho_{p_0})$ contains non-identifiable parameters. It can be shown that $g(x|\theta,\beta)$ is a proper density function if $\beta\in\mathcal{B}$ with
\begin{eqnarray*}
\mathcal{B}&=&\{\beta: \lambda_i\ge 0, \eta_i\in \bR\times \bR^+, 1\le i\le p-p_0; \delta_l\in\bR\times\bR^+,\rho_l\in\bR,  1\le l\le p_0;\\
&& \sum_{i=1}^{p-p_0}\lambda_i+\sum_{l=1}^{p_0}\rho_l=0; 
\sum_{i=1}^{p-p_0}\lambda_i^2+\sum_{l=1}^{p_0}\rho_l^2+\sum_{l=1}^{p_0}||\delta_l||^2=1\},
\end{eqnarray*}
where $n(\beta)$ is a normalisation factor such that $E_0[(\frac{\partial g(X|0,\beta)}{\partial\theta})^2]=1$. 
Letting $l(x|\theta,\beta)=\log(g(x|\theta,\beta),$ we have 
$\frac{\partial l(x|0,\beta)}{\partial \theta}=\frac{\partial g(x|0,\beta)}{\partial \theta}/g_0(x)$.  Note that $E_0[\frac{\partial l(X|0,\beta)}{\partial \theta}]=0$. Letting $\bigtriangledown$ denote the gradient operator,  we have the directional Fisher information 
$I(0,\beta)^2=E_0[(\frac{\partial l(x|0,\beta)}{\partial \theta})^2] 
=-E_0[\frac{\partial^2 l(x|0,\beta)}{\partial \theta^2}]=1
$
when the squared normalising factor $n(\beta)^2$ satisfies
\[
n(\beta)^2=E_0\left[\left(\sum_{i=1}^{p-p_0}\lambda_i\phi(X|\eta_i)+\sum_{l=1}^{p_0}\rho_l\phi(X|\eta^0_l)+\sum_{l=1}^{p_0}\pi^0_l\delta_l^T\bigtriangledown\phi(X|\eta^0_l)\right)^2g_0(X)^{-2}\right].
\]
Assume that

(C1): For $1\le i\le p$, $\eta_l\in\Gamma$, a compact set of $\bR\times\bR^+$, and
$\sigma^{2}_l$ is uniformly bounded below from $0$.

Following Dacunha-Castelle and Gassiat (1999) and Keribin (2000), under Conditions (C0) and (C1), for $g\in \mathcal{G}_p$, we restrict its conic parameters $\beta$ to a compact set. Then we find a small interval $[0,\theta_g]$ for $\theta$ and define a conic neighborhood of $g_0$, 
$\{g(x|\theta,\beta): \theta\in [0,\theta_g]\} $. Define  $\hat{p}_x=\argmax_{1\le p\le p_{max}}(W_{np}(X)-0.5\log(n)(3p-1) $, a BIC estimator of order $p_0$.
Keribin (2.30) proved that  as the sample size $n$ tends to infinity,
$
\hat{p}_x\to p_0.
$
Define $\mathcal{D}_p$ as the set of functions of form
\[
d(x|\beta)=\frac 1{n(\beta)g_0(x)}\left(\sum_{i=1}^{p-p_0}\lambda_i\phi(x|\eta_i) 
+\sum_{l=1}^{p_0}\rho_l\phi(x|\eta^0_l)+\sum_{l=1}^{p_0}\pi^0_l\delta_l^T\bigtriangledown\phi(x|\eta^0_l)\right), \beta\in \mathcal{B}.
\]
Define the log-likelihood ratio $W_{np}(X)=\sup_{g\in\mathcal{G}_p}\sum_{i=1}^n\log(g(X_i)/g_0(X_i)).$
Let $W_1(d)$ be a Gaussian process indexed by $\mathcal{D}$ with covariance defined by the usual $L_2$ product. Let $I(\cdot)$ be an indicator. Then, we have
$$
(W_{np}(X))_{1\le p\le p_{max}} \to 0.5(\sup_{d\in \mathcal{D}_p}W_1(d)^2I(W(d)\ge 0))_{1\le p\le p_{max}}
$$
in distribution as $n$ tends to infinity. Using Slutsky's theorem, we have, as $n$ tends to infinity,
$$
W_{n\hat{p}_x}(X) \to 0.5\sup_{d\in \mathcal{D}_{p_0}}W_1(d)^2I(W_1(d)\ge 0)= 0.5\sup_{d\in \mathcal{D}_{p_0}}W_1(d)^2
$$ 
in distribution. Note that the last equality follows from the fact that $ \mathcal{D}_{p_0} $ is a symmetric set.

\subsubsection{Two-sample test}
 Suppose that we have two samples $\bX=(X_1,\ldots,X_n)$ and $\bY=(Y_1,\ldots,Y_n)$ generated from $g(x|\psi_x)$ and $g(x|\psi_y)$ respectively. We want to test the null hypothesis 
$H_0: g(x|\psi_x)=g(x|\psi_y)$. Define the following two-sample log-likelihood test statistic
\begin{eqnarray*}
W_{np}(\bY,\bX)=\sup_{g\in \mathcal{G}_p}\sum_{i=1}^n\log(g(X_i)g(Y_i))-W_{np}(\bX)-W_{np}\b(Y)
\end{eqnarray*}
Let
\begin{eqnarray*}
\hat{p}_y&=&\argmax_{1\le p\le p_{max}}W_{np}(\bY)-0.5\log(n)(3p-1),\\
\hat{p}_{x,y}&=&\argmax_{1\le p\le p_{max}}W_{np}(\bY,\bX)-0.5\log(2n)(3p-1),\\
W_{n\hat{p}}(\bY,\bX)&=&\sup_{g\in \mathcal{G}_{\hat{p}_{x,y}}}\sum_{i=1}^n\log(g(X_i)g(Y_i))-W_{n\hat{p}_x}(\bX)-W_{n\hat{p}_y}(\bY).
\end{eqnarray*}
Similar to before, we can show that $\hat{p}_x, \hat{p}_y$ and $\hat{p}_{x,y}$ all converge to $p_0$ in probability. Furthermore, let $\{(W_1(d),W_2(d)): d\in\mathcal{G}_p\}$ denote two independent Gaussian process with covariance matrix defined by $L_2$ product as before. Let $W_{1p_0}$ denote $0.5\sup_{d\in \mathcal{D}_{p_0}}(W_{1}(d)+W_2(d))^2$ $
-\sup_{d\in \mathcal{D}_{p_0}}W_1(d)^2-\sup_{d\in \mathcal{D}_{p_0}}W_2(d)^2.$
Then, we have
\begin{proposition}
Under the conditions (C0) and (C1),
 as $n$ tends to infinity,
\begin{eqnarray*}
(W_{np}(\bY,\bX),W_{np}(\bX),W_{np}(\bY)) &\to& 0.5 (\sup_{d\in \mathcal{D}_{p_0}}(
W_{1}(d)+W_{2}(d))^2I(W_1(d)+W_2(d)\ge 0)), \\
&&\qquad\sup_{d\in \mathcal{D}_{p_0}}
W_1(d)I(W_1(d)\ge 0),\sup_{d\in \mathcal{D}_{p_0}}W_2(d)I(W_1(d)\ge 0)),
\end{eqnarray*}
and
\begin{eqnarray*}
W_{n\hat{p}}(\bY,\bX)&\to& 0.5\sup_{d\in \mathcal{D}_{p_0}}(
W_{1}(d)+W_2(d))^2I(W_1(d)+W_2(d)\ge 0)\\
&&-\sup_{d\in \mathcal{D}_{p_0}}W_1(d)^2I(W_1(d)\ge 0)-\sup_{d\in \mathcal{D}_{p_0}}W_2(d)^2I(W_2(d)\ge 0))=W_{1p_0}
\end{eqnarray*}
which depends on $p_0$.
\end{proposition}
{\bf Proof:} It follows from Dacunha-Castelle and Gassiat (1999), Keribin (2000) and Slutsky's theorem.

\subsubsection{OK test}
In a single-case study, we aim to test a single subject again $m$ controls. The case density and control densities are modelled by normal mixtures $g(x|\psi)$ and $g(y|\psi_k)$, $1\le k\le K,$ respectively, where $g(y|\psi_k)$ is assumed to have the order $p_k\sim$ $ \pi(q),$ $1\le q\le p_{max}$.
 Suppose that we have samples of size $n$ for the case and controls, say $\bY$, $\bX_1$, $\ldots$, $\bX_K$. The null hypothesis $H_0$ is that the case comes from the control group. For each pair $(\bY,\bX_k)$, we construct a likelihood ratio test statistic 
$W_{n\hat{p}_k}(\bY,\bX_k)$. For any $c_0$, count the number of times that $W_{n\hat{p}_k}(\bY,\bX_k)$ is larger than or equal to $\log(1-c_0)$, and define a p-value by
$
\hat{p}(c_0)= \sum_{k=1}^K I(W_{n\hat{p}_k}(\bY,\bX_k)\ge\log(1-c_0))/K.
$
 We have
\begin{proposition}
Under the conditions (C0) and $(C1)$, for large $K$, as the sample size tends to infinity,
\begin{eqnarray*}
\hat{p}(c_0)&\to&\sum_{k=1}^K I(W_{p_k}\ge\log(1-c_0))/K\approx \int P(W_{q}\ge\log(1-c_0))d\pi(q)
\end{eqnarray*}
in probability.
\end{proposition}

{\bf Proof:} As in the previous subsections, under Conditions (C0) and (C1), we show that under $H_0,$
$(W_{n\hat{p}_k}(\bY,\bX_k))_{1\le k\le K}$ converges to $(W_{p_k})_{1\le k\le K}$ in distribution. The result follows straightforward.

\subsubsection{Bootstrap cross-validation }
Quantifying uncertainties in the estimated p-value $\hat{p}(c_0)$ is important in determining the tuning constant $c_0$. A common approach to such an uncertainty quantification is using bootstrap samples to estimate how extreme the estimated p-value is compared to its bootstrapped null distribution. 

To derive the bootstrapped null distribution, we need to modify Conditions (C0) as follows:

(C0a): There is a small Kullback-Leibler neighborhood of $\phi(x|\psi_0)$ in which the normal mixture
$\phi(x|\psi)$ is identifiable.

\begin{proposition}
Under Conditions (C0a) and (C1), the bootstrap p-value $\hat{\mbox{cp}}(c_0)$ will convergences to a $c_0$-dependent limit in probability.
\end{proposition}

{\it Proof:} It follows from the uniform convergence theorem of empirical processes. See van der Vaart (1998). 

Propositions 4 and 5 imply that $\arg\min_{c_0}(\hat{p}(c_0)+\hat{cp}(c_0))$ will converges to its theoretical value under certain regularity conditions.

 \section {Discussion and conclusion}
Modelling and testing complex resting-state MEG scan data for abnormality in an mTBI patient is challenging due to high subject-variability and nonspecificity of posttraumatic symptoms, for example when differentiating between mild cognitive impairment and normal aging-induced cognitive decline. There has been a significant surge in using MEG source imaging to find abnormal regions in an mTBI patient (Huang et al., 2014;  It\"{a}linna et al., 2023; Allen et al., 2021, among others).
  Commonly used hypothesis tests for finding diagnostic biomarkers of mTBI are based on group means, regarding individual differences as errors or noises. These statistical tests, implicitly assuming homogeneity within the case-control groups, are fundamentally oriented to comparing the 'average case’ against 'average control'.  In particular, in a recent group study, Huang et al. (2023) reached  a sensitivity of $95\%$ and a specificity of $90\%$  in pediatric mTBI when combining delta and gamma band
specific features under a traditional case-control framework. However, the findings from their study may not be generalisable to single-subject studies, where a single case is compared to a group of potentially heterogeneous controls as demonstrated in this paper.  

 Here, we have proposed a double-mixture based likelihood ratio testing procedure by using the existing R-software such as Mclust (Scrucca et al., 2023) along with a modified pairwise Anderson-Darling type test. To alleviate the effect of heterogeneity on statistical significance of a test, we have introduced a cross-validation based calibration  of the resulting p-values and a hiearchical clustering based visualisation and correction for subject-heterogeneity. To understand the behavior of the proposed procedure, we have also established an asymptotic theory for the test statistic.

 By a real data application and simulation studies, we show a strong performance of the proposed testing procedure. In particular, we have demonstrated that the proposed likelihood ratio test can substantially outperform the conventional nonparametric tests such as the Anderson-Darling tests  in a wide range of scenarios. By hiearchical clustering based test visualisation, we have demonstrated why the proposed FLR performs better than the PAD in terms of ability in separating a case from heterogeneous controls.  Based on the MEG source localisation in the delta- and gamma-band,  using the proposed likelihood ratio test and the modified Anderson-Darling test, we have shown that abnormal brain areas in mTBI patients can be detected when compared to healthy controls with an overall accuracy, $F_1$ score around $82\%$, even in the presence of data skewness,  multimodality and subject-heterogeneity in the case and controls. 
 In the real data analysis, we have demonstrated that the proposed likelihood ratio test is more sensitive in finding abnormal areas in the brain than the other methods such as the pairwise Anderson-Darling test, the permuted Anderson-Darling test and the Anderson-Darling test on mean-shifts.  The significant regions found in our study to be located in the frontal, occipital, paracentral,  parietal and temporal lobes and in cingulate gyrus and cuneus of the brain.

 We have shown that it is likely that the control group includes subjects whose brain activities in some areas are barely distinguishable from those of an mTBI patient.
 This implies that visualising the inter-individual variability in cases and controls is very important for improving the accuracy of diagnosis for an mTBI subject.
Note that  increased neurral oscillatory activities in the delta- and gamma-bands is most frequent finding in mTBI patients (see Allen et al., 2021 and reference therein).
The results obtained from the real data analysis are thus in line with the previous literature. Note that interpretability of a prediction in a medical context is important due to safty concern: clinicians want to minimise possible errors in the prediction. So, revealing heterogeneity in controls highlights the need to focus on abnomalities at an individual-level rather than a typical mTBI patient as in a traditional case-control paradigm.

While our simulation studies have assumed that a sample for each subject has been generated from an inverse MEG imaging, in practical situations these samples are obtained by a preprocessing step:  Estimate the source spectrum data from the MEG scan. An important point of future research in this field will account for uncertainty in this preprocessing step when applying the OK test for making a clinical decision.

\section*{Data and Code Availability}

Codes to reproduce simulation results are in https://github.com/zhangjsib/OKtests. A software for source magnitude imaging has been developed by the Innovision IP Ltd for its business and not publicly available. While the controls in the real data come from the Cambridge
Centre for Ageing and Neuroscience (Cam-CAN) dataset (Shafto et al., 2014), the data for the case subject are private.

\section*{Author Contributions}
G.G. conceptualisation, methodology, data collection and source imaging, review \& editing of original draft;
J.Z. conceptualisation, methodology, coding, data analysis, writing and editing of original draft,  funding acquisition and project administration. 
\section*{Funding}
The research of J.Z. is supported by the EPSRC grant (EP/X038297/1) and the Innovate UK
KTP grant (with reference 13481). 

\section*{Declaration of Competing Interests}
 J.Z. has no competing interests.
G.G. is an employee of Innovision IP Ltd which provides commercial reports on
individuals who may have had a head injury.

\section*{Acknowledgements}
We are grateful to the University of Kent for setting this project as an impact case study and to Innovision IP Ltd for sharing the data and the software Flimal with us.

\section*{Supplementary Material}
Online Supplementary Materials are available in https://github.com/zhangjsib/OKtests.
In the Online Supplementary Materials, a pdf file contains extra figures derived from the numerical analysis and real data analysis and {\it  the Desikan-Killiany Atlas}  referred in the paper.


\section{Appendix}

\Bigskip
The Supplementary Materials are organised as follows. In Section 1,  we provide a list of brain areas of interest in terms of  the Desikan-Killiany Atlas. In Section 2, we presented extra simulation results for sample size $N=150$. In Section 3, we present the results of the real data analysis by applying the FLR-HC and PAD-HC to the abnormal areas claimed by the FLR and PAD respectively. In Section 4, we presented simulation results on 
the abilities of the PAD-HC and the FLR-HC in capturing heterogneity in controls and in seperating the case from controls.


\newpage

\section*{Brain areas of interest:  the Desikan-Killiany Atlas}
\begin{verbatim}
Index  Region Name
1       ctx-lh-bankssts
2       ctx-lh-caudalanteriorcingulate
3       ctx-lh-caudalmiddlefrontal
4       ctx-lh-cuneus
5       ctx-lh-entorhinal
6       ctx-lh-frontalpole
7       ctx-lh-fusiform
8       ctx-lh-inferiorparietal
9       ctx-lh-inferiortemporal
10     ctx-lh-insula
11     ctx-lh-isthmuscingulate
12     ctx-lh-lateraloccipital
13     ctx-lh-lateralorbitofrontal
14     ctx-lh-lingual
15    ctx-lh-medialorbitofrontal
16    ctx-lh-middletemporal
17    ctx-lh-paracentral
18    ctx-lh-parahippocampal
19    ctx-lh-parsopercularis
20    ctx-lh-parsorbitalis
21    ctx-lh-parstriangularis
22    ctx-lh-pericalcarine
23    ctx-lh-postcentral
24    ctx-lh-posteriorcingulate
25    ctx-lh-precentral
26    ctx-lh-precuneus
27    ctx-lh-rostralanteriorcingulate
28    ctx-lh-rostralmiddlefrontal
29    ctx-lh-superiorfrontal
30    ctx-lh-superiorparietal
31    ctx-lh-superiortemporal
32    ctx-lh-supramarginal
33    ctx-lh-temporalpole
34    ctx-lh-transversetemporal
35    ctx-rh-bankssts
36    ctx-rh-caudalanteriorcingulate
37    ctx-rh-caudalmiddlefrontal
38    ctx-rh-cuneus
39    ctx-rh-entorhinal
40    ctx-rh-frontalpole
41    ctx-rh-fusiform
42    ctx-rh-inferiorparietal
43    ctx-rh-inferiortemporal
44    ctx-rh-insula
45    ctx-rh-isthmuscingulate
46    ctx-rh-lateraloccipital
47    ctx-rh-lateralorbitofrontal
48    ctx-rh-lingual
49    ctx-rh-medialorbitofrontal
50    ctx-rh-middletemporal
51    ctx-rh-paracentral
52    ctx-rh-parahippocampal
53    ctx-rh-parsopercularis
54    ctx-rh-parsorbitalis
55    ctx-rh-parstriangularis
56   ctx-rh-pericalcarine
57   ctx-rh-postcentral
58   ctx-rh-posteriorcingulate
59   ctx-rh-precentral
60  ctx-rh-precuneus
61  ctx-rh-rostralanteriorcingulate
62  ctx-rh-rostralmiddlefrontal
63  ctx-rh-superiorfrontal
64  ctx-rh-superiorparietal
65  ctx-rh-superiortemporal
66  ctx-rh-supramarginal
67  ctx-rh-temporalpole
68  ctx-rh-transversetemporal
\end{verbatim}


\section*{Simulation results}
\begin{center}
\begin{figure}[htp]
\includegraphics[height=1.3in,width=0.4\textwidth]{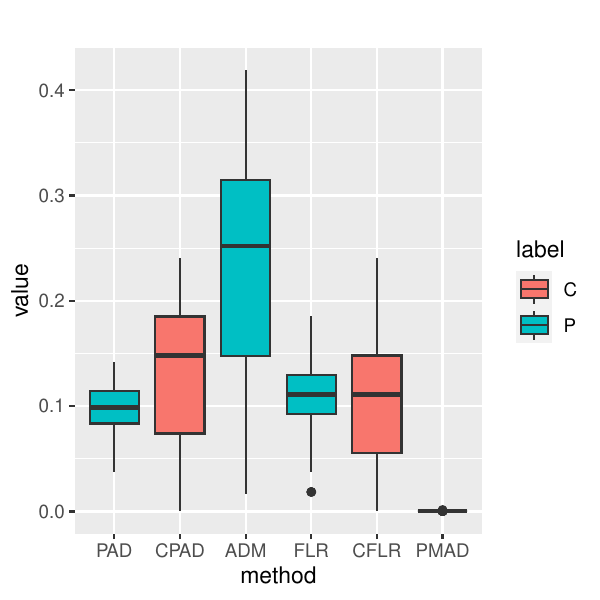}\hfill
\includegraphics[height=1.3in,width=0.6\textwidth]{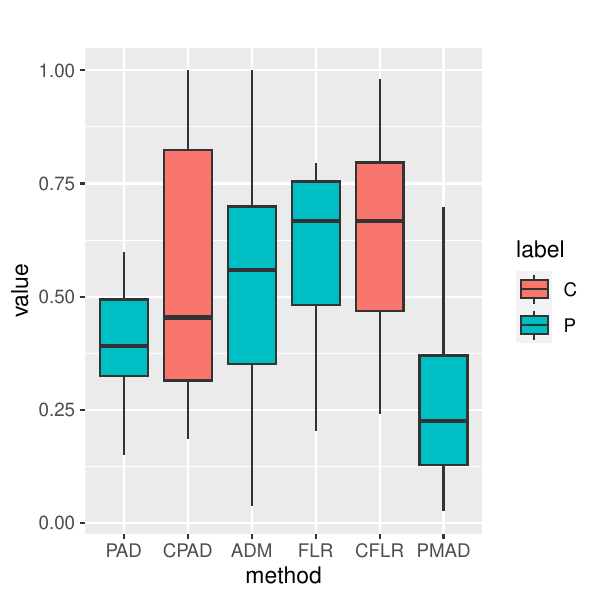}\\
\includegraphics[height=1.3in,width=0.4\textwidth]{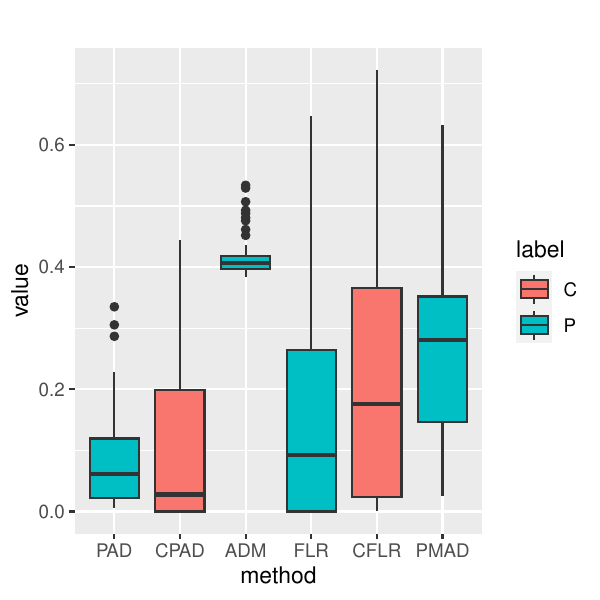}\hfill
\includegraphics[height=1.3in,width=0.6\textwidth]{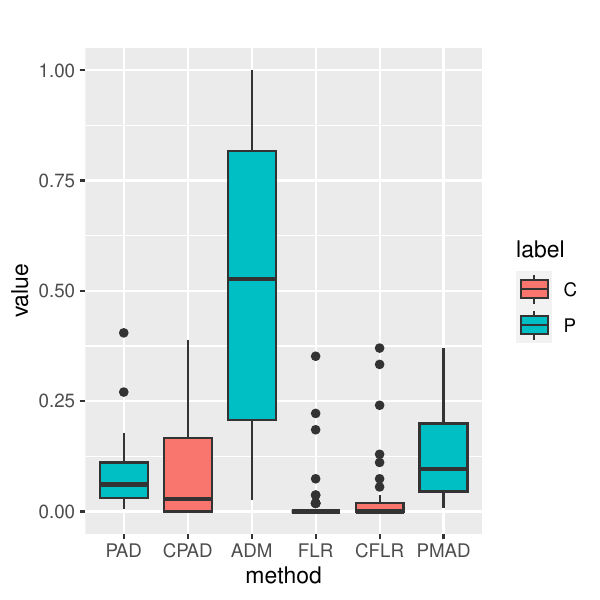}\\
\includegraphics[height=1.3in,width=0.4\textwidth]{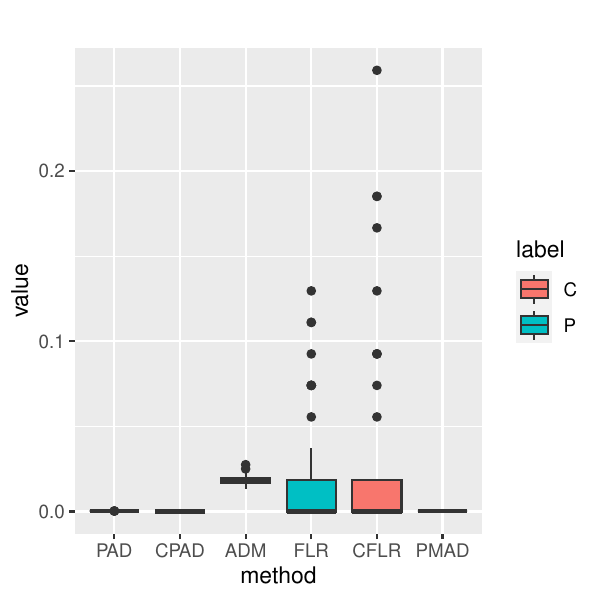}
\hfill
\includegraphics[height=1.3in,width=0.6\textwidth]{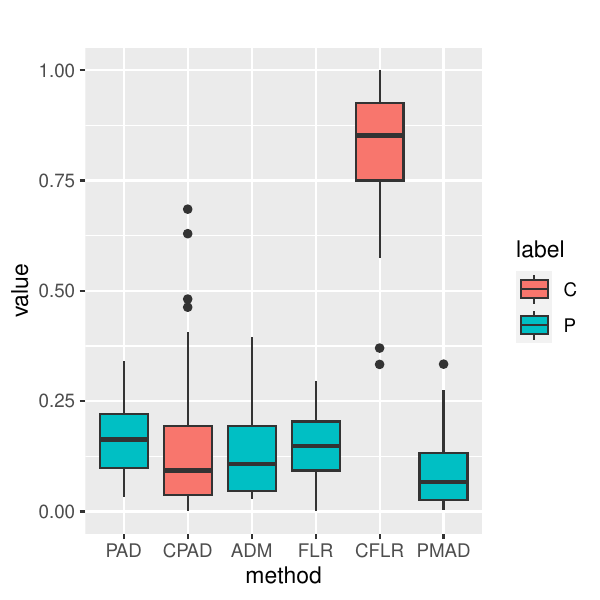}\\
\includegraphics[height=1.3in,width=0.4\textwidth]{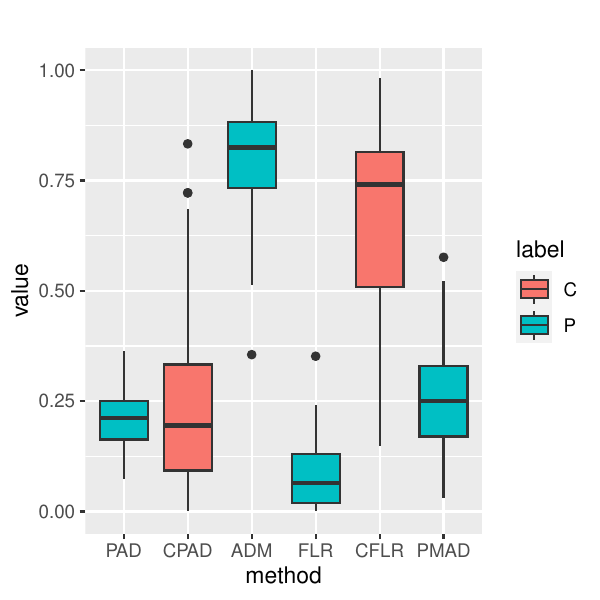}
\hfill
\includegraphics[height=1.3in,width=0.6\textwidth]{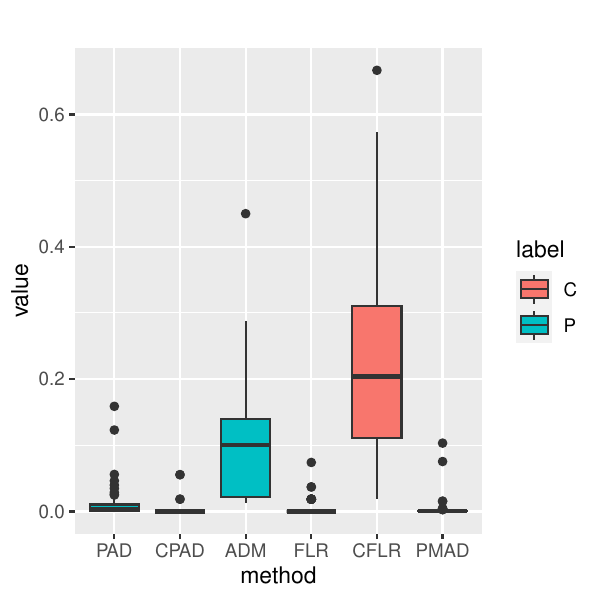}\\
\includegraphics[height=1.3in,width=0.4\textwidth]{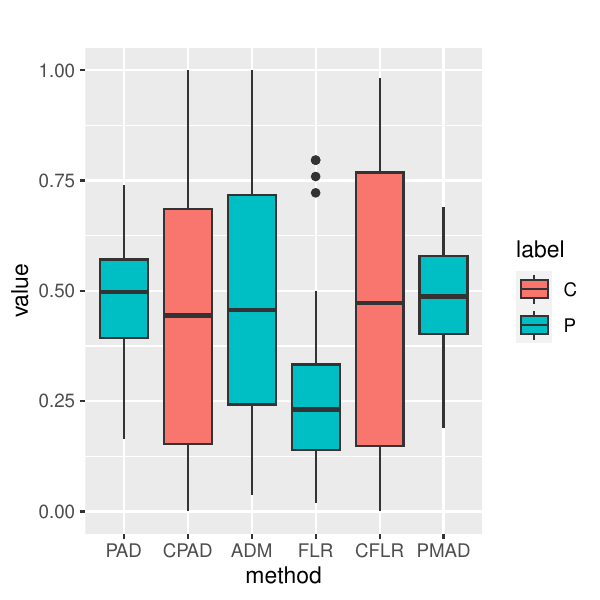}
\hfill
\includegraphics[height=1.3in,width=0.6\textwidth]{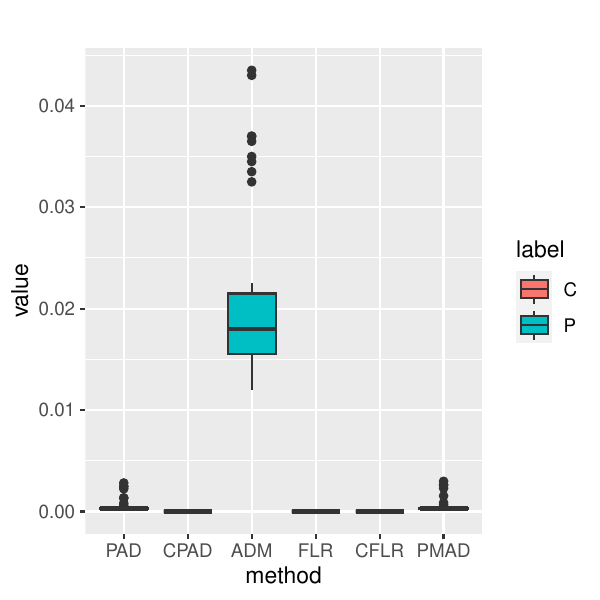}\\
\includegraphics[height=1.3in,width=0.6\textwidth]{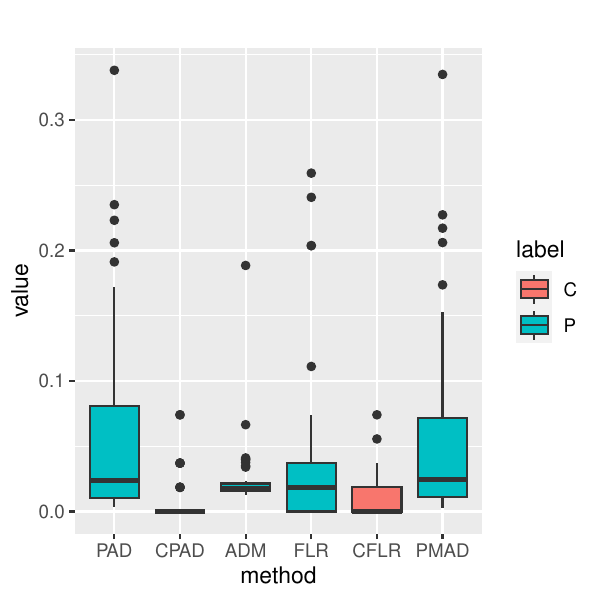}\\
\caption{\small{\noindent Settings 1, 2.3 for $N=150$. P-value plots from left to right and from top row to down row respectively. Rows 1 to 5 are corresponding to testing each of 5 cases against the Controls respectively. 
}}
\label{Set1:150}
\end{figure}
\end{center}

 Table \ref{tab2:simulation} and Figure \ref{Set1:150} show simulated results for the sample size $N=150.$

\centerline { [Put Table \ref{tab2:simulation} and Figure \ref{Set1:150}  here.]}

\begin{table}[htbp]
    \centering
    \caption{Percentages of instances with p-value less than $0.05$ and metrics: $N=150$. }
    \begin{tabular}{cccccccc}
        \toprule
   Setting/Metrics    & $ $ &  &  & Method &  &   &\\
      \cline{1-8}
        &  & FLR & CFLR  & PAD & CPAD & PMAD  & ADM\\
      \cline{3-8}
        \multirow{1}{*}{1.1} 
        &  & 0.06 &  0.2 &  0.06 &  0.16 &  1&0.02\\
        \cmidrule{3-8}
        \multirow{1}{*}{1.2}
        &  & 0 &  0 &  0.04 &  0  &  0& 0.08\\
        \cmidrule{3-8}
        \multirow{1}{*}{1.3}
        &  & 0.42 &  0.32&  0.46 &  0.56 & 0.04& 0\\
        \cmidrule{3-8}
        \multirow{1}{*}{1.4}
        &  & 0.92 &  0.86 &  0.44 &  0.56 &  0.26&0.1\\
       \cmidrule{3-8}
        \multirow{1}{*}{1.5}
        &  & 1    &  1    &  1         &  1         &  1        & 1\\
 \cmidrule{3-8}
Precision &  & {\bf 0.98} & 0.92  & 0.97 & 0.93 & 0.57  & 0.92\\
Recall &  &{\bf 0.78} & 0.73  & 0.63 & 0.71 & 0.43  & 0.37\\
$F_1$ &  & {\bf 0.87} & 0.81  & 0.76 & 0.81 & 0.49  & 0.53\\
$F_{0.5}$ &  & {\bf 0.93} & 0.72  & 0.87 & 0.87 & 0.53  & 0.71\\
$F_{2}$ &  & {0.81} & {\bf 0.91}  & 0.68 & 0.75 & 0.45  & 0.37\\
   \cline{1-8}
        \multirow{1}{*}{2.1}
        &  & 0.12 &  0&  0.06 &  0.3 & 0.44& 0.28\\
        \cmidrule{3-8}
        \multirow{1}{*}{2.2}
        &  & 0.36&  0 &  0 &  0.14 &  0.04&0\\
       \cmidrule{3-8}
        \multirow{1}{*}{2.3}
        &  & 0.98       &  0.1   &  0.94     &  0.96    &  0.96        & 0.28\\
 \cmidrule{3-8}
Precision &  &{\bf  0.92} & 1  & 0.94 & 0.79 & 0.69  & 0.50\\
Recall &  & {\bf 0.67} & 0.05  & 0.47 & 0.55 & 0.50  & 0.14\\
$F_1$ &  & {\bf 0.78} & 0.1  & 0.63 & 0.65 & 0.58 & 0.22\\
$F_{0.5}$ &  & {\bf 0.86} & 0.21  & 0.78 & 0.73 & 0.71  & 0.33\\
$F_{2}$ &  & {\bf 0.71} & 0.06  & 0.52 & 0.59 & 0.53  & 0.16\\
 \cmidrule{1-8}
        \multirow{1}{*}{3.1}
        &  & 0.12&  0.06&  0 &  0.04 & 0& 0.02\\
        \cmidrule{3-8}
        \multirow{1}{*}{3.2}
        &  & 1&  1&  1 &  1 &  1&1\\
       \cmidrule{3-8}
        \multirow{1}{*}{3.3}
        &  & 0.8        &  0.96   &  0.4     &  0.96     &  0.68        & 0.96\\
 \cmidrule{3-8}
Precision &  &{ 0.94} & 0.97  & {\bf 1} & 0.98 & 1  & 0.99\\
Recall &  & 0.82 & 0.94  & 0.70 &{\bf 0.98} & 0.72  & 0.90\\
$F_1$ &  & 0.88 & 0.95  & 0.82 & {\bf 0.98} & 0.91 & 0.98\\
$F_{0.5}$ &  & 0.91 & { 0.96}  & 0.92 &{ 0.98} & 0.96  & {\bf 0.99}\\
$F_{2}$ &  & 0.84 & 0.95  & 0.74 & {\bf 0.98} & 0.74 & 0.98\\
        \bottomrule
    \end{tabular}
\label{tab2:simulation}
\end{table}

\section*{Comparison of PAD-HC and FLR-HC}
\begin{center}
\begin{figure}[htp]

 \begin{subfigure}[b]{0.5\textwidth}
      \caption{Setting 1.1}
\includegraphics[height=3.5in,width=\textwidth]{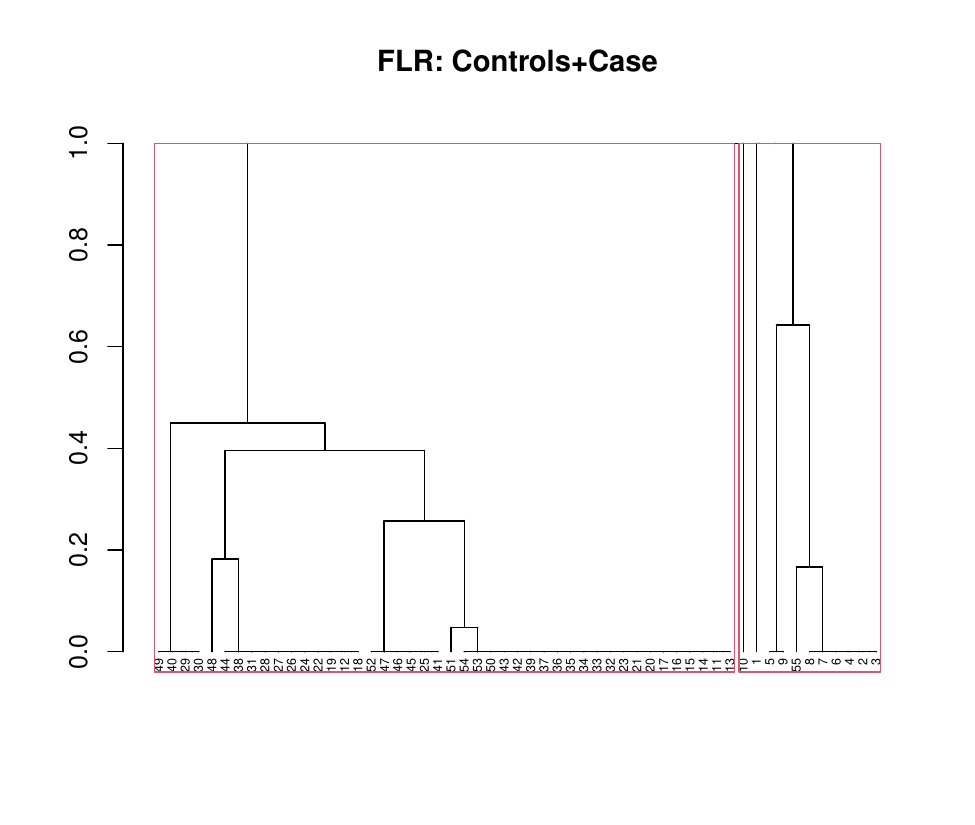}
    \end{subfigure}
 \begin{subfigure}[b]{0.5\textwidth}
      \caption{Setting 1.1}
\includegraphics[height=3.5in,width=\textwidth]{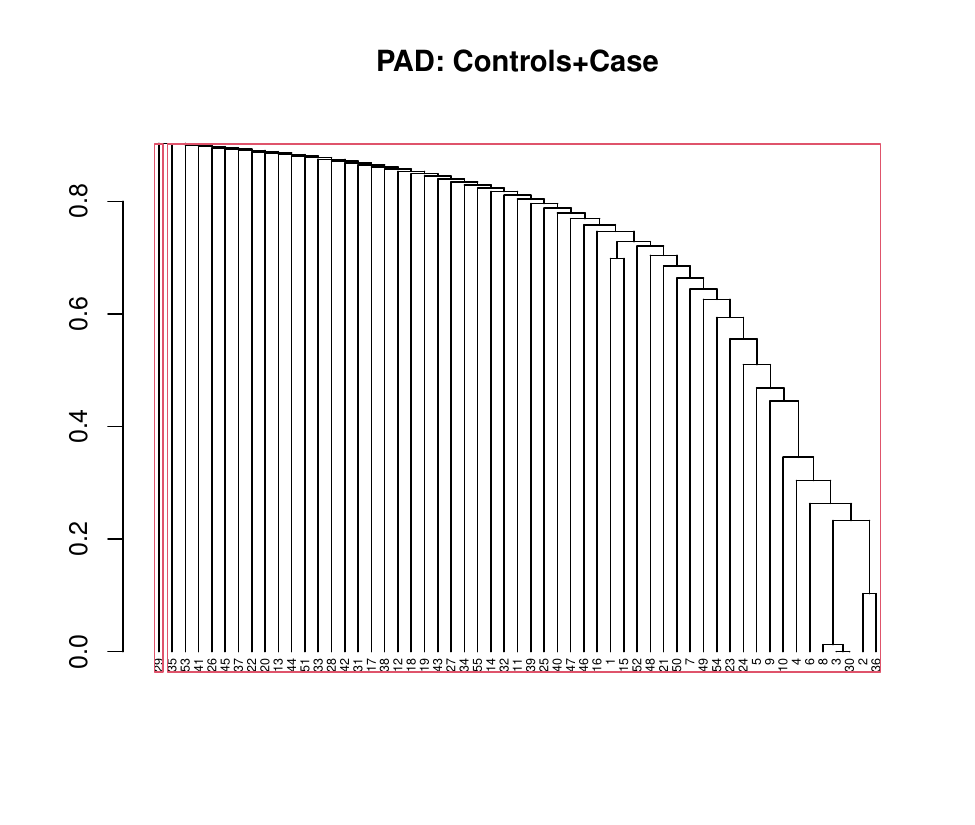}
    \end{subfigure}
 \begin{subfigure}[b]{0.5\textwidth}
     \caption{Setting 1.3}
\includegraphics[height=3.5in,width=\textwidth]{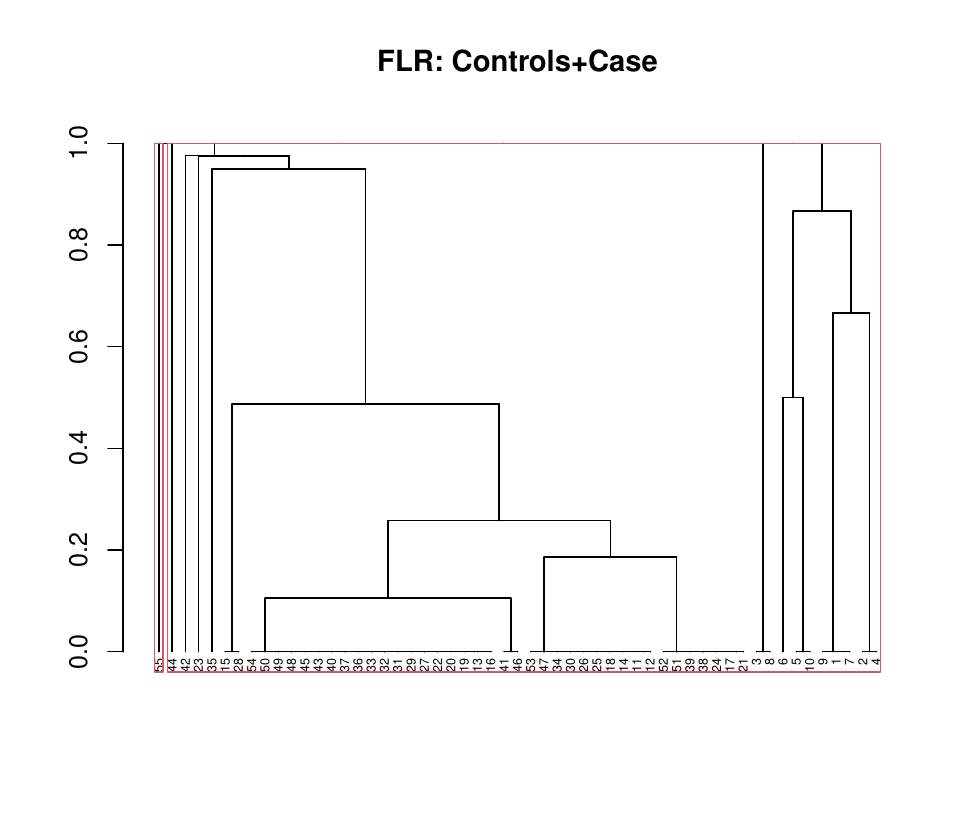}
    \end{subfigure}
 \begin{subfigure}[b]{0.5\textwidth}
      \caption{Setting 1.3}
\includegraphics[height=3.5in,width=\textwidth]{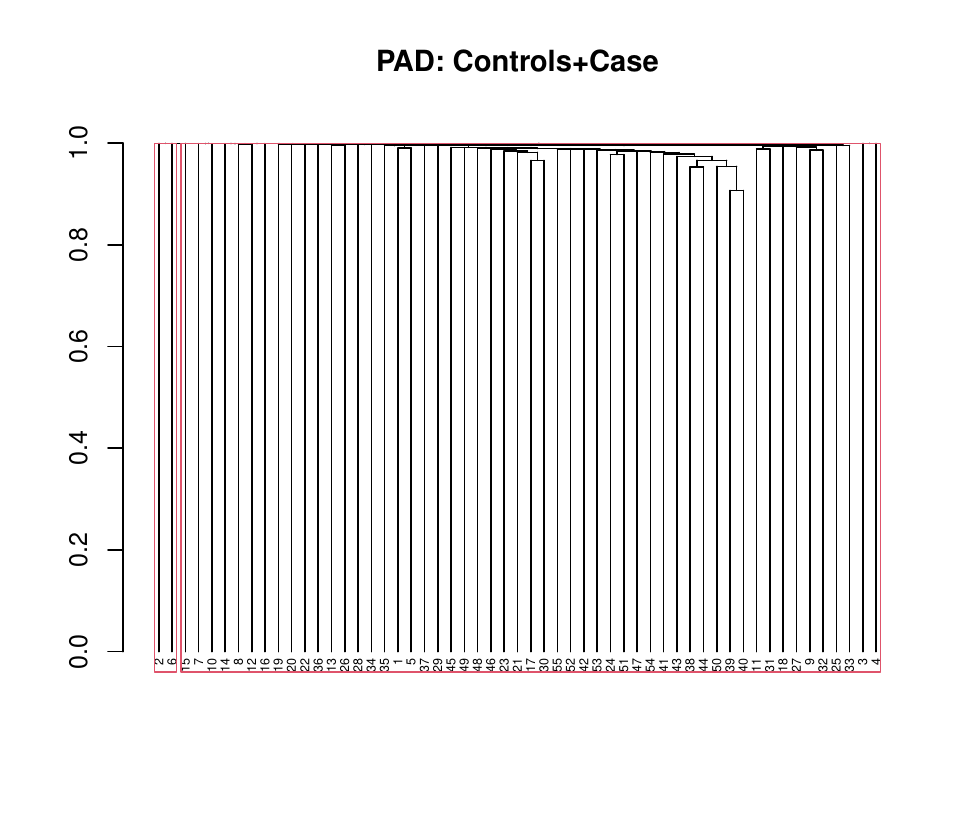}
    \end{subfigure}
\caption{\small{\noindent Examples of the FLR-HC dendrograms for the simulated Settings 1.1 and 1.3.  Red boxes indicate the cluster borders if we partition $55$ subjects into two clusters.
 }}
\label{FLR_PAD_HC1}
\end{figure}
\end{center}

\begin{center}
\begin{figure}[htp]
 \begin{subfigure}[b]{0.5\textwidth}
     \caption{Setting 1.3}
\includegraphics[height=3.5in,width=\textwidth]{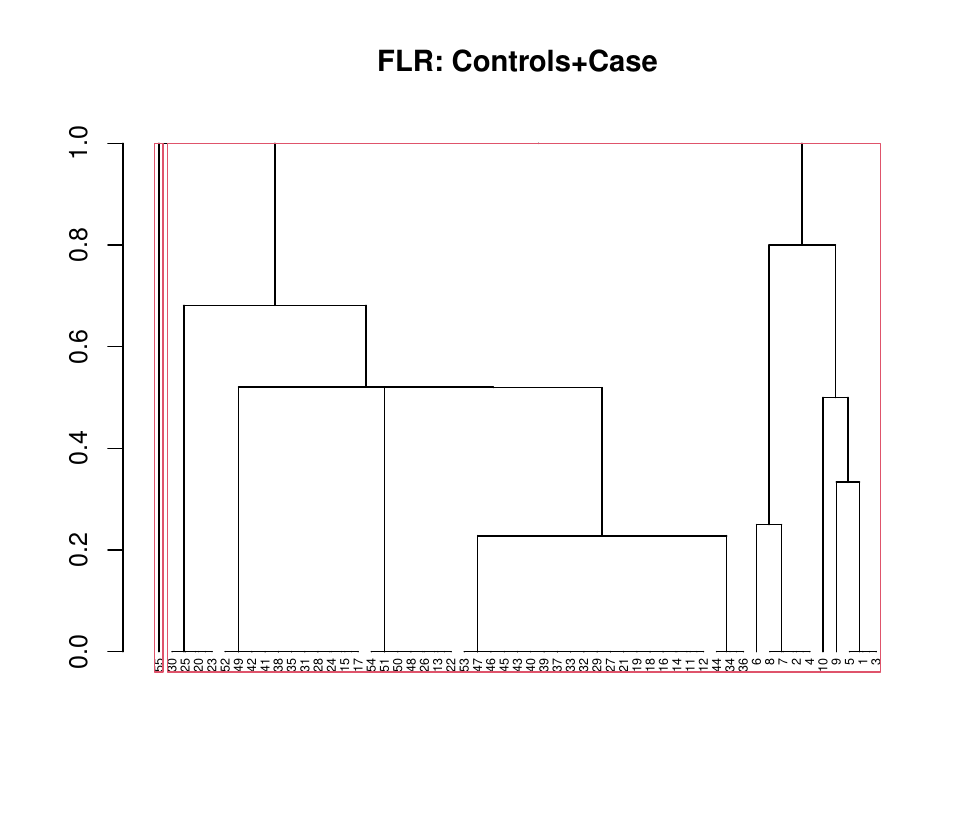}
    \end{subfigure}
 \begin{subfigure}[b]{0.5\textwidth}
      \caption{Setting 1.3}
\includegraphics[height=3.5in,width=\textwidth]{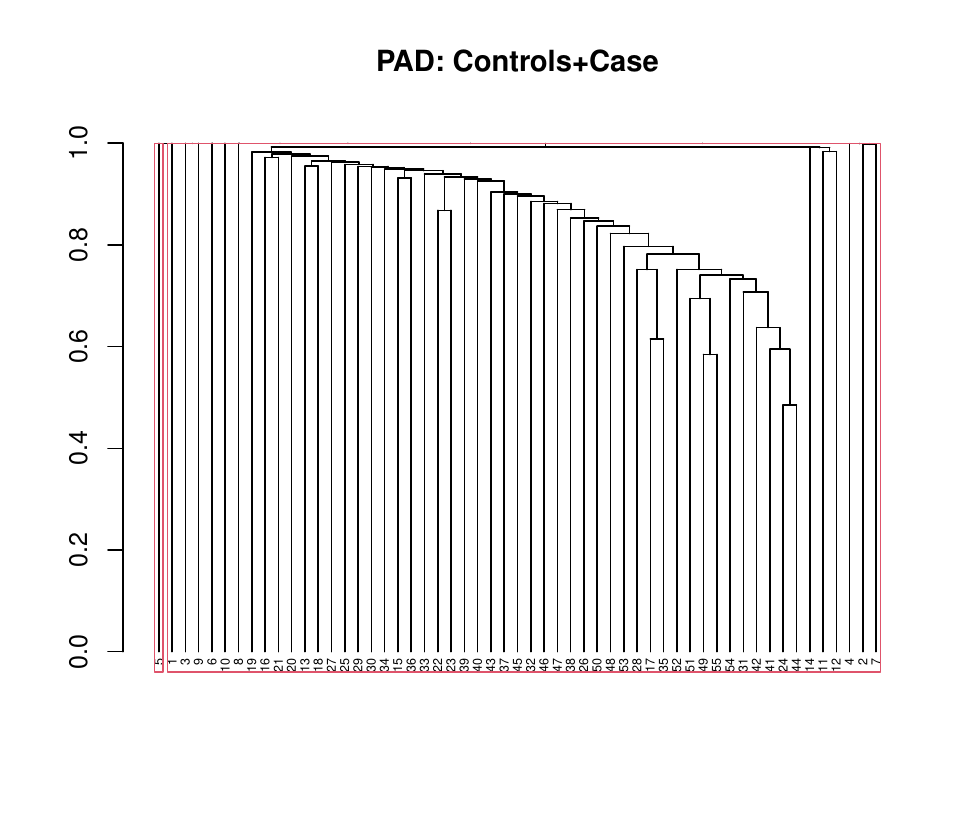}
    \end{subfigure}
 \begin{subfigure}[b]{0.5\textwidth}
      \caption{Setting 1.2}
\includegraphics[height=3.5in,width=\textwidth]{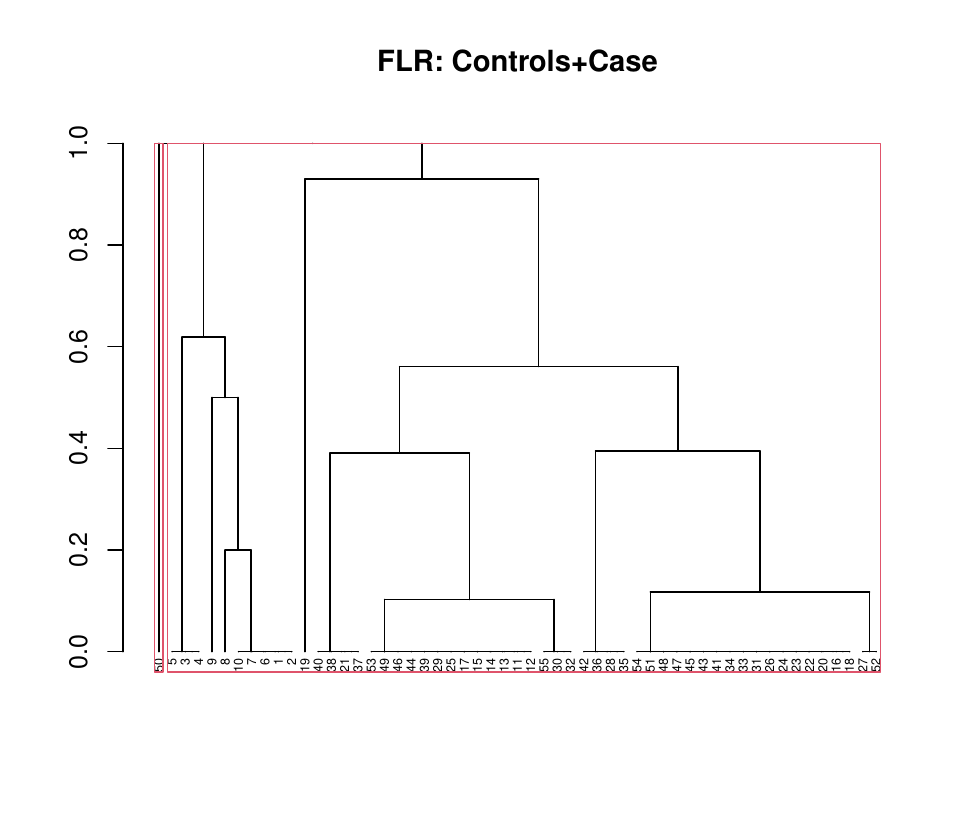}
    \end{subfigure}
 \begin{subfigure}[b]{0.5\textwidth}
      \caption{Setting 1.2}
\includegraphics[height=3.5in,width=\textwidth]{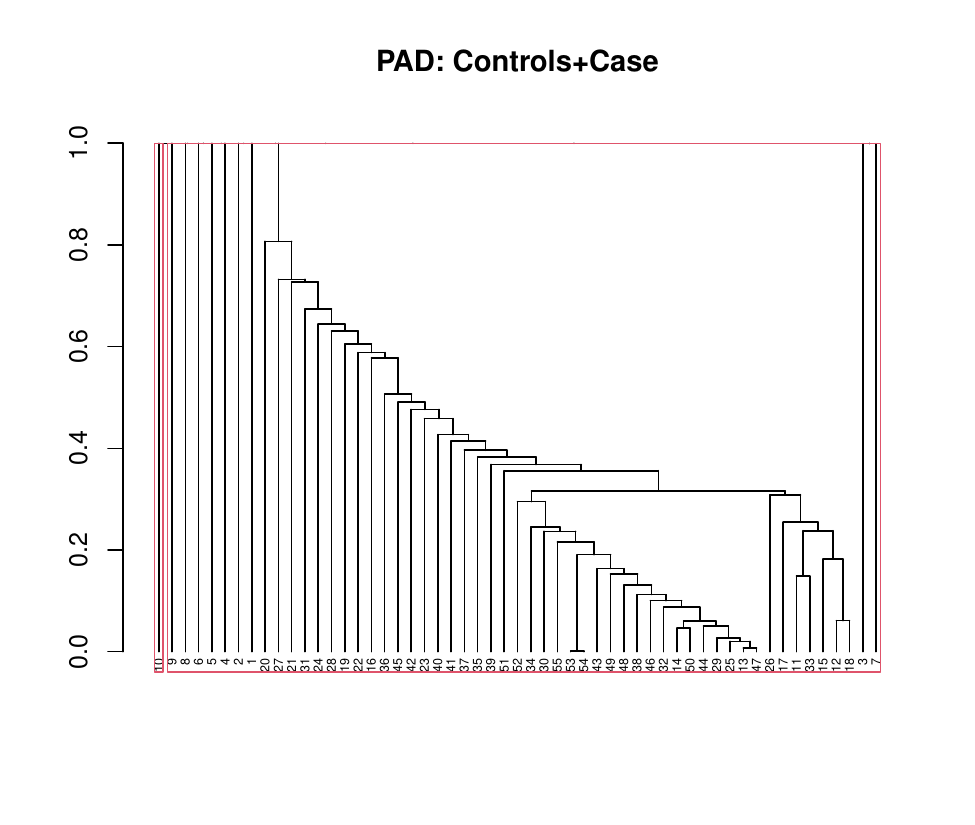}
    \end{subfigure}
\caption{\small{\noindent Examples of the FLR-HC dendrograms for the simulated Settings 1.2 and 1.3.  Red boxes indicate the cluster borders if we partition $55$ subjects into two clusters.
 }}
\label{FLR_PAD_HC2}
\end{figure}
\end{center}

In Figures \ref{FLR_PAD_HC1} and  \ref{FLR_PAD_HC2}, we demonstrate the abilities of the PAD-HC and the FLR-HC in separating the case from heterogeneous controls by using simulated Settings 1.1, 1.2 and 1.3, where there are two subgroups in controls. It is clearly shown that the FLR is more capable than the PAD to capture heterogeneity in controls and thus to distinguish the case from the controls.

\centerline { [Put Figures \ref{FLR_PAD_HC1} and \ref{FLR_PAD_HC2}  here.]}

\section* {Results on hierarchical clustering}
\begin{center}
\begin{figure}[htp]
 \begin{subfigure}[b]{0.5\textwidth}
     \caption{Delta}
\includegraphics[height=3.2in,width=\textwidth]{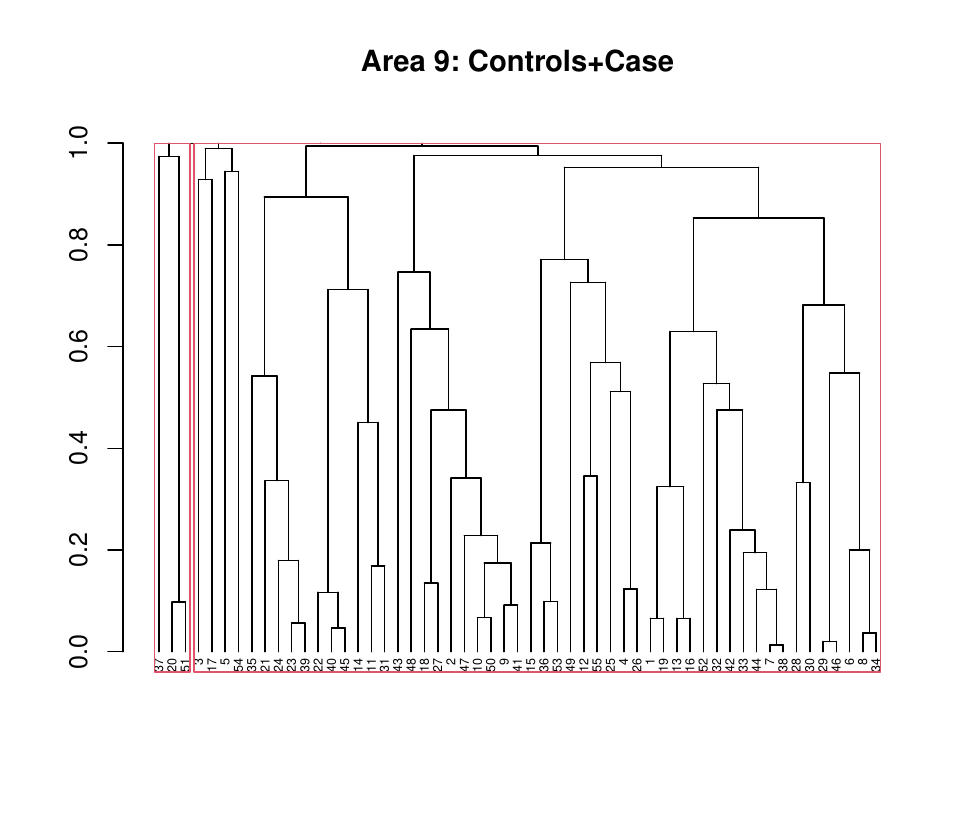}2
    \end{subfigure}
 \begin{subfigure}[b]{0.5\textwidth}
      \caption{Gamma}
\includegraphics[height=3.2in,width=\textwidth]{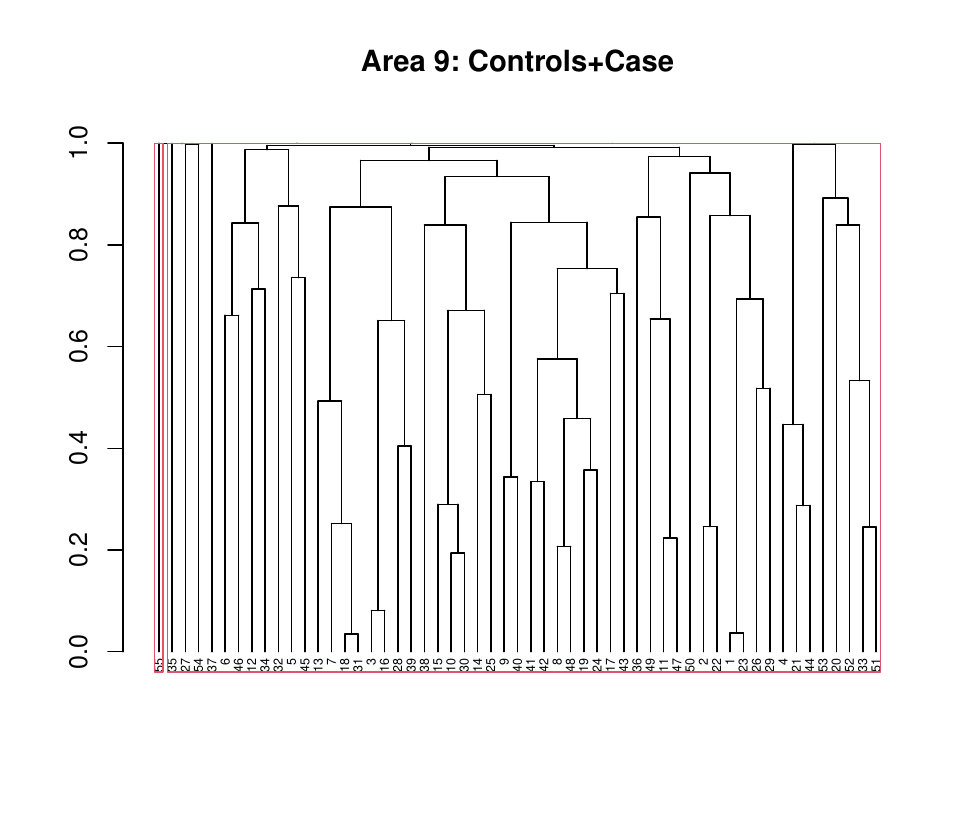}
    \end{subfigure}
 \begin{subfigure}[b]{0.5\textwidth}
      \caption{Delta}
\includegraphics[height=3.2in,width=\textwidth]{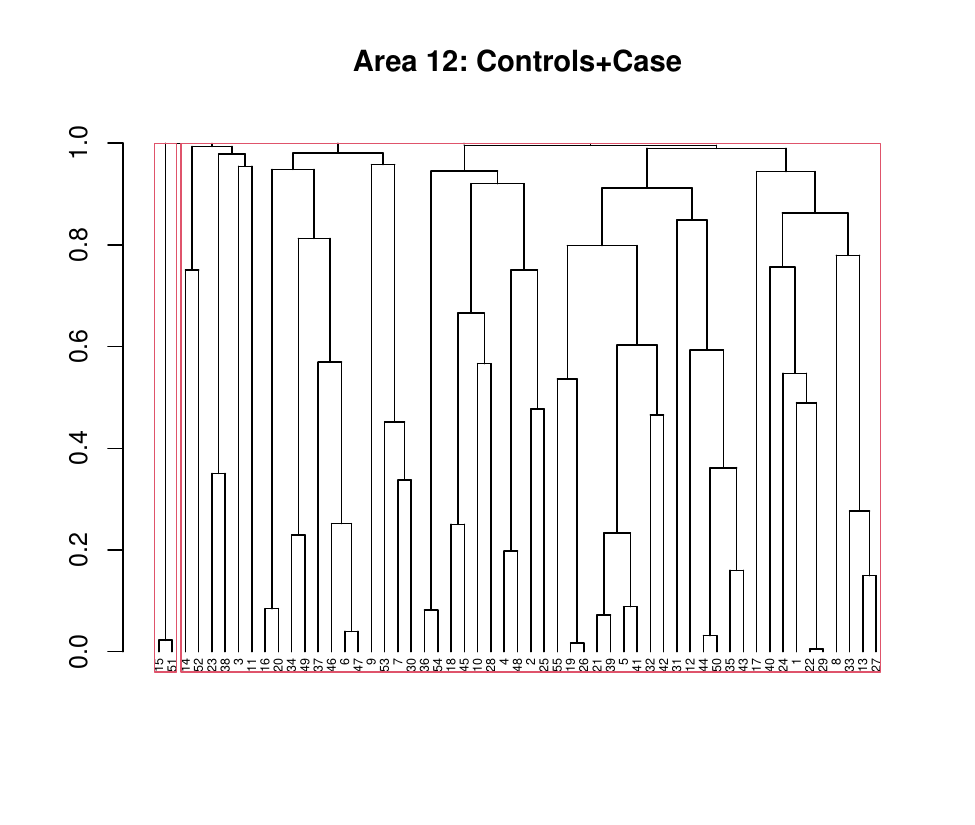}
    \end{subfigure}
 \begin{subfigure}[b]{0.5\textwidth}
      \caption{Gamma}
\includegraphics[height=3.2in,width=\textwidth]{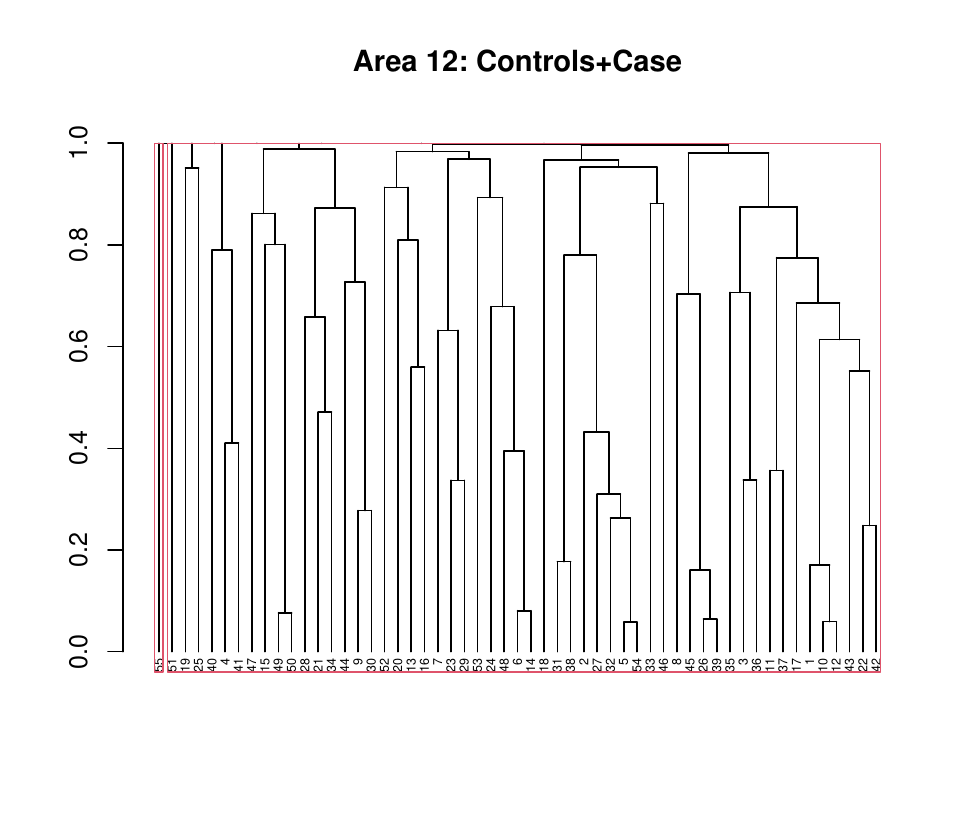}
    \end{subfigure}
\caption{\small{\noindent Average linkage hierarchical clustering of $55$ subjects using p-values derived from pairwise Anderson-Darling tests. The controls are indexed by $1$ to $54$ while the case is indexed by $55$. Rows 1 to 2 are respectively for areas 9 and 12.  Each row displays the resulting dendrograms for the delta and gamma bands from the left to the right. Red boxes indicate the cluster borders if we partition $55$ subjects into two clusters.
 }}
\label{ADhc}
\end{figure}
\end{center}

In the following,  in addition to Figures $8$ and $9$ in the main text,  we present the dendrograms for the abnormal areas claimed by the FLR and PAD in the delta- and gamma-band respectively. We select these areas, where the case-subject $55$ is the last one to be merged into the dendrogram, as the FLR-HC and PAD-HC adjusted areas. See Figures~\ref{ADhc}$\sim$\ref{FLRpairwise_gamma_suppl5} .

\centerline { [Put Figures \ref{ADhc}$\sim$\ref{FLRpairwise_gamma_suppl5}  here.]}

\newpage

\begin{center}
\begin{figure}[htp]
 \begin{subfigure}[b]{0.5\textwidth}
     \caption{Delta}
\includegraphics[height=3.5in,width=\textwidth]{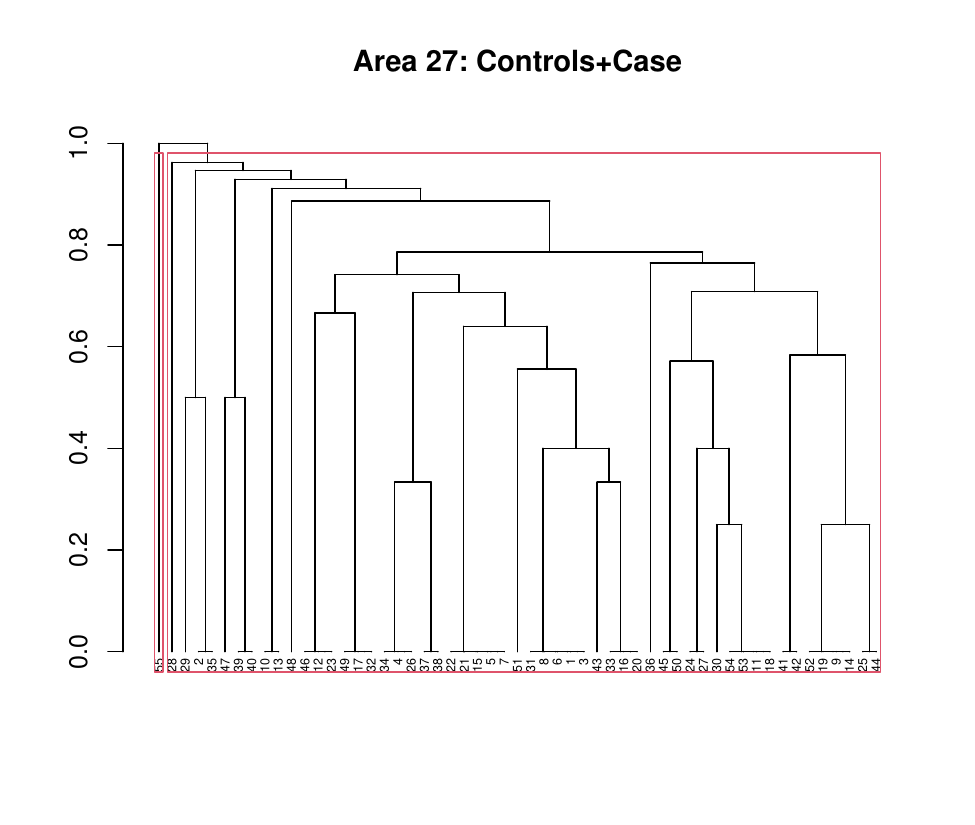}
    \end{subfigure}
 \begin{subfigure}[b]{0.5\textwidth}
      \caption{Delta}
\includegraphics[height=3.5in,width=\textwidth]{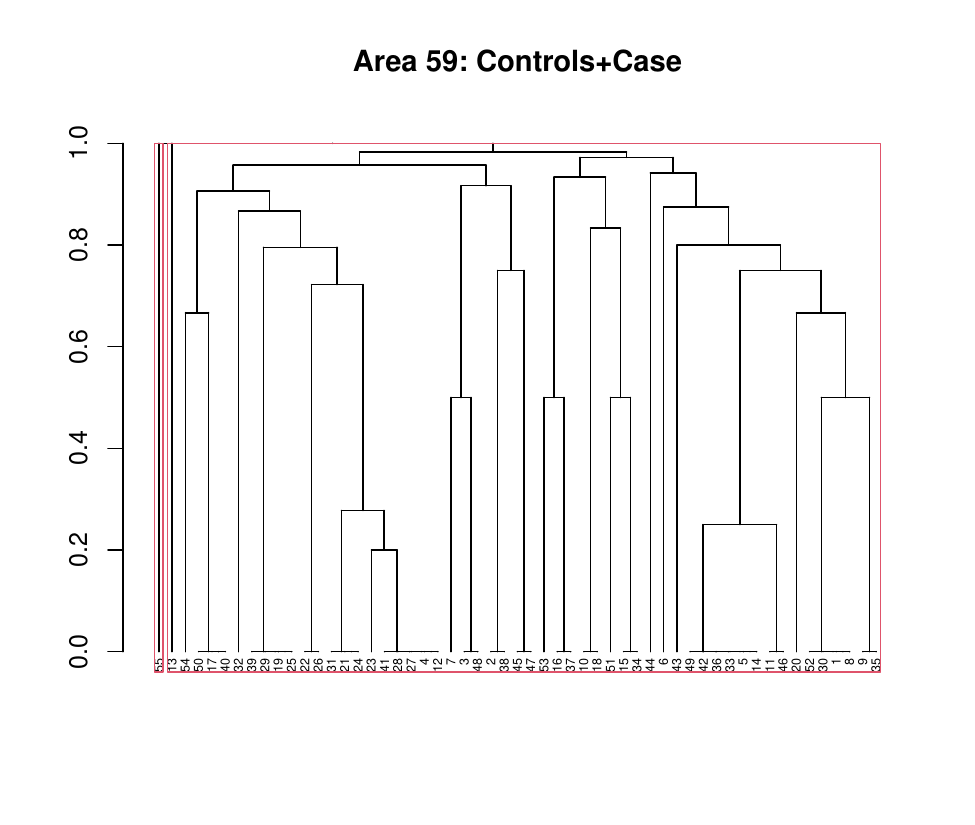}
    \end{subfigure}
 \begin{subfigure}[b]{0.5\textwidth}
      \caption{Gamma}
\includegraphics[height=3.5in,width=\textwidth]{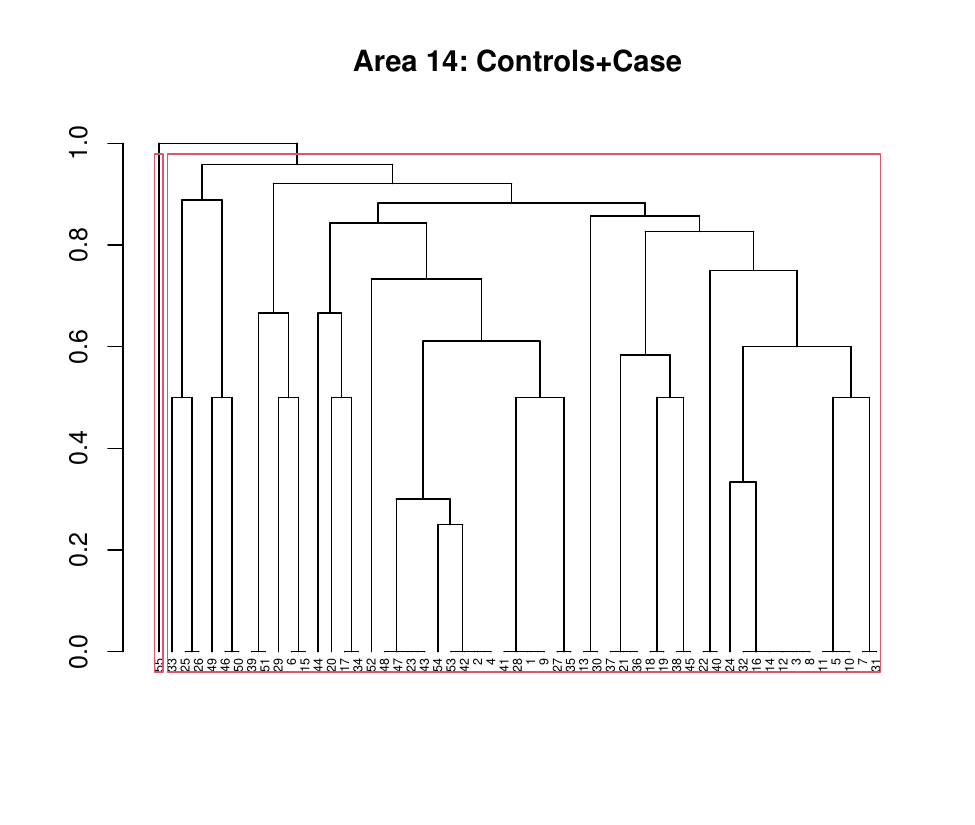}
    \end{subfigure}
 \begin{subfigure}[b]{0.5\textwidth}
      \caption{Gamma}
\includegraphics[height=3.5in,width=\textwidth]{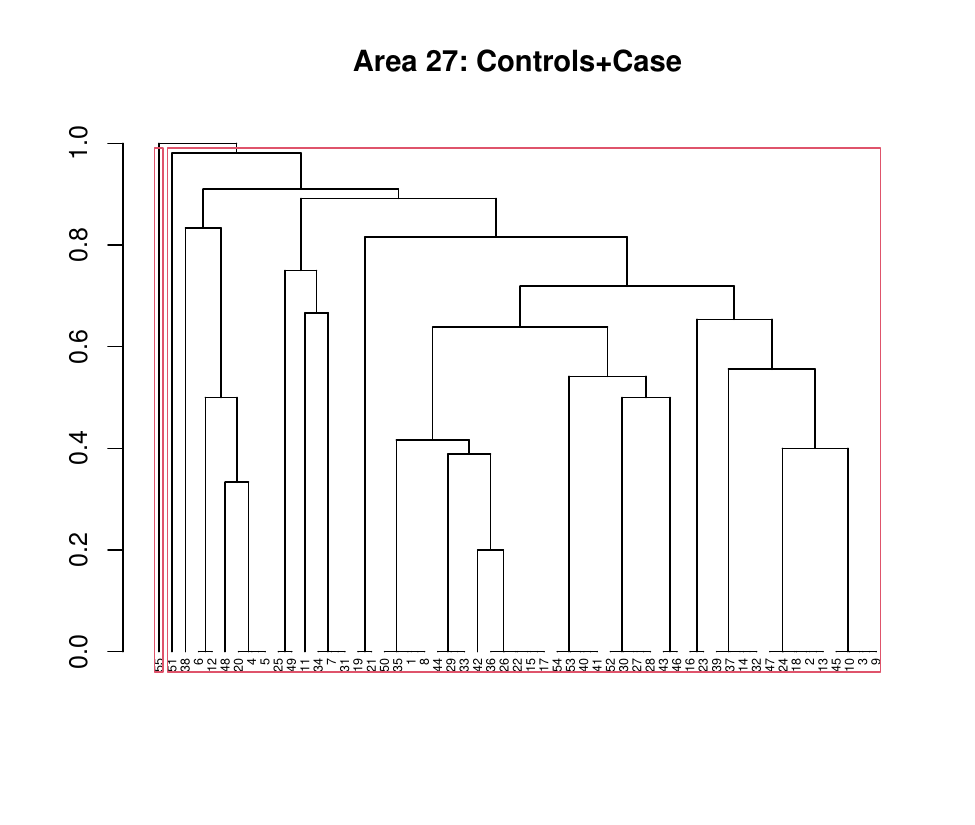}
    \end{subfigure}
\caption{\small{\noindent Examples of the FLR-HC dendrograms for the delta- and gamma- band data.  Red boxes indicate the cluster borders if we partition $55$ subjects into two clusters.
 }}
\label{FLRpairwise_deltagamma1}
\end{figure}
\end{center} 

\newpage

\begin{center}
\begin{figure}[htp]
 \begin{subfigure}[b]{0.5\textwidth}
     \caption{Delta}
\includegraphics[height=3.5in,width=\textwidth]{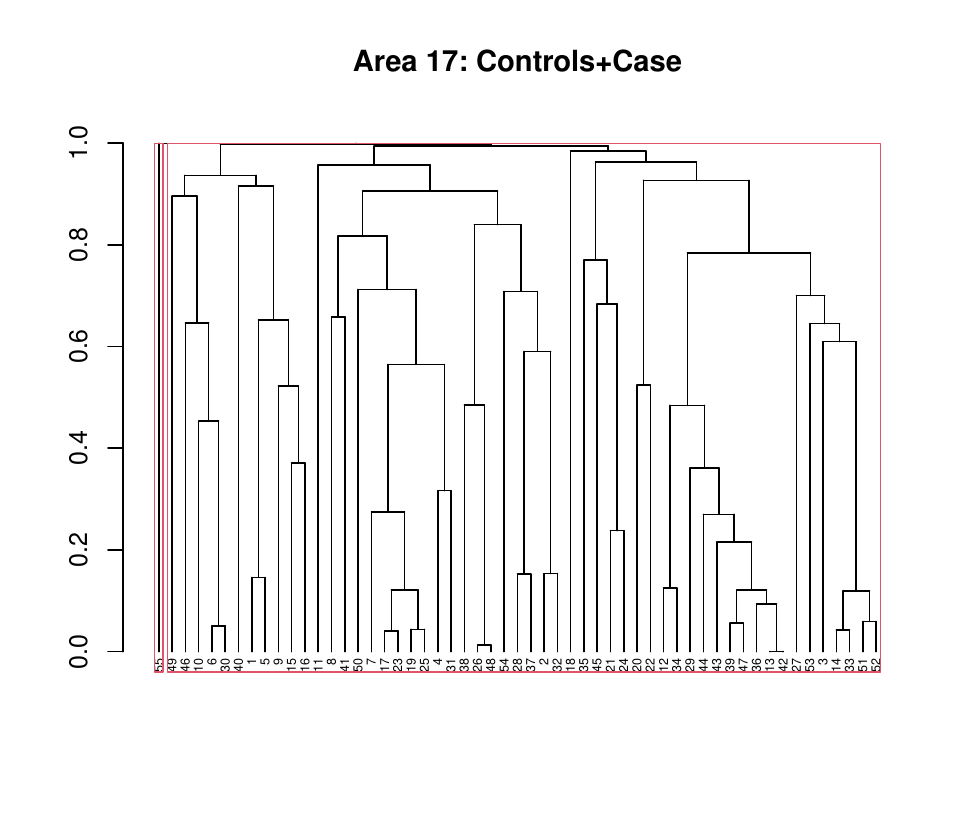}
    \end{subfigure}
 \begin{subfigure}[b]{0.5\textwidth}
      \caption{Delta}
\includegraphics[height=3.5in,width=\textwidth]{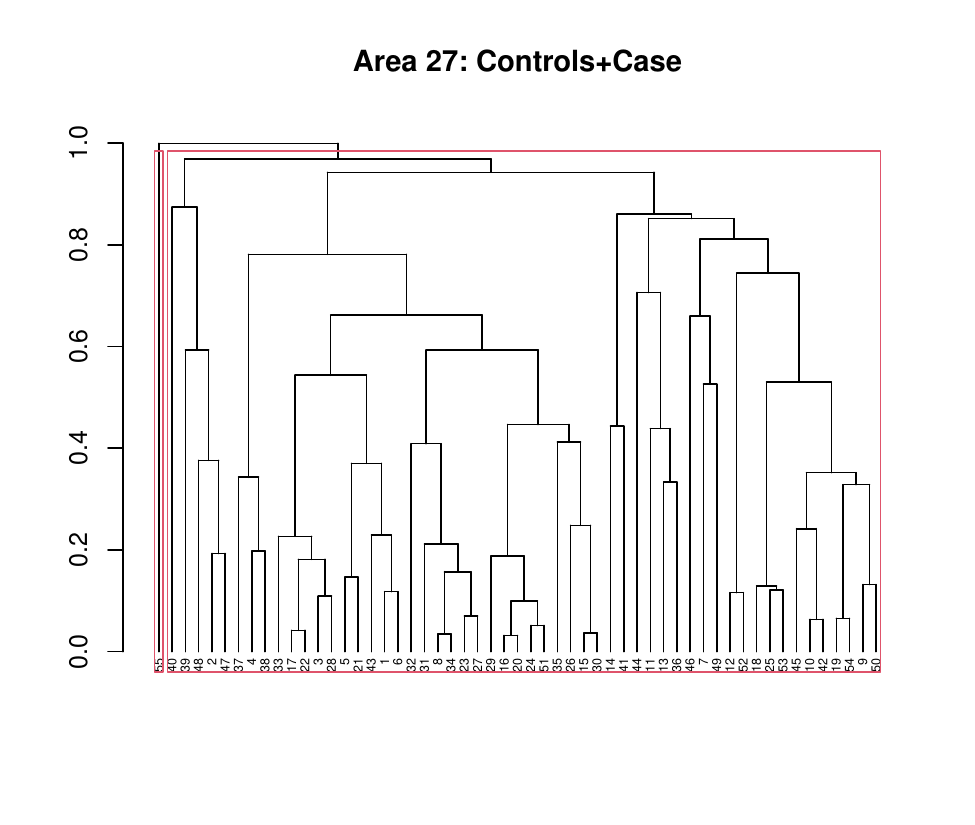}
    \end{subfigure}
 \begin{subfigure}[b]{0.5\textwidth}
      \caption{Gamma}
\includegraphics[height=3.5in,width=\textwidth]{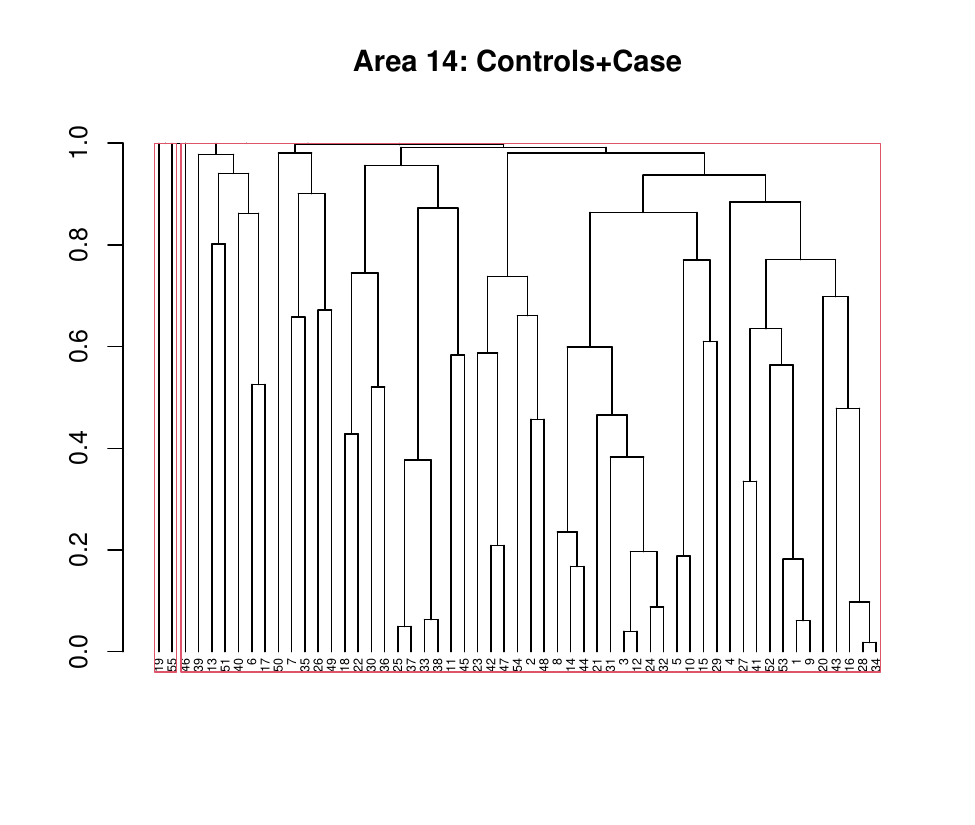}
    \end{subfigure}
 \begin{subfigure}[b]{0.5\textwidth}
      \caption{Gamma}
\includegraphics[height=3.5in,width=\textwidth]{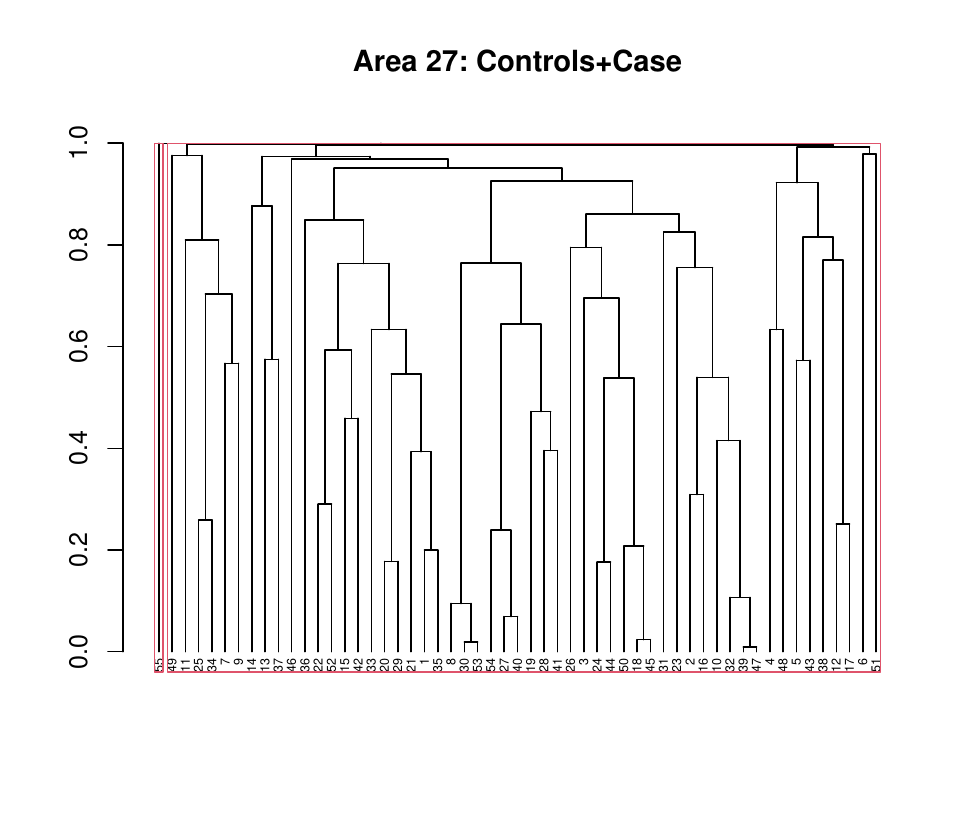}
    \end{subfigure}
\caption{\small{\noindent Examples of the PAD-HC dendrograms for the delta- and gamma- band data.  Red boxes indicate the cluster borders if we partition $55$ subjects into two clusters.
 }}
\label{PADpairwise_deltagamma1}
\end{figure}
\end{center} 

\newpage

\begin{center}
\begin{figure}[htp]
 \begin{subfigure}[b]{0.5\textwidth}
     \caption{Delta}
\includegraphics[height=3.3in,width=\textwidth]{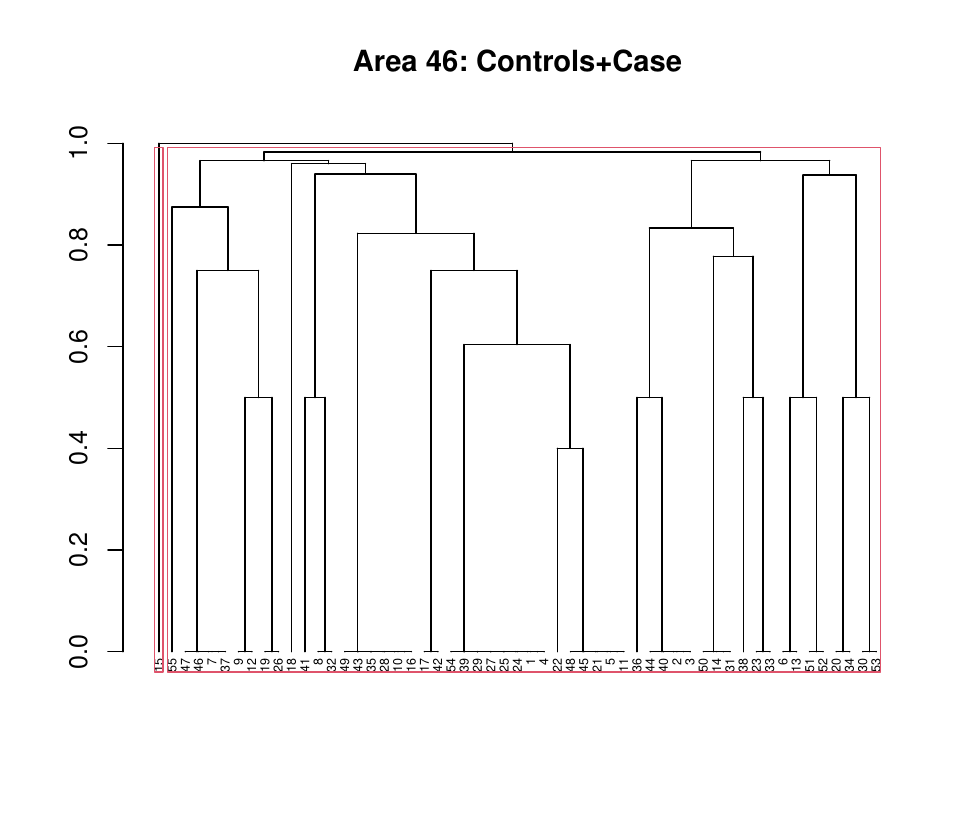}
    \end{subfigure}
 \begin{subfigure}[b]{0.5\textwidth}
      \caption{Delta}
\includegraphics[height=3.3in,width=\textwidth]{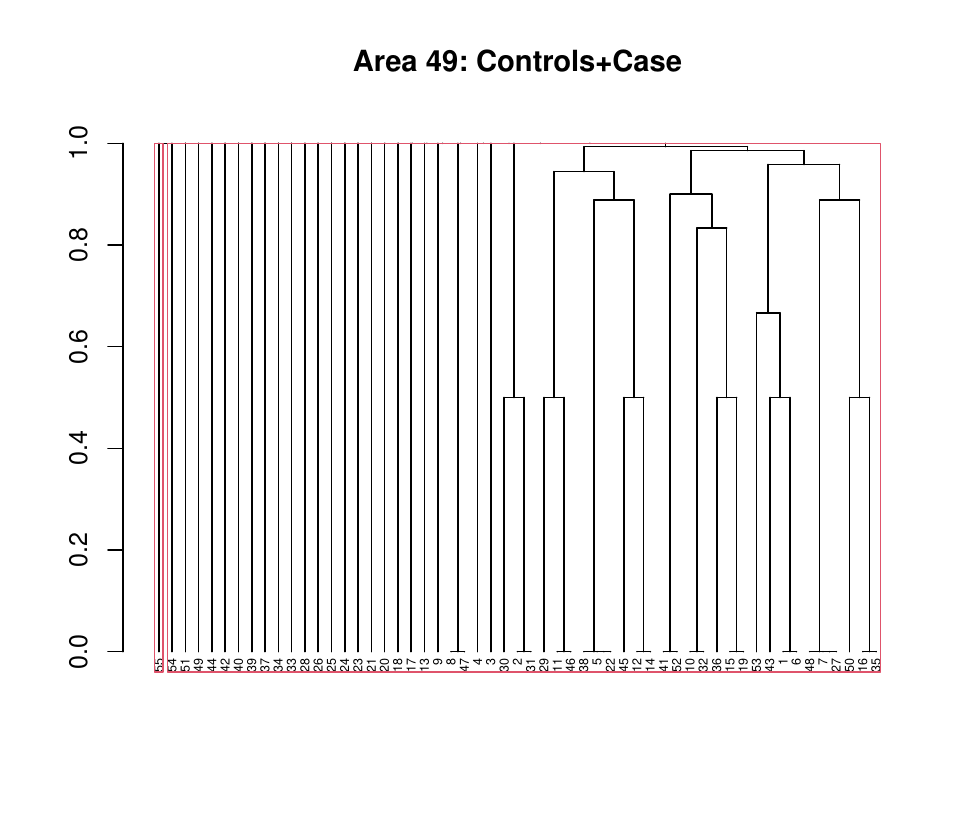}
    \end{subfigure}
 \begin{subfigure}[b]{0.5\textwidth}
      \caption{Delta}
\includegraphics[height=3.3in,width=\textwidth]{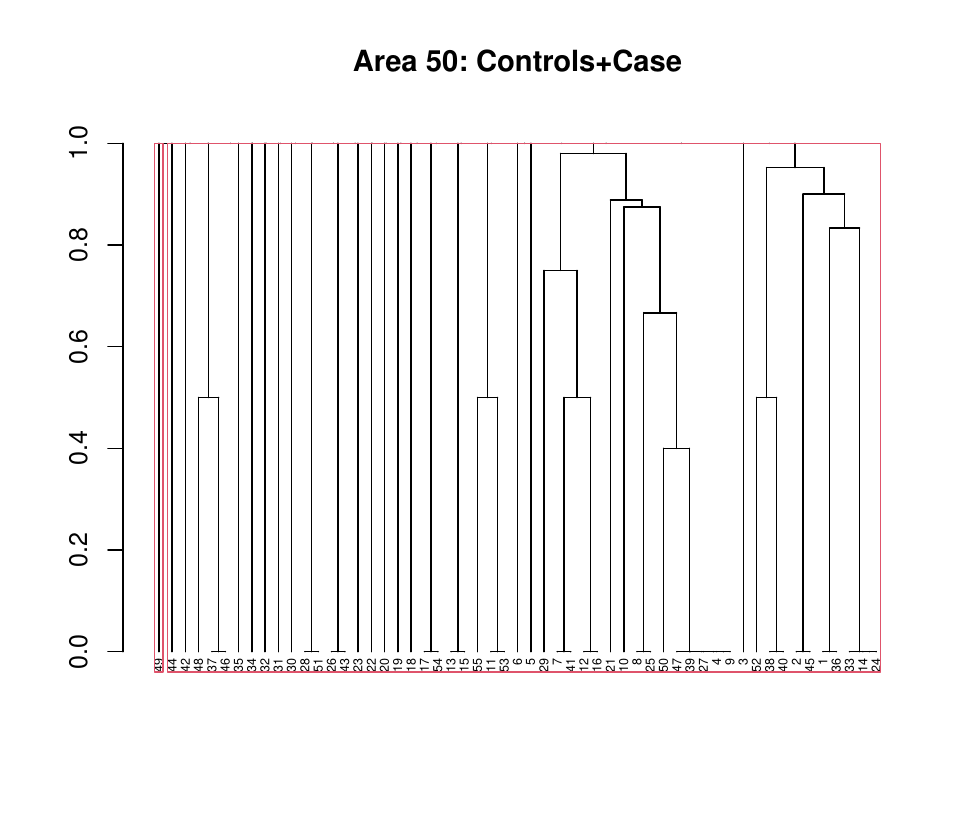}
    \end{subfigure}
 \begin{subfigure}[b]{0.5\textwidth}
      \caption{Delta}
\includegraphics[height=3.3in,width=\textwidth]{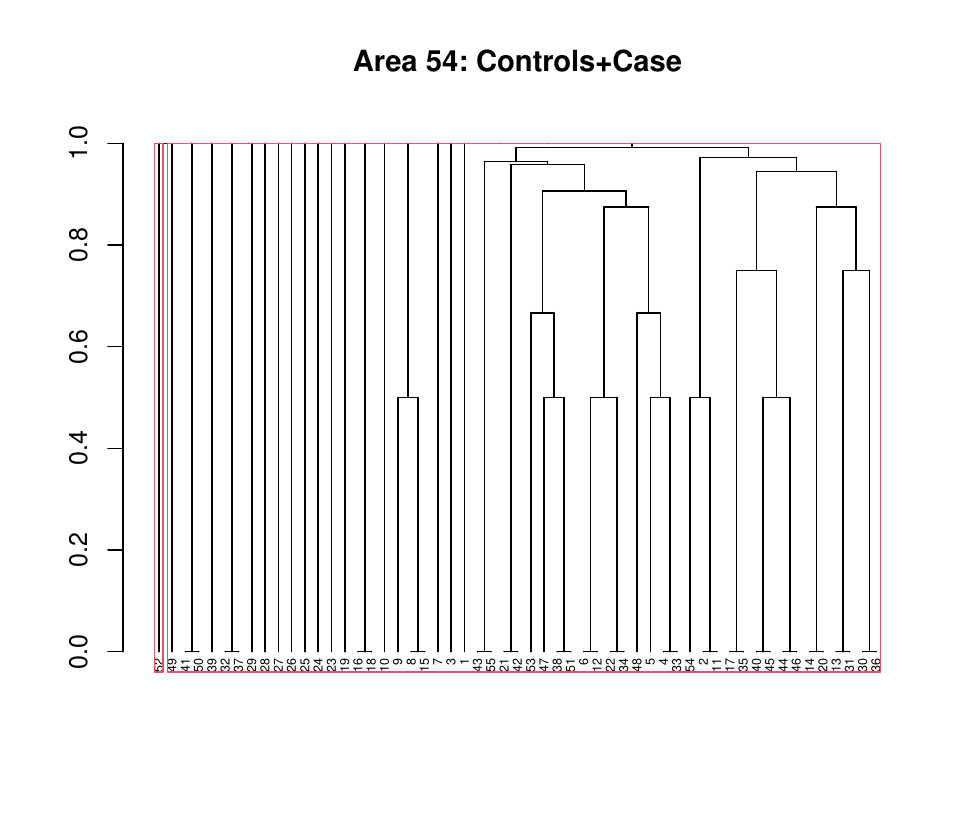}
    \end{subfigure}
\caption{\small{\noindent FLR-HC dendrograms for the areas identified as positive by the FLR in the delta band.  Red boxes indicate the cluster borders if we partition $55$ subjects into two clusters.
 }}
\label{FLRpairwise_delta_suppl3}
\end{figure}
\end{center} 

\newpage
\begin{center}
\begin{figure}[htp]
 \begin{subfigure}[b]{0.5\textwidth}
     \caption{Gamma}
\includegraphics[height=3.2in,width=\textwidth]{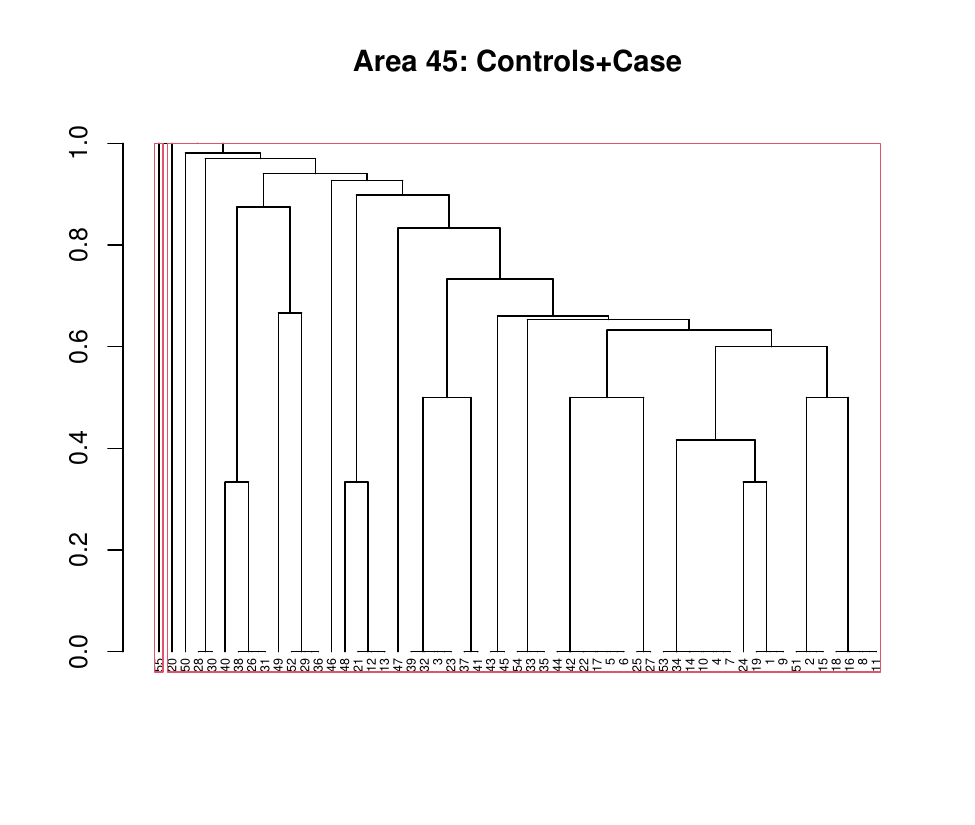}
    \end{subfigure}
 \begin{subfigure}[b]{0.5\textwidth}
      \caption{Gamma}
\includegraphics[height=3.2in,width=\textwidth]{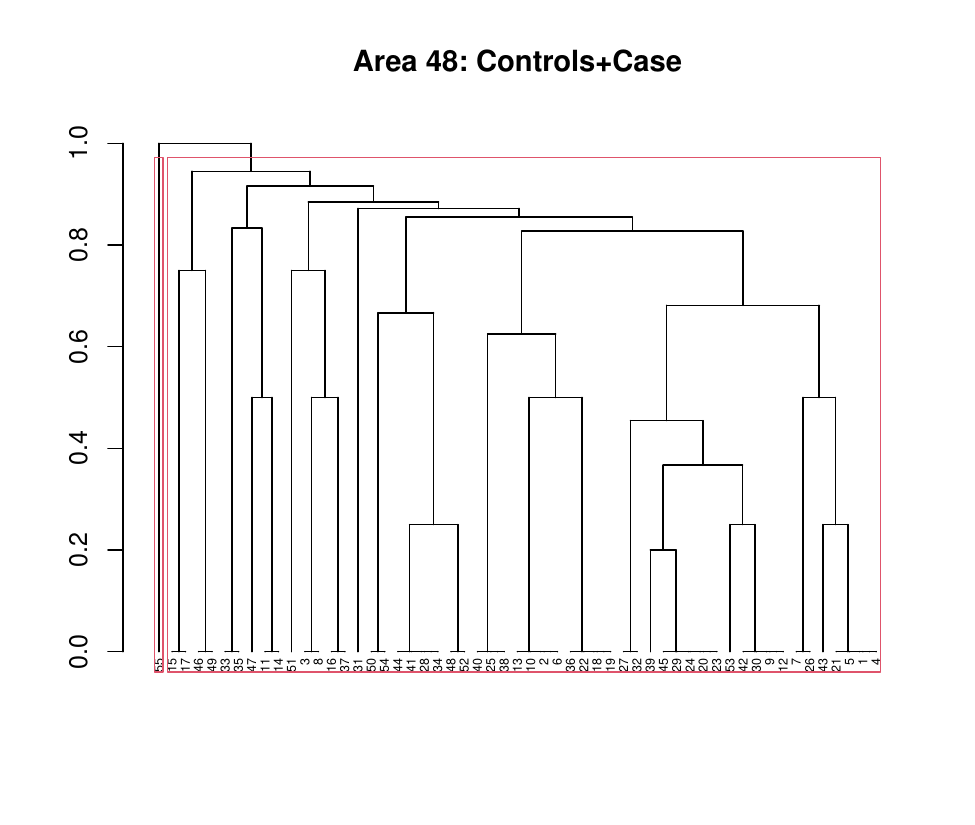}
    \end{subfigure}
 \begin{subfigure}[b]{0.5\textwidth}
      \caption{Gamma}
\includegraphics[height=3.2in,width=\textwidth]{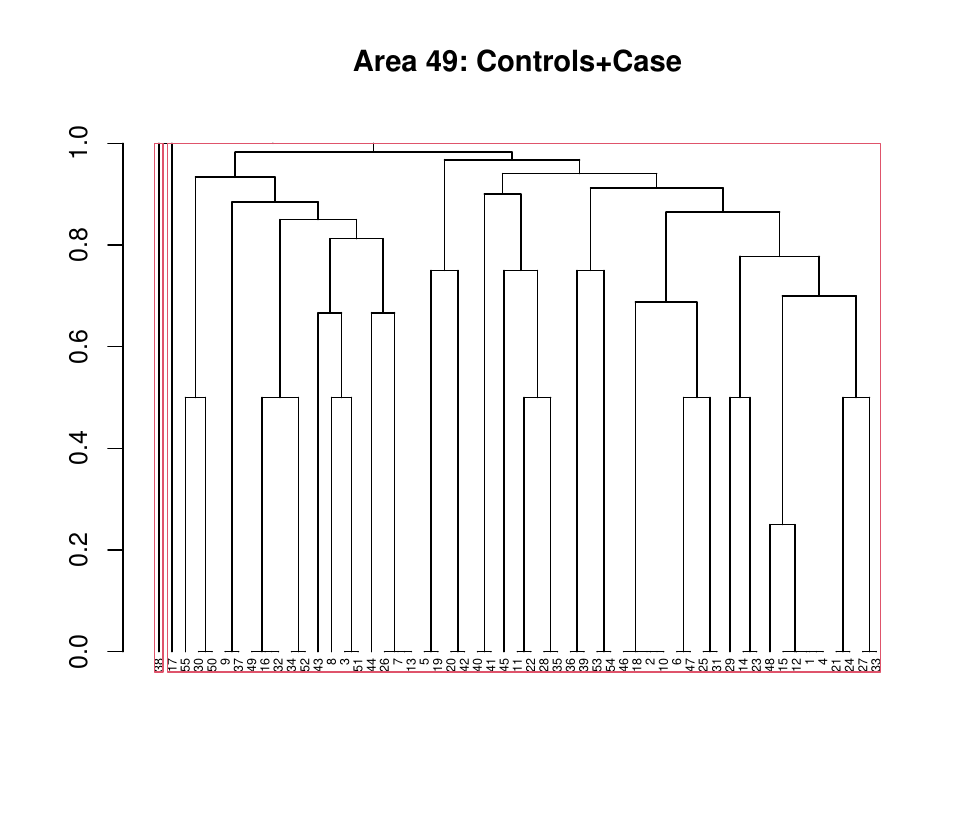}
    \end{subfigure}
 \begin{subfigure}[b]{0.5\textwidth}
      \caption{Gamma}
\includegraphics[height=3.2in,width=\textwidth]{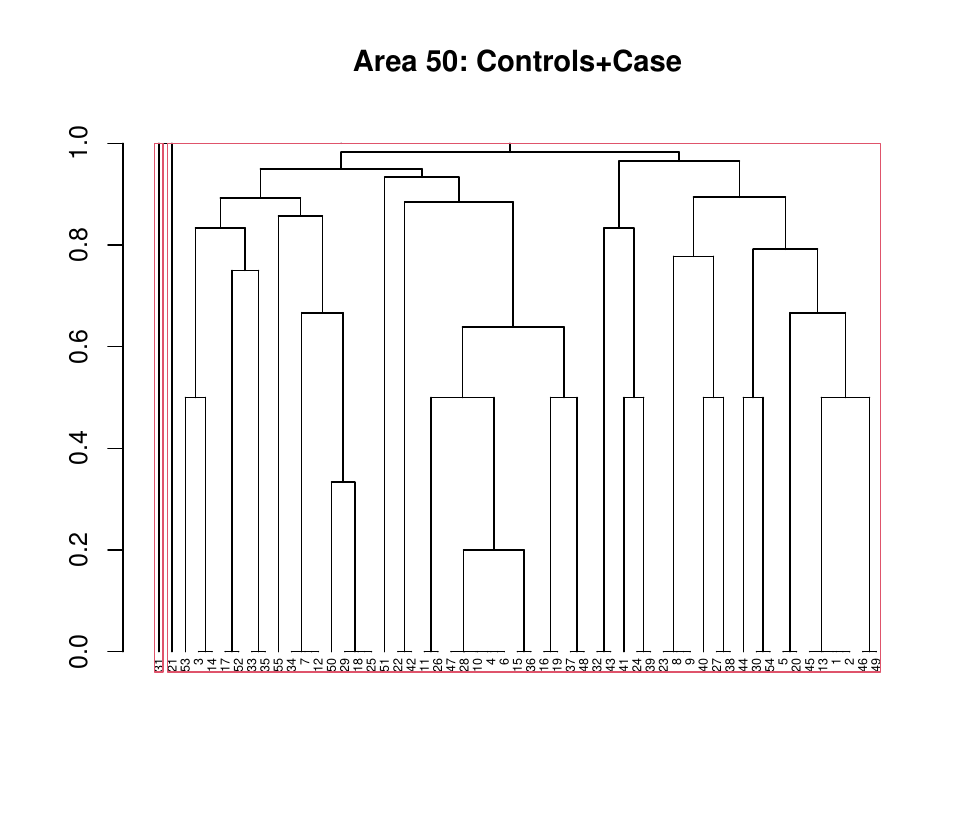}
    \end{subfigure}
\caption{\small{\noindent FLR-HC dendrograms for the areas identified as positive by the FLR in the gamma band.  Red boxes indicate the cluster borders if we partition $55$ subjects into two clusters.
 }}
\label{FLRpairwise_gamma_suppl5}
\end{figure}
\end{center}

\label{lastpage}


\end{document}